\documentstyle[prl,eqsecnum,aps]{revtex}

\begin{document}

\title{
 DECOHERENCE IN THE QUANTUM DYNAMICS OF A "CENTRAL SPIN" COUPLED TO A SPIN
ENVIRONMENT }

\author{N. V. Prokof'ev$^{1,2}$  and P. C. E. Stamp$^{2}$}
\address{
$^{1}$ Russian Science Center "Kurchatov Institute", Moscow 123182, Russia\\
$\;\;\;$ \\
$^{2}$ Physics Department, University of British Columbia, 6224 Agricultural
Rd.,\\
 Vancouver B.C., Canada V6T 1Z1 \\ }
\maketitle

\vspace{1cm}
\begin{abstract}
We consider here the problem of a "central spin", with spin quantum number $S
\gg 1$,
interacting with a set of microscopic spins. Interactions between
the microscopic spins are ignored. This model describes magnetic grains or
magnetic macromolecules (ferromagnetically or
antiferromagnetically ordered) interacting with nuclear spins and with any
surrounding
paramagnetic electronic spins. It has also been used to describe $Si:P$ near
the
metal-insulator transition, and quantum spin glasses.

We investigate this model in zero external field. We first set up the
general formalism required to analyse problems in which a macroscopic
degree of freedom (here, a central spin) interacts with a set of
microscopic spins. This is done by reducing the model to
an effective low-energy
Hamiltonian, and then calculating the correlation function $\langle {\vec S}
(t) {\vec S} (0) \rangle $  using instanton methods.
Three physical effects come into play; we call these "topological decoherence"
(coming from the
phase randomization by the spin environment), "orthogonality blocking" (coming
from mismatch between initial and final environmental states) and "degeneracy
blocking" (whereby the spin environment destroys degeneracy between the initial
and final central spin states). In this paper we consider the possible
coherent motion of the central spin. We find that all 3 of the above mechanisms
suppress coherence. Orthogonality blocking and topological decoherence
destroy phase coherence, and degeneracy blocking prevents most central spins
from tunneling at all. The general solution for $\langle \vec{S}(t)
\vec{S}(0) \rangle $ is given for the unbiased case, for all relevant values
of the couplings. Under certain conditions, nuclear spin diffusion also plays a
role.

The results are then used to calculate the spectral absorption $\chi^{\prime
\prime }
(\omega )$ for 2 experimental systems, viz., $TbFe_3$ grains, and ferritin
molecules.
For $TbFe_3$ (as for most systems), coherence is completely destroyed. However
ferritin
is very unusual, in that only the degeneracy blocking mechanism is effective -
consequently a resonance may still exist, but with a peculiar lineshape.

Our results have general implications for the observation of mesoscopic and
macroscopic quantum coherence, and for the foundations of quantum mechanics.
They show that a spin environment usually has a very destructive effect on
coherence,
which cannot be understood using the conventional "oscillator bath" models of
quantum environments.
\end{abstract}

\vspace{2cm}

\section{Introduction}
\label{sec:in}

In this paper we solve a model problem which, we believe, should be of
interest to both condensed matter physicists working on quantum spin problems,
and to physicists outside this field who are interested in the foundations of
quantum mechanics (particularly the "measurement problem", and the physics of
measurements and "decoherence"). It should also be of interest to physicists
and chemists working on quantum dissipation and models of the environment,
for one of the central points in what follows is the elucidation of the
properties of a new kind of quantum environment, quite different from the
usual "oscillator bath" models.

The model we discuss was originally introduced to deal with the effects of
environmental spins on the coherent tunneling of magnetic grain
magnetisation \cite{1,2,3}. It consists of a central "giant spin", having
spin quantum number $S \gg 1$, coupled to a set $\{ {\vec \sigma}_k \}$ of
"microscopic spins"; the microscopic spins are not coupled to each
other (Fig.1). The model is then completely described by a Hamiltonian
$H_o( {\vec S} )$ for the central spin, a Hamiltonian $H_{env} ( \{ {\vec
\sigma}_k \} ) $ for the microscopic spins, and a coupling
$H_{int} ( {\vec S} , \{ {\vec \sigma}_k \} )$ between the two. One can also
introduce
an external field ${\vec H}_o $ which couples to these spins - the effect of
external fields will be considered in detail in a $2nd$ paper.

In most practical applications of this "central spin model", the environmental
spins will be either nuclear spins (which may be either inside the object
carrying the giant spin, or outside it, in some substrate, or solvent, or
surrounding matrix) or else paramagnetic electronic "spin impurities", which
may also be inside or outside the giant spin. The central spin may be
a magnetic grain \cite{4} or a magnetic macromolecule such as ferritin \cite{5}
or, on a smaller scale \cite{6}, $ Mn_{12}O_{12}$; or it may be one of the
large superparamagnetic "spin clusters" which are believed
to exist in many disordered magnets at low temperature, such as
$Si:P$ near the metal-insulator transition \cite{7}, or "giant magnetic
polarons" \cite{8}. Similar spin clusters exist in "quantum spin glasses"
\cite{9}.

The crucial point which has come out of our study of this model,
and which we believe is of great importance for both quantum magnets and
for discussions of decoherence and the theory of measurement, is as follows.
The effect of microscopic spins on the quantum coherence of the central spin
is akin to the effect of an invasion of microscopic viruses on a host
bacterium, i.e., in most cases complete destruction! Environmental spins act
as a far more potent suppressor of quantum coherence than any other kind
of quantum environment hitherto discussed - in particular,
the conventional bosonic oscillator baths \cite{10,11,12} seem relatively
benign.

Now although  the spin
environment we consider constitutes an example of an "unconventional
environment"  ref.\cite{13}, a moment's thought shows that such spin
environments
will be the rule  rather than the exception, when considering
macroscopic or mesoscopic quantum coherence. This is because most large
quantum objects will have at least a coupling to their own nuclei. Thus in the
oft-cited example of the superconducting SQUID ring \cite{12}, there is
an electromagnetic coupling of the tunneling flux coordinate (and the
associated ring supercurrent) to all of the nuclear spins within a penetration
depth of the surface (either in the junction or in the ring), as well as
to any paramagnetic impurity spins that happen to be present \cite{14}.
While the effect of these interactions on {\it tunneling} may not be serious,
except under special circumstances \cite{15}, their effect on coherence
(MQC) will be quite drastic.

This means that our results are not only of practical importance for any
experiments which attempt to test quantum coherence on the mesoscopic
or macroscopic scale. They also show why coherence is so
easily destroyed at the mesoscopic or macroscopic level (in particular,
in measuring devices). This point is not obvious in the previous discussions
of, e.g.,  MQC in the conventional models, in which bosonic oscillators
models of the environment are used - in fact, it is predicted,
using such models, that MQC {\it ought to be visible} in SQUID rings
\cite{12,16} and in magnetic grains \cite{4,5}, using presently
achievable values of the {\it classical} dissipation in these systems.
Whilst no observation of MQC has yet been claimed in SQUIDs, it {\it has}
been claimed that MQC has been detected in samples of ferritin magnetic
macromolecules \cite{5,17}. Our central spin model can be directly applied
to this case - as we shall see here, the case of ferritin is very unusual, in
that
most of the decoherence mechanisms we shall discuss are inoperative. In fact
only one mechanism, which we call "degeneracy blocking", operates on ferritin,
leading to an unusual lineshape which is not obviously inconsistent with the
experiments
(the ferritin experiments have also been criticized on other grounds,
mostly related to the power absorption and the behaviour as a
function of field \cite{18}). In section \ref{sec:6} we will give
a quantitative discussion of spin environment effects on ferritin, and also
$TbFe_3$ grains.

Our general thesis is that the almost universal presence of spin
environments is going to make it far harder to see quantum coherence on the
macroscopic scale than has previously been realized. Even on
the mesoscopic scale it will be difficult. There are general
strategies that one may adopt in looking for suitable candidates for MQC
(or mesoscopic coherence). For example, it helps to choose a
{\it quantum soliton} as one's macroscopic quantum object \cite{19,20},
since quantum solitons, by definition, do not couple linearly to
other excitations of their own field. An example of current interest
concerns the quantum tunneling of macroscopic magnetic domain walls - recent
experiments have apparently discovered this in $Ni$ wires \cite{21}, and
it will be interesting to see how experiment compares with theory in this case.
It is also useful to make  the quantum object {\it neutral} - this
removes all electromagnetic couplings to the environment (although these are
often not important, since they are infra-red weak \cite{20}). However
the crucial step is clearly one of getting rid of any environmental spins -
this suggests at the very least a system for which the nuclear spin is zero.
There is in fact an almost ideal candidate \cite{14}, viz., superfluid
$^4$He. Here the macroscopic quantum object (whose size is defined by the
geometry of the container) is a vortex (a special kind of topological
quantum soliton). The system is neutral, and the nuclear spin is zero;
moreover, the system can be purified to a quite extraordinary degree \cite{22},
and the experimental cell walls coated with solid $^4$He to remove any
influence from impurities, bumps, etc. In fact the only couplings left are
 to phonons; in common with other quantum vortex problems, such as magnetic
vortices \cite{20}, this coupling has both a local and a non-local component.
However, just as with photons, that part of this coupling  which is
responsible for decoherence is infra-red weak (both of these couplings can be
described by a Caldeira-Leggett spectral function $J(\omega ) \sim \omega^3$).

In this paper we will not be able to discuss all of these issues, concerning
the physics of spin environments. In particular, we intend to discuss elsewhere
some of the consequences for both mesoscopic quantum devices and quantum
coherence experiments. These applications of the theory are at the forefront
of present attempts by physicists to push the "F.A.P.P. barrier", between
the quantum world and the classical world \cite{23}, ever further into the
mesoscopic and even macroscopic domain. The implications for the measurement
problem will also be dealt with elsewhere.

In what follows we begin (in section \ref{sec:2}) by describing the
central spin model, and giving some discussion of the relevant physical
coupling energies in it, with the aid of physical examples (ranging
from magnetic grains to "spin clusters" in Si:P).
In section \ref{sec:3} we show how an initial microscopic Hamiltonian,
describing the giant central spin and its microscopic environmental
satellites, can be reduced to a low-energy effective Hamiltonian description
\cite{2}. The particular way this reduction is done, and the terms that are
left at the end, is partly a matter of theoretical choice. We have done
it in a way which displays most conveniently the important physical processes
that operate at low energies. To help the reader who finds this reduction
a little abstract, we sketch how this reduction is done in practice
for a real system.

In section \ref{sec:4} we proceed to the discussion of 3 solvable
limits of our model Hamiltonian \cite{2,3}. Each of these limits brings
out a different physical effect. The simplest of these to
understand is "topological decoherence", in which the {\it topological
phase} of the spin environment becomes entangled with the coherent phase
of the giant spin, and destroys the phase coherence of the latter. This
decoherence mechanism is completely novel, and has many amusing
consequencies - it leads to decoherence without dissipation or energy exchange
of any kind, and the decay of the correlation functions of the central spin,
caused by it, is most unusual \cite{2,3}. The second effect \cite{3} is of
"orthogonality blocking"; it arises when there is a mis-match
between the final-state spin wave-functions in the environment, and their
initial states. Although this effect is reminiscent of  the
"orthogonality catastrophe" first discussed by Anderson \cite{24}, which is at
the heart of decoherence in the conventional oscillator bath models \cite{12},
it differs in important details (for example, instead of exponential
suppression of tunneling rates, one gets power low suppression; and the
effect on the correlation functions of the central spin ${\vec S}$ is most
unusual). Finally we study the limit appropriate to pure "degeneracy
blocking" \cite{2,3}, which arises because the coupling between the central
spin and the environmental spins lifts that degeneracy, between initial
and final states of the central spin, which is essential for coherence.

In this paper we will largely concentrate on the coherence properties
of the central spin, rather then its relaxation from any particular state.
As we shall see, this usually requires consideration only of processes in
which the total polarization of the environmental spin bath is unchanged
when $\vec{S}$ flips. Other processes necessarily destroy coherence.

Having thoroughly understood the 3 limits, we proceed in section \ref{sec:5}
to the discussion of the "generic case", i.e., having  arbitrary
values of the parameters in the effective Hamiltonian, so that all 3
mechanisms are operating.  At this point we
also include the weak nuclear spin diffusion effects, which can also play
a role in decoherence.

In section \ref{sec:6}, we return to two of the
physical examples described in section \ref{sec:3}.C (i.e., $TbFe_3$ and
ferritin), and show how their
behaviour is determined by the solutions derived in sections \ref{sec:4} and
\ref{sec:5}. In the case of $TbFe_3$, which is fairly typical, we
find a catastrophic destruction of coherence - no trace of any resonance is
left.
On the other hand ferritin has very few nuclear spins, and one finds that only
degeneracy blocking play any role in the final behaviour. As a result there is
a significant reduction in the spectral weight of the resonance, but the
lineshape
still shows a strong but highly asymmetric peak.
 Finally in section \ref{sec:7}, we
summarize our results.

An important omission from this paper is the effect of an external field.
This field will strongly
affect  the giant spin dynamics (with usually a much lesser affect
on the microscopic spins). Moreover,
dissipation can play an essential role in the problem, whereas in the unbiased
case we
consider, decoherence arises without dissipation. A full discussion
of the relaxation of $\langle {\vec S} (t) {\vec S} (0) \rangle $
(as opposed to coherence) also demands that we consider incoherent transitions
involving a change in the spin bath polarization - moreover, since the
relaxation of both the central spin and the spin bath is determined  by
coupling to phonons and possibly electrons, these must also be included.
 In view of the experimental importance of the field
effects, we have reserved discussion of them to a second paper.

For practical applications of our results, it is important to note that
there will also be an influence on coherence coming from
 phonons or electrons. These can interfere in interesting ways with the
effect of the environmental spins \cite{25}. We hope to address this aspect
in more detail at a later time.

Many of the results appearing in sections \ref{sec:2},\ref{sec:3}
 and \ref{sec:4} of the present paper
have appeared previously in short communications \cite{1,2,3,26}. Here we give
detailed derivations for the first time. The results in section
\ref{sec:5} (the generic case) are new here, as is the detailed analysis
for the various applications. Some of the mathematical derivations
are quite lengthy, and are relegated to Appendices.

\section{The model, and some simple results}
\label{sec:2}

In this section we give a physical discussion of the central spin problem,
with reference to a number of different examples, including magnetic
grains, "spin clusters", etc. The various physical interactions in these
problems are described and quantified.

\subsection{The central spin}

A typical central spin is made up of a very large number of microscopic spins,
whose motion is locked together by nearest-neighbour exchange couplings (or, in
the case of superparamagnetic spin complexes, by indirect spin-spin
couplings such as the RKKY interaction). The exchange coupling can be very
large, as much as $1-2eV$; this is much larger than other individual spin
energy scales such as those coming from anisotropy, so that in principle
one can have large monodomain magnetic particles, with many spins (up to
perhaps $10^8$, depending on the particular systems) lined up to form
a "giant spin". Another more subtle example is the antiferromagnetic
grain, in which the sign of the nearest-neighbour exchange is {\it positive},
so that nearest neighbours are anti-parallel, and one obtains a "giant
N\'{e}el vector" \cite{27}. The superparamagnetic spin complexes are more
delicate still, and typically arise at low temperatures when paramagnetic
electronic spins in a disordered magnet "lock together" via indirect
exchange, RKKY interactions, etc.

Now it is clear that a realistic Hamiltonian for such a giant spin will
be very complex. Let us consider first the simplest case of a
ferromagnetic grain. An enormous literature exists on the properties of such
grains, ranging from simple elemental grains, such as $Ni$, $Fe$, $Tb$,
or $Ho$, to more complex magnetic compounds such as $FeCo_5$, $Fe_2O_3$
($\gamma$-haematite), $Mn_{12}O_{12}$ and its derivatives, right up to
organic magnetic molecules. With present techniques it is possible to
prepare many grains of very uniform size \cite{28} (or else rely on
nature for the preparation of magnetic macromolecules).

An obvious first step in analysing such a system is to treat the entire
spin complex as a rigid quantum rotator ${\vec S}$, with
dynamics governed by the single-spin anisotropy field in the
particle/grain; i.e., we write a central spin "bare Hamiltonian"
$H_o( {\vec S} )$, with $\mid {\vec S} \mid = S$ a constant. Early studies
of the quantum dynamics of this model were stymied by the constraint of
constant $\mid {\vec S} \mid$, which implies a kinetic term in the Lagrangian
which
is linear in time derivatives - this leads to well-known difficulties
when one tries to apply path-integral methods to the problem \cite{29}.
A number of crucial advances allowed the solution of the problem,
most particularly the proper treatment of the kinetic term, which
leads to the "Haldane phase" \cite{30}, and also  a correct
formulation of WKB theory for the tunneling dynamics of ${\vec S}$, which
yields
the correct semi-classical solution for tunneling of an angular momentum
through the barrier \cite{31}. This semi-classical solution was also later
derived using the instanton calculus \cite{4,32} (although there still
seem to be differences between the 2 methods, concerning the tunneling
prefactor \cite{31}).

As a simple example of the bare Hamiltonian $H_o( {\vec S} )$ we may take
(we write ${\vec S} =S {\vec s}$, where $S = \mid {\vec S} \mid $)
\begin{equation}
H_o( {\vec S} ) = -K_{\parallel}\:s_z^2 + K_{\perp}s_y^2 \;,
\label{1}
\end{equation}
which has two degenerate classical minima at $S_z =\pm S$; the
semi-classical paths between these 2 minima move on or near the easy
XZ-plane (note that $K_{\parallel}$ and $K_{\perp}$ are proportional to $S$).
Notice that if $K_{\perp} =0$, there is no tunneling because
$S_z$ is conserved; this point is obviously connected with the fact that
we are dealing with angular momenta and not particles, and caused some
confusion in the literature before 1986. Application of a magnetic field
adds a term $-{\vec M} \cdot {\vec H}_o = -\gamma {\vec S} \cdot {\vec H}_o $
to (\ref{1});
if ${\vec H}_o$ is directed along ${\hat {\vec z}}$, this {\it biases} the
symmetric 2-well problem in (\ref{1}), but if it is applied along
${\hat {\vec x}}$, it lowers the barrier height and displaces the degenerate
minima towards each other in the XZ-plane. Application of ${\vec H}_o$ along
${\hat {\vec y}}$ distorts the semi-classical path and lowers the barrier.
A further effect of an applied  ${\vec H}_o$ is mostly nicely seen in the
instanton language; since the 2 possible paths between the degenerate minima
involve opposite Haldane topological phase \cite{33}, this phase can
be changed by an external field \cite{34}, causing the tunneling splitting
to change and even oscillate.

In fact the original papers of van Hemmen and Suto analyzed a very general
class of central spin bare Hamiltonian, given by
\begin{equation}
H_o( {\vec S} ) = -K_{\parallel}\mid s_z \mid^l -1/2 \sum_{r=1}^{n} \:
^{\perp}K_r\:(s^r_+ +s^r_- ) \;,
\label{2}
\end{equation}
and a considerable number of more recent papers have concerned themselves
with re-analyzing special cases of (\ref{2}).

Consider now what has been left out of (\ref{1}) and (\ref{2}). One obvious
omission is the infra-red weak coupling of ${\vec S}$ to photons (via the
magnetic
dipolar interaction \cite{19,20})   and to phonons. Another omission
is coupling to higher excited states of the particle, involving spin flips
of spins inside the grain (internal magnons); the neglect of these is usually
justified  by appeal to the large value $\mid J \mid$ of spin exchange
\cite{4,20}, in models like equation (\ref{1}). Incorporation of such
processes implies relaxation of the constraint  $\mid {\vec S} \mid$=constant.
 Another omission is that of
surface excitations of the grain - in any real grain, there can be
low-energy magnetic excitations involving "loose spins", i.e., spins
which, because of imperfect preparation or the inevitable defects,
dislocations, etc., have a coupling energy $\ll J$ to their neighbours.
The effect of these is hard to quantify, but will be included below in our
description of environmental spin effects.

Some of the above additional physics has been discussed already in the
literature. The effect of phonons has been analyzed in the usual
Caldeira-Leggett instanton scheme by Garg \& Kim \cite{36}  and Chudnovsky
\cite{35}. These calculations
indicate
 very small corrections to tunneling rates coming from phonons, which is
not surprising given the  infra-red
weakness of phonon environments. Notice however that these calculations
ignore barrier fluctuation effects \cite{37,38,39}.
 From now on we will ignore phonon effects since
they are so small. For the same reason we will ignore the even smaller
dissipative effects of photons (calculated in ref.\cite{19}). However we
note again that in quantitative discussion of
relaxation experiments, as opposed to coherence experiments it will
 be necessary
to include phonon effects.

We shall also ignore coupling of  ${\vec S}$ to electrons.
Actually, in contradistinction to domain walls \cite{40,41},
electrons have a very powerful effect on the dynamics of $\vec{S}$. Here we
simply assume the grain is insulating.

We now turn to the really important omission from (\ref{1}) and (\ref{2}),
viz, the coupling of ${\vec S}$ to other spin degrees of freedom - as already
noted, these will include nuclear spins and paramagnetic electronic
impurities, both inside and outside the grain. It will {\it not} include
the low-energy excitations of {\it coupled} spins inside or outside the
grain - these are generally describable in terms of a bath of oscillators
(e.g. magnons) with a very weak super-Ohmic coupling to ${\vec S}$. Thus it is
crucial that the environmental spins be weakly coupled to each other - then
the variety of effects we study in this paper become possible.

The most ubiquitous of such environments is the nuclear spin environment -
this is well understood. A nucleus with a finite spin $I$, and spin moment
$-\gamma_N \hbar I$, will interact both locally with the currents of electron
clouds at the same ionic site, via the contact hyperfine
interaction, and also non-locally with other ions via the dipolar field.
There may also be other residual interaction such as
"transferred" hyperfine interactions (as in, e.g., $Fe^{19}$ in $MnF_2$
or $CoF_2$), or quadrupolar couplings if $I>1/2$, or nuclear spin-phonon
interaction. Finally, one may have interactions between the different
nuclear spins, via, e.g., spin waves (the Suhl-Nakamura interaction),
or dipolar couplings - these are very weak
(Suhl-Nakamura interactions are $\le 10^{-5} \: K$, and nuclear dipole-dipole
interactions $\sim 10^{-7} \: K$) but, it turns out, can play a role in
the dynamics of $\vec{S}$; this is discussed in Section \ref{sec:5}.E. The most
important interactions are the contact hyperfine interaction and the
dipolar interaction between the nuclei and electrons. Thus the nuclear
moments in nonmagnetic hosts (e.g., $H^1$ in $H_2O$, or in $CuCl_2 \cdot H_2O$)
will be described by a Hamiltonian
\begin{equation}
H_I=-\gamma_N \hbar {\vec I} \cdot \left\{ {\vec H}_o - \gamma_e \hbar
\sum_j {1 \over r_j^3 } \big( \langle {\vec S}_j(T) \rangle -3 {{\vec r}_j
({\vec r}_j \cdot \langle {\vec S}_j(T) \rangle ) \over r_j^2 } \big) \right\}
\;,
\label{1.3}
\end{equation}
where ${\vec H}_o$ is an applied field, $\hbar \gamma_e = \mu_B g_e$, where
$g_e$ is
the electronic g-factor, $\sum_j$ sums over all ionic sites and
$\langle {\vec S}_j(T) \rangle$ is the temperature-dependent expectation  value
of spin polarization at the sites (for weak fields, $\langle {\vec S}_j(T)
\rangle =
\chi_j {\vec H}_o $, where $\chi_j $ is the susceptibility tensor). On the
other hand
a nucleus in a magnetic ion will have a Hamiltonian
\begin{equation}
H_I=-\gamma_N \hbar {\vec I} \cdot \left\{ {\cal A} \cdot \langle {\vec S}_o
\rangle
 +{\vec H}_o + H_{dip}  \right\} \;,
\label{1.4}
\end{equation}
\begin{equation}
H_{dip}=- \gamma_e \hbar
\sum_{j\ne 0} {1 \over r_j^3 } \big( {\vec S}_j -3 {{\vec r}_j
({\vec r}_j \cdot {\vec S}_j ) \over r_j^2 } \big) \;.
\label{1.5}
\end{equation}

The strength of these interactions is measured directly in NMR experiments;
the hyperfine coupling range from $1-3\:MHz$ (i.e., $5-15\times 10^{-5}K$)
for protons in $H^1$, up to values greater than $5000\:MHz$ ($0.25\:K$) for
some rare-earth magnetic nuclei (e.g., $Tb^{159}$, $Dy^{163}$); these
latter correspond to local fields acting on the nuclei as high as $500 \:
Tesla$, and come overwhelmingly from the contact interaction. Thus when
a central spin rotates, with all internal electronic spins rotating in unison,
the contact interaction tries to force the nuclei on the magnetic sites to
follow. If the hyperfine interaction strength is $\omega_o$, then it is
clear that the ratio $\omega_o /\Omega_o$, with $\Omega_o^{-1}$ the
timescale of central spin rotation, is going to be crucial to the
nuclear spin dynamics. If the electronic spin frequencies $\Omega_o \gg
\omega_o$, then the nuclear spins will experience a "sudden" perturbation, and
few of them will follow the central spin; conversely, if $\omega_o \gg
\Omega_o$, the nuclear spins will follow adiabatically.
In fact, as we shall see in more detail later in the paper, the ratio
$\omega_o /\Omega_o$ is usually somewhere between $0.01-0.05$ (although for
rare earth nuclei like $Tb^{159}$ or $Dy^{163}$ it can be $\sim O(1)$,
depending on the host). On the other hand for nuclei in non-magnetic
ions, $\omega_o /\Omega_o$ may be much smaller, and will of course depend
strongly on the host (in a magnetic host, dipolar interactions from permanent
moments, and also possible transfer hyperfine interactions
$ -\gamma_N \hbar {\vec I} \cdot  {\cal A}_{0j} \cdot {\vec S}_j $
from nearby magnetic moments ${\vec S}_j$, can greatly increase $\omega_o$;
for example,
$\omega_o$ for $F^{19}$ in $MnF_2$ is $160\: MHz$).

A large central spin will also interact significantly with nuclear spins in a
surrounding  medium, such as a substrate or solvent, via the dipolar
interaction generated by the central spin dipolar field, i.e.,
\begin{equation}
H_{int}=-\hbar^2 \gamma_e \gamma_N
\sum_{i} {\vec I}_i \cdot \big( \sum_{j}
{1 \over r_{ij}^3 } \big[ {\vec S}_j -3 {{\vec r}_{ij}
({\vec r}_{ij} \cdot {\vec S}_j ) \over r_{ij}^2 } \big] \big) \;,
\label{1.6}
\end{equation}
where $r_{ij}$ is the distance between some nuclear moments ${\vec I}_i$
in the surrounding matrix and a local moment ${\vec S}_j$ which is
incorporated in the central spin. This interaction is not
negligible; for a central spin with $S=10^7$, the dipolar coupling to a nucleus
at a distance of $10^3 \AA$ is already $\sim 1\:MHz$, rising to
$30\: MHz$ at a distance of $300 \AA$; and there is clearly a very large
number of nuclear spins within $1000 \AA$ of the central spin!

This brings us to another potentially very serious source of
decoherence for the central spin, viz., paramagnetic impurity spins in the
surrounding matrix. The
Hamiltonian has the same form as above except that $\gamma_N {\vec I}_i $ is
replaced by $\gamma_e {\vec S}_l$, describing an impurity electronic spin
${\vec S}_l$
at some distance $r_{il}$ from the central spin. In a very pure
surrounding matrix this is a problem because $\gamma_e \sim 2 \times
10^3\gamma_N $;
for the central spin with $S=10^7$, this dipolar coupling is still
$>2\:MHz$ for a paramagnetic impurity $10^4 \AA$ distant! A sphere of this
radius will contain over $ 10^{11}$ ions, so unless we have a matrix of, e.g.,
solid ultrapure $^4$He (or perhaps very high grade $Si$), there will be
many such impurities with appreciable coupling to ${\vec S}$.

Finally, as we approach the surface of the giant spin, we may expect a number
of electronic spins to have "weak exchange" coupling to the giant spin. This
may
arise from imperfections or defects or even dirt at the surface (or near it, as
in the coating around ferritin molecules). Typically we expect some kind
of "superexchange" coupling to apply, with energy anywhere between
$10-1000\:K$ (i.e., roughly in the range $10^{10} - 10^{13} \:Hz$). We shall
refer to such spins in the spin environment as "loose spins".
 How many of them there are will depend very much on the materials and material
preparation used.

A somewhat similar picture may well apply to $Si:P$ near the metal-insulator
transition, where it is believed that the ESR lineshape \cite{42} may
well be explained in terms of a Hamiltonian
\begin{equation}
H_{eff} = A{\vec S} \cdot \sum_{k=1}^N c_k {\vec I}_k + {\vec H}_o \cdot {\vec
S} \;,
\label{2.66}
\end{equation}
in which a central spin "superparamagnetic cluster" interacts with nuclear
spins.
This may well be so, provided there are insignificant decohering
interactions between the spin complexes. We hope at some time to give a
detailed analysis
of this case, and also of similar quantum spin glass systems.

We may summarize this survey of the various components of the
spin environment surrounding a ferromagnetic grain by the diagram in Fig.2.
It is important to notice \cite{1,2,3} how completely the range of
frequencies from $10^9\:Hz$ downwards is covered by the various coupling
strengths; and even above $10^9\:Hz$, for very large grains and/or
including "loose spins".

\subsection{Some examples}

The reader of this paper who is not a specialist in magnetism may also find it
helpful to get a feel for specific numbers, for different materials in which
different interactions are important. Thus in this sub-section we will give
some
"vital statistics" for a number of real magnetic systems.

Before doing so, it will be useful to glance at Table I, which lists some
nuclear hyperfine coupling energies, in various host systems. For those who
like to
think of energy in temperature units, note that
$1\:K \sim 21.6\: GHz$.  It will be noticed that the couplings
range over a factor $\sim 10^3$,
and that even the nuclei from non-magnetic ions can have reasonably large
couplings, via
transferred hyperfine interaction ($F^{19}$ in the Antiferromagnet (AFM)
$MnF_2$ being
a good example). For the transition metal Ferromagnets (FM), $\omega_k$ ranges
from
$50-500\:MHz$, but the rare-earth FM's can have $\omega_k$ an order of
magnitude greater. Where more then one value  of $\omega_k$ is given, there is
either
a quadrupolar splitting of the lines (e.g. $Tb$, $Dy$) or nuclei are at
inequivalent lattice sites (as in $YIG$).

The internuclear dipole interactions are way down from this (typically $\sim
O(1kHz)$);
and even the indirect Shul-Nakamura interactions have energy scale $\sim
100-200 kHz$ (see Table \ref{t2}).
Of course these interactions are always there, and they lead to an inevitable
minimum spreading $\delta \omega_k$ in the couplings; but $\delta \omega_k
/\omega_k $
is clearly very small ($10^{-5}-10^{-3}$) and so our assumption of independent
nuclei is very accurate (note that the temperature $T$ will always be far
larger than this linewidth $\delta \omega_k$). However we notice that $\delta
\omega_k$ {\it is
not necessarily small compared to} $\Delta_o$; this will be important for the
magnetic relaxation at long times, but is irrelevant for coherence, as we shall
see later.

We have already seen (cf. Fig.2) how the coupling to external nuclei and
external electronic
spins is also important. As a rough guide, this coupling can be estimated as
$\omega_k
\sim 10^8 S/r^3 \: Hz$, where $S$ is the spin quantum number of the central
spin, and
$r$ is the distance of a nuclear spin from ${\vec S}$ in $\AA$. For an
electronic spin,
$\omega_k \sim 2\times 10^{11} S/r^3\:Hz$ (obviously the prefactor is modified
depending on the spin quantum numbers and $g$-factors of the environmental
spin).

Consider now a few experimental systems, by way of example:

(a) \underline{$MnF_2$ grain}: This is an AFM; however it will have an
uncompensated
excess spin moment which may couple to external nuclei or electronic spins,
depending on the host. The size of $\Omega_o$ will depend on many factors for a
grain,
mostly uncontrollable strain fields - in principle one expects $\Omega_o$ to
range from
$\sim 0.1 \:K$ to $1\:K$ for such a system. There are two hyperfine couplings;
$\omega_k = 160\:MHz\;(F^{19})$ and $680\:MHz\;(Mn^{55})$. If $\Omega_o =
0.1\:K$ this
gives values of $\alpha_k=(\pi /2 ) \omega_k /\Omega_o$ equal to $0.125$ and
$\sim 0.5$,
respectively; if $ \Omega_o = 1\:K$, they are ten times smaller.

(b) \underline{$YIG$ grain}: One can make very pure grains of Yttrium Iron
Garnet. Since it is
a soft magnet, a well-prepared system would have $\Omega_o < 0.1\:K$, even
perhaps
as low as $0.01 \:K$. In the former case, the $Fe^{57}$ hyperfine coupling has
$\alpha_k \sim 0.05$; in the latter case $\alpha_k \sim 0.5$.
Note that only $2.19 \% $ of
$Fe$ nuclei have a moment.

(c) \underline{Ferritin}: Ferritin is a magnetic macromolecule made in all
eukaryotic cells;
a typical molecule contains $4500\;Fe$ ions in a ferritin structure, which are
antiferromagnetically ordered, although there is  an excess uncompensated
moment of size
somewhere between $200$ and $600\:\mu_B$. The $2.19 \% $ of
$Fe^{57}$ ions (roughly $100$ of them) have $\omega_k = 64\:MHz$. In the
experiments
of Awschalom et al.\cite{17}, an anisotropy field $\sim 1.72\:Tesla$ was found,
indicating $\Omega_o \sim 2\:K$, which indicates a value for $\alpha_k \sim 2.2
\times  10^{-3}$.

(d) \underline{$TbFe_3$ grain}: Here $\Omega_o $ is typically $3\:K$; this is
very high
indeed. However so are the  $\omega_k $ since $Tb$ is a rare-earth magnet. With
the
quadrupolar split hyperfine lines, one finds $\alpha_k \sim 0.08,\:0.07,\:0.05$
for the
$Tb$ nuclei. The $2\%$ of $Fe^{57}$ nuclei will have $\alpha_k$ some 50 times
smaller than this;
the result of this for coherence of ${\vec S}$ will be, in effect, to simulate
a spread
$\delta \omega_k$ in the $\omega_k$, rather as though we had a spread $\delta
\omega_k \sim
50\:MHz$ in the system.

{}From these examples we see that typical values of $\alpha_k$ are between
$10^{-3}$ and
$O(1)$, usually something like $0.05$. Moreover the presence of other nuclear
species such as $H^1$ (in $H_2O$, as in ferritin) will cause a small spread in
the
$\alpha_k$. Finally, of course, there will be a much larger spread in the
$\omega_k$
coming from spins outside the grain or molecule.

In section \ref{sec:6} we shall return to examine two of these examples,
namely $TbFe_3$ and ferritin, in greater detail.

\section{The effective Hamiltonian}
\label{sec:3}

In this section we make the crucial step of going from "bare Hamiltonian"
like those in Section \ref{sec:2}, to a low energy "effective Hamiltonian".
This effective Hamiltonian is our starting point for calculating the
dynamics of $\vec{S}$ at low $T$. It contains a variety of effective couplings,
which already contain all the high-energy physics. We emphasize that in
realistic cases one may actually calculate these renormalised couplings,
starting with the original bare Hamiltonian - they do {\it not} have to be
treated as phenomenological couplings to be determined from experiment. To
illustrate this we actually do the truncation explicitly for a typical
hyperfine interaction.

\subsection{Reduction to $H_{eff}$}

Let us go back to our "bare" Hamiltonian $H_o({\vec S} )$ for ${\vec S}$,
before
either truncation to a low energy effective Hamiltonian, or coupling to any
environment. There are two
important energy scales in $H_o({\vec S} )$, viz, $\Omega_o$ and $\Delta_o$;
the former describes the "fast" motion of the central spin in the effective
potential in $H_o({\vec S} )$ (small oscillations near the potential
minima, or the "bounce frequency" involved in tunneling through
the barrier), whereas the latter characterizes the much slower coherence
between
tunneling events, since $\pi /2\Delta_o$ is the average time between
successive tunneling events. These 2 time scales are seen very clearly
if we show a "typical trajectory" for ${\vec S} (\tau )$, i.e., a typical
path that would contribute to a path integral evaluation of the
dynamics of ${\vec S}$.
This is shown in Fig.3, for a case in which ${\vec S}_1$ and ${\vec S}_2$ are
not antiparallel.
There will be other even more rapid time scales which are buried in this path,
corresponding to transitions to very high excited states of $H_o({\vec S} )$;
but their amplitude is very small.

The truncation of $H_o({\vec S} )$ is standard \cite{30,31,33}. Consider, e.g.,
the bare Hamiltonian (\ref{2}); truncation gives
\begin{equation}
 H_0( {\vec S} ) \to 2 \Delta_o  {\hat \tau }_x \cos \pi S = H_o(
 \hat{ \vec{\tau} } )\;,
\label{3.0}
\end{equation}
where the Pauli matrix ${\hat \tau }_x $ flips ${\vec S}$ between the
degenerate configurations ${\vec S}_1 =\hat{\vec{z}} S$ and
${\vec S}_2 = - \hat{\vec{z}} S$, the factor $2 \cos \pi S  $ is an extression
of Kramer's theorem. The bare splitting
$\Delta_o$ is given in terms of $\Omega_o$ by
\begin{equation}
 \Delta_o \sim \Omega_o e^{-A_o} \;,
\label{3.02}
\end{equation}
\begin{equation}
 A_o \sim 2(K_{\parallel}/K_{\perp})^{1/2}S \;,
\label{3.03}
\end{equation}
showing the exponential ratio between $\Delta_o$ and $\Omega_o$ (more
accurate expression for $\Delta_o$ appear in Refs.\cite{4,20,31}); and for this
model
\begin{equation}
\Omega_o \sim {2 \over S} (K_{\parallel}K_{\perp})^{1/2}  \;.
\label{3.04}
\end{equation}
Notice that although $\Omega_o$ is absent from (\ref{3.0}), it is nevertheless
implicitly contained via the "WKB renormalization" of (\ref{3.02}).

Let us now consider how one should make the same truncation in the presence of
a large number of environmental spins. In general the same
argumentation will apply - we look first at how the environmental spins will
affect the "fast" physics of the central spin (on a time scale
$\Omega_o^{-1}$),
and absorb this into a new effective Hamiltonian describing a
"renormalized instanton" for the central spin. We then deal with the "slow"
physics (on time scales $> \Omega_o^{-1}$) using this renormalized instanton,
and derive a new effective Hamiltonian for this slow, low-energy physics.

However the same is certainly not true if we have $N$ environmental
spins. How can we handle this case? The method we have adopted \cite{2,3}
starts by asking how we should write the effective action to be
associated with the instanton. In the absence of the spin environment,
the transition amplitude  matrix from
$\mid {\vec S}_1 \rangle $ and $\mid {\vec S}_2 \rangle $ is given by
\begin{equation}
K_o^{\pm}={\hat \tau }_x \Omega_o \exp \{ -A_o \pm i\pi S \} = \Delta_o {\hat
\tau }_x
e^{\pm i \pi S} \;,
\label{3.1}
\end{equation}
where $\Delta_o = \Omega_o e^{-A_o}$, and $A_o
\sim (K_{\parallel}/K_{\perp})^{1/2}S$, in accordance with equations
(\ref{3.02})
and (\ref{3.04}); the $\pm$ refers to the 2 different
(clockwise or anticlockwise) paths between
$\mid {\vec S}_1 \rangle $ and $\mid {\vec S}_2 \rangle $.
Now when we add the spins, the instanton will act as an operator
in the subspace of each environmental spin (able to cause transitions in
each subspace, in line with our discussion in section \ref{sec:2}.B). The
general form now must be
\begin{equation}
{\hat K}^\pm_{\eta}(k) = \Delta_o  {\hat \tau }_{\eta} e^{\delta_k } \exp
\left\{ [\overline{\beta}_k{\vec u}_k \pm \xi_k {\vec v}_k ] \cdot {\hat {\vec
\sigma }}_k
\pm i \eta [\pi S + \phi_k + \alpha_k {\vec n}_k  \cdot {\hat {\vec \sigma }}_k
] \right\}
\;,\;\;\;(\eta =\pm )\;,
\label{3.2}
\end{equation}
in the subspace of the
$k$-th spin - here ${\hat \tau }_{\eta } \equiv {\hat \tau }_{\pm} = {\hat \tau
}_x \pm i{\hat \tau }_y$,
${\vec u}_k$, ${\vec n}_k$, and  ${\vec v}_k$ are unit vectors in some
direction, and $\delta_k$ is a number. The instanton from
$\mid {\vec S}_2 \rangle $ to $\mid {\vec S}_1 \rangle $ is Hermitian
conjugated to that
in (\ref{3.2}).
 The role of the new terms in (\ref{3.2}) is as follows:

(i) The adiabatic correction $\delta_k$ is due to the high-frequency
($\gg \Omega_o$) fluctuations of ${\vec \sigma }_k$, which renormalize the bare
instanton
action $A_o$. This correction is usually very small, and is of course
unobservable because it does nothing  but renormalize $A_o$. It increases
the "moment of inertia" of ${\vec S}$.

(ii) The "barrier fluctuation" terms
$(\overline{\beta}_k{\vec u}_k \pm \xi_k {\vec v}_k) \cdot {\hat {\vec \sigma
}}_k $
come from fluctuations
of ${\vec \sigma }_k$ which are of frequency $\sim \Omega_o$ or less. As
discussed in
refs.\cite{37} and \cite{38} (see pp. 53-55 of ref.\cite{38}), the effect
of these fluctuations is, amongst other things, to cause the potential barrier
to fluctuate in height. In Section \ref{sec:3} B
 we discuss in more detail how this
leads to a term of this form for our problem. Here we simply note that
(a) the effect of $\overline{\beta}_k$ is to {\it increase} the tunneling rate,
because barrier fluctuations occasionally lower the barrier , thereby creating
a "window of opportunity" for ${\vec S}$ to tunnel; and (b)
 this term can be entirely absorbed into a renormalization of the
"orthogonality blocking" term that we will have in our final effective
Hamiltonian. In the most general case one has to
allow different couplings for the clockwise and counter-clockwise instantons,
e.g., when the environmental spin is oriented along  the
${\hat {\vec x }}$ direction it influences the two instanton trajectories
in the $XZ$-plane   differently, thus giving
a non-zero value of $\xi_k$.

(iii) The topological term $\alpha_k {\vec n}_k  \cdot {\hat {\vec \sigma }}_k
+\phi_k $ describes the dynamical effect on ${\vec \sigma }_k$ when
$\vec{S}$ flips.
 Its origin is two fold: one contribution comes from
a transfer matrix (note that $\alpha_k \ne \overline{\alpha}_k $)
\begin{equation}
{\hat T}_k^{\pm}= e^{i\int_{\pm} d\tau H_{int} (\tau ) } =
e^{ \pm i (\overline{\alpha}_k {\vec n}_k  \cdot {\hat {\vec \sigma }}_k +
\phi_k )} \;,
\label{3.3}
\end{equation}
in the spin subspace of the $k$-th spin, where in the strong coupling case the
phase $\phi_k$ approaches the value $\phi_k = \phi_k^B -\pi/2$;
 $\phi_k^B$ is the adiabatic Berry phase accumulated by the nuclear spin.
The other contribution $(\alpha_k - \overline{\alpha}_k ) $
is closely related to the barrier preparation effect, but has no analog in the
standard description of
the particle tunneling through the barrier;  it
arises from the influence of the environmental spins on the topological
phase (see below).

This summarizes the mutual effects of ${\vec S}$ and $\{ {\vec \sigma }_k \}$
during the tunneling of ${\vec S}$. However after this tunneling is over,
there will be a residual interaction between ${\vec S}$ and the $\{ {\vec
\sigma }_k \}$. This can be described as follows. Suppose the 2 relevant
quasiclassical states of $\vec{S}$ are again oriented along
${\vec S}_1$ and ${\vec S}_2$, and that the total effective fields acting
on ${\vec \sigma }_k$ are correspondingly ${\vec \gamma }_k^{(1)}$ and
${\vec \gamma }_k^{(2)}$, defined in units such that their interaction energy
with ${\vec \sigma }_k$ is ${\vec \gamma }_k \cdot {\vec \sigma }_k$. Now
define the {\it sum} and the {\it difference} terms
\begin{eqnarray}
\omega_{\parallel}{\vec l}_k & =& {\vec \gamma }_k^{(1)} -
 {\vec \gamma }_k^{(2)} \nonumber \\
\omega_{\perp} {\vec m}_k & =& {\vec \gamma }_k^{(1)} +
 {\vec \gamma }_k^{(2)}\;.
\label{3.88}
\end{eqnarray}
where the ${\vec l}_k$ and ${\vec m}_k$ are unit vectors. In the truncated
Hilbert space of $\vec{\tau}$ and $\{ {\vec \sigma }_k \}$, the residual
interaction term takes the form
\begin{equation}
 H_{static}= {1 \over 2} \bigg\{
{\hat \tau }_z \sum_{k=1}^N \omega_k^{\parallel} \:
 {\vec l}_k \cdot {\hat {\vec \sigma }}_k  + \sum_{k=1}^N
\omega_k^{\perp}\: {\vec m}_k \cdot {\hat {\vec \sigma }}_k \bigg\} \;,
\label{3.94}
\end{equation}
i.e., a term which changes when $\vec{S}_1 \to \vec{S}_2$, and an extra
term which does not. This latter term can arise in various ways. Most
commonly it will arise because the spin-$1/2$ variable ${\vec \sigma }_k $ is
produced by truncating some environmental spins $\vec{I}_k$, in a bare
Hamiltonian, with $I_k >1/2$. It will also arise if, e.g., $\vec{S}_1$ and
$\vec{S}_2$ are not exactly antiparallel (e.g., in an imperfect grain), or
simply because there are small stray fields acting on the
${\vec \sigma }_k$. Usually $ \omega_k^{\perp} \ll \omega_k^{\parallel} $
(at least for nuclei in a magnetic system).

 We are now ready to write the full low-energy effective Hamiltonian for our
central spin model, in the absence of an external field, but
incorporating all the mutual effects between ${\vec S}$ and its spin
environment \cite{2,3}
\begin{equation}
 H_{eff}=2{\tilde \Delta}_o \big\{ {\hat \tau }_+\cos \big[ \Phi   +
\sum_{k=1}^N
( \alpha_k {\vec n}_k -i \xi_k {\vec v}_k ) \cdot {\hat {\vec \sigma }}_k
\big] +H.c. \big\} +
{\hat \tau }_z \sum_{k=1}^N { \omega_k^{\parallel} \over 2}\:
 {{\vec l}_k \cdot {\hat {\vec \sigma }}_k } + \sum_{k=1}^N
{ \omega_k^{\perp} \over 2} \: {{\vec m}_k \cdot {\hat {\vec \sigma }}_k } \;.
\label{3.4}
\end{equation}
The form of this effective Hamiltonian is central to all further discussions.
We have dropped the barrier fluctuation term $\beta_k {\vec u}_k \cdot {\hat
{\vec \sigma }}_k $,
since as already noted its effects will simply consist in a renormalization
of ${\tilde \Delta}_o $ and the effective ratio $\omega_k^{\perp}
/\omega_k^{\parallel} $,
whose effects appear in orthogonality blocking
(see the next section). The phase
$\Phi =(\pi S +\sum_k \phi_k )$, i.e., we absorb the Berry phase into our total
value of the topological phase.  Finally,
 ${\tilde \Delta}_o =\Delta_o exp(-\sum_k \delta_k )$,
absorbing the adiabatic corrections.

Before going on, it is worth asking (a) how general is $H_{eff}$, and (b)
what, if anything, has been left out of it?

Consider first the general structure of $H_{eff}$. We see that in our
$2^{2N+1}$-dimentional Hilbert space, it contains not only all possible
{\it pairwise} interactions between $\vec{\tau}$ and each $\vec{\sigma}_k$,
but also a set of higher couplings, involving as many as $N$
environmental spins, produced by expanding the cosine; i.e., we have couplings
of form
\begin{equation}
H_{dyn} \sim \hat{\tau}_i A^i_{\alpha \beta \gamma \delta \cdots }
\hat{\sigma}_{k_1}^{\alpha} \hat{\sigma}_{k_2}^{\beta}
\hat{\sigma}_{k_3}^{\gamma} \hat{\sigma}_{k_4}^{\delta} \cdots  \;.
\label{3.111}
\end{equation}
Some readers might like to think of (\ref{3.111}) as being an infinite set
of terms in a perturbation expansion of the bare coupling to all orders,
but we feel that this is mistaken - the coefficients in (\ref{3.111}) are
fixed precisely by the cosine, and in any way there is nothing perturbative
about the form of $H_{eff}$ - it is valid even when the coupling between
the $\vec{\sigma}_k$ and $\vec{\tau}$ is very strong.

In fact the only terms left out of $H_{eff}$ are {\it direct} interactions
between the $\{ \vec{\sigma}_k \}$ (i.e., ones not generated, like
(\ref{3.111}), via coupling to $\vec{S}$).
For nuclei these are the dipolar and  Nakamura-Suhl
interactions, and for paramagnetic impurities they could also be  RKKY
interactions between widely-separated spins. Thus the only terms left
out of $H_{eff}$ are terms of the form
\begin{equation}
\hat{V}( \{ \vec{\sigma}_k \} ) = \sum_{k\ne k^\prime } V_{k k^\prime }^{ij}
 \hat{\sigma}_k^i   \hat{\sigma}_{k^\prime }^j \;.
\label{3.112}
\end{equation}
As we shall see later in this paper, these terms can also be included
in our analysis. Their effect on coherence is usually small (although
under some circumstances they can effect coherence in large grains).
However in a second paper, dealing with {\it relaxation} of $\vec{S}$ in
a bias, we will see that $\hat{V}$ does play a very important role.

Readers not used to the philosophy underlying the derivation of low-energy
effective Hamiltonians may also suspect that somehow the form of $H_{eff}$
depends in some way on the semiclassical analysis we use of the dynamics
of $\vec{S}$. We emphasize this is not so - the {\it form} of  $H_{eff}$
depends in no way on any semiclassical or WKB approximation. What {\it does}
depend on the semiclassical analysis is the exact value of each of the
renormalised parameters in $H_{eff}$. If $\vec{S}$ were a microscopic spin,
it would not be appropriate to use WKB theory to evaluate, e.g.,
$\Delta_o$ in $H_o (\hat{\vec{\tau}} )$, or the $\{ \alpha_k \}$ in
$H_{eff}$. However  when $S \gg 1$, the use of WKB/instanton methods
becomes extremely accurate, both in determining $H_o (\hat{\vec{\tau}} )$
(cf. Refs.\cite{29,31,32,33}), and (as we shall see in the next sub-section)
in determining the values of the parameters in $H_{eff}$.

Finally, note how differently $H_{eff}$ describes the "spin bath" environment
from the usual oscillator bath \cite{10,11,12} models. Not surprisingly,
the physical effects of the spin bath will also turn out to be very different -
from this point of view it is useful to think of the spin bath as an example
of an "unconventional environment", having unusual effects (whether it
interacts with a central spin, or SQUID, or any other collective coordinate).
This theme has been elaborated elsewhere \cite{13}.

\subsection{ Instanton Derivation of $H_{eff}$}

 As a first step we will present
general arguments and the physical ideas involved in such a derivation, and
proceed then with the more specific calculation for the Hamiltonian
(\ref{1}).

Suppose we are interested in the transition amplitude from
$\mid {\vec S}_1 \rangle$ to $\mid {\vec S}_2 \rangle$ in the WKB approximation
for the central spin only, ignoring for the moment the
initial and final states of the environment.
Without any interaction between the ${\vec S}$  and $ \{ {\vec \sigma }_k \}$
we have, of course,
\begin{equation}
^oK_a^{\pm}={\hat \tau }_a \Omega_o \exp \{ -A_o \pm i\pi S \}
\;  ,\;\;\;\;(a=\pm )\;,
\label{a.1}
\end{equation}
Now if we calculate the transition amplitude with all the interactions
included,
according to
\begin{equation}
{\hat K}^{\pm}
\mid \{ \chi_k^{(in)} \} \rangle = \langle {\vec S}_1 \mid  e^{-i\int
d\tau^\prime
[H_o (\tau^\prime ) +H_{int} (\tau^\prime )]}
\mid {\vec S}_2 \rangle \: \mid \{ \chi_k^{(in)} \} \rangle \;,
\label{a.2}
\end{equation}
it will obviously depend on the initial state of the $ \{ {\vec \sigma }_k \}$,
and, when
projected on the particular final state of the environmental spins, it
will depend on the spin flips in the environment. Thus \cite{2,3}
in the most general
way we have to consider the instanton from $\mid {\vec S}_1 \rangle$ to
$\mid {\vec S}_2 \rangle$ as an operator in the subspace of $ \{ {\vec \sigma
}_k \}$.
Eq.(\ref{3.2})
is the most general form for this operator one can write down no matter what
are
the interactions in the system (the only omission is the simplified form
for the adiabatic contribution $\delta_k$ which in a more general form should
read $\delta_k \pm \overline{\delta}_k$, applying   to the
case when the central spin is in an external
magnetic field so that the adiabatic
contribution is asymmetric for the clockwise and anticlockwise instantons
\cite{25}).
Thus the real question to ask is not why  we write the instanton in this form,
but rather - what is the physical meaning of these parameters, and what is
their relation to
the original parameters characterizing $H_o $ and $H_{int}  $?

We start by describing the adiabatic  contribution to the
action due to $\delta_k$. This comes from the environmental spins
strongly coupled to the central spin, i.e., with couplings
 $\omega_k \gg \Omega_o$. The slow (on a time scale defined
by $\Omega_o^{-1}$) rotation of the central spin can not cause any transitions
from the ground state for these spins (in quantum tunneling problems we
deal with temperatures $T \ll \Omega_o$ - otherwise the "classical"
over-barrier transitions will prevail), and they have to follow the
direction of the local field acting on them. It is appropriate to
include these spins in the "combined central spin". In general,
we increase the moment of
inertia which will result in a correction to the instanton action. Let us
calculate this correction using the particular form
of $H_o({\vec S} )$ given by (\ref{1}). The corresponding Lagrangian written in
terms of angles $(\theta , \varphi )$ has the form
\begin{equation}
{\cal L} (\theta , \varphi ) = iS^\prime \varphi \sin \theta \:\dot{ \theta }
+ U(\theta , \varphi )
\label{a.4}
\end{equation}
\begin{equation}
U(\theta , \varphi ) = K_{\parallel} \big[
\sin ^2\theta (1+\lambda_{_A} \sin ^2 \varphi ) -
2h \sin \theta \cos \varphi \big] \;,
\label{a.5}
\end{equation}
where $\lambda_{_A} = K_{\perp}/K_{\parallel} $
is the anisotropy parameter, and
$S^\prime -S =1/2 \cdot N_{ad} $ gives the number of
environmental spins locked in a joint rotation with the central spin. For
future discussion we also include a weak magnetic field acting in the
${\hat {\vec x}}$-direction $h= \gamma_e H_x S/(2K_{\parallel} ) \ll 1$.
As usual \cite{4,20}, we perform integration over the small deviations of the
angle $\varphi $ around zero, and obtain the effective Lagrangian for
$\theta $ as
\begin{equation}
{\cal L} (\theta ) = { (S^\prime )^2 \over 4 K_{\parallel} }
{\sin \theta \over \lambda_{_A} \sin \theta +h } \dot{\theta }^2 +
K_{\parallel} \big[ \sin ^2\theta  -2h \sin \theta  \big] \;,
\label{a.6}
\end{equation}

The classical  equation of motion starting at $\theta = \pi /2 $ at $t=0$,
and approaching the angle $\: \arcsin ( h )$ as $t \to \infty$ has the form
\begin{equation}
 \dot{\theta }= \Omega_o {S  \over S^\prime} (h-\sin \theta)
\big( 1-{ h \over \lambda_{_A} \sin \theta } \big) ^{1/2} \;,
\label{a.7}
\end{equation}
where $\Omega_o$ is the bare bounce frequency (\ref{3.04}). For small $h$ and a
large value of the anisotropy parameter (the action is proportional to
$2S/\lambda_{_A}^{1/2} $, so that very large values of $\lambda_{_A}$ are
{\it necessary} to make any discussion of mesoscopic spin tunneling
physically meaningful) we may drop the last factor in the r.h.s. of equation
of motion. The solution is now
\begin{equation}
\sin \theta (\tau ) ={1+h\cosh \Omega \tau \over \cosh \Omega \tau +h }
\label{a.8}
\end{equation}
\begin{equation}
 \Omega = \Omega_o { S \over S^\prime} \sqrt{1-h^2} \;.
\label{a.9}
\end{equation}
Substituting this expression back into  the effective action we have
\begin{equation}
A=2\sqrt{K_{\parallel}/K_{\perp}} S^\prime (\sqrt{1-h^2} -h \arccos (h) )\;.
\label{a.10}
\end{equation}
{}From this we derive the correction to the instanton action due to the
increase of the moment of inertia
\begin{equation}
\delta A_{in} =\sqrt{K_{\parallel}/K_{\perp}} N_{ad} \;,
\label{a.11}
\end{equation}
which is proportional to the number of spin-$1/2$ "satellites"
added to the "bare" central spin. The result is quite general, although the
coefficient in front of
$N_{ad}$ will depend of course on the particular form of $H_o({\vec S} )$.

For the hyperfine interaction  we do not expect
any variation in the potential energy. In a more
general case, i.e., for the dipolar coupling between the $\vec{S}$ and
$\{ {\vec \sigma }_k \}$, or when
there are  interactions acting on $\{ {\vec \sigma }_k \}$ other then their
interaction
with ${\vec S}$, the energy $U(\theta ,\varphi )$ will be modified. This
problem
is hard to deal with exactly because now the potential energy will depend
on the orientations ${\vec r}_{ij} /r_{ij}$ between the environmental spins
and those forming the central spin. However to estimate the effect we do not
need the exact solution. It is sufficient to look at the correction
to the action due to weak applied field, which also changes the
potential energy, by amount $\delta U \sim \gamma_e HS$. It follows
from (\ref{a.10}) that for small $h$ the correction is
\begin{equation}
\delta A_{pot} \sim \pi {\delta U \over \Omega_o} \:,
\label{a.12}
\end{equation}
and may be quite strong if $\delta U$ is large. Still we assume that only
a small number of "loose spins" on the surface of the grain $N_{ad} \ll S$
have dipolar couplings strong enough to move adiabatically with the grain
magnetisation; thus the total change $\delta A_{pot} $ is much less then $A_o$.

To summarize,  the adiabatic parameter $\delta_k$ can
vary from a very small value $\sim \lambda_{_A}^{-1/2}$ up to $\pi \delta U
/\Omega_o $ depending on the model ( for nuclear spins inside the grain the
 former case is the most relevant one since the dipolar coupling is much less
than
the contact hyperfine interaction).

Now we proceed to the discussion of the terms given by $\overline{\beta }_k$
and $\xi_k$ in (\ref{3.2}). Suppose we have some environmental spins which are
coupled to the central spin  weakly,
$\omega_o /\Omega_o \ll 1$. These spins can not follow the central spin
adiabatically - they rather see the rotating ${\vec S}$ as a sudden
perturbation, so that most of them simply conserve their spin
orientations during the time $\Omega_o^{-1}$. On the other hand their
coupling to the
central spin slightly changes the actual instanton path. We may start by
considering
this influence as a weak {\it static}
magnetic field due to the $\{ {\vec \sigma }_k \}$ acting on
${\vec S}$, i.e., we write the interaction Hamiltonian as $H_{int} = {\vec S}
\cdot
\delta {\hat {\vec H}}$, where $\delta {\hat {\vec H}}$ is  an operator
in the environmental spin subspace, and can be easily found for the
hyperfine interaction (\ref{1.4}) and dipolar interaction (\ref{1.5}).
Using the fact that ${\vec S}$ is rotating so fast that
$\delta {\hat {\vec H}}$ hardly
changes during the transition time, we may neglect its time dependence
(which will be accounted for later in our discussion of the parameters
$\alpha_k$ and $\phi_k$). The rest is simple now, for we have to find a
linear in $\delta {\hat {\vec H}}$ correction to the action, i.e., the
variation of $A$ due to the extra potential energy ${\vec S}_\alpha \cdot
\delta {\hat {\vec H}}_\alpha $. In fact this was already done for the
$H_o ({\vec S} )$ described by (\ref{1}), and $\delta {\hat {\vec H}}$ along
the ${\hat {\vec x}}$-direction. The linear in $h$ correction is easy
to obtain from (\ref{a.10})
\begin{equation}
\delta A_x = -\pi {\gamma_e\delta {\hat H}_x S \over \Omega_o } = -{
\pi \omega_o \over 2 \Omega_o } {\hat \sigma }_x \;,
\label{a.13}
\end{equation}
where  $\omega_o$ is the contact hyperfine interaction.
In this particular case
we find
\begin{equation}
\xi_k = -{\pi \omega_k \over 2 \Omega_o } \;; \;\;\;\;{\vec v}_k = {\hat {\vec
x}}\;.
\label{a.14}
\end{equation}

One might try to argue that the  calculation of the first order correction
to $A$ can be done
for arbitrary $H_o ({\vec S})$ and $H_{int} ( \{ {\vec \sigma }_k \},{\vec S} )
$ as follows.
In the expression for the instanton action
\begin{equation}
A= -\int d\tau ({\cal L}_o (\tau ) +H_{int} (\tau ))\;,
\label{a.15}
\end{equation}
the first order correction is given simply by
\begin{equation}
\delta A= -\int d\tau H_{int} ({\vec S}_o (\tau ))\;,
\label{a.16}
\end{equation}
where ${\vec S}_o (\tau )$ is evolving according to the unperturbed
semi-classical solution. For the case of hyperfine interactions
considered above this expression takes the form
\begin{equation}
\delta A= - {\omega_k \over 2S } {\hat {\vec \sigma }}_k \int d\tau {\vec S}_o
(\tau )\;,
\label{a.17}
\end{equation}
Substituting here  (\ref{a.8}) (for $h=0$)
we reproduce the result (\ref{a.14}).
The symmetry of the problem in zero external field implies that that
the corrections due to $\delta {\hat {\vec H}}_{z,y} $ are zero, which means
that here $\overline{\beta}_k \approx 0$.

This would be a correct way of calculating things if we were dealing with
 ordinary particles but it is {\it not} correct for  the angular momenta. The
formula (\ref{a.16})
takes into account the change in the potential energy {\it only}, while
for the constrained rotation of ${\vec S}$ any time-reversal symmetry breaking
field can also influence the topological phase. To see how it works we apply
the magnetic field along ${\hat {\vec y }}$-direction, which results in a term
\begin{equation}
-2K_{\parallel}h \sin \theta \sin \varphi \approx -2K_{\parallel}h \sin \theta
\: \varphi \;. \nonumber
\end{equation}
Integrating over small $\varphi$ as before we obtain the effective Lagrangian
for the angle $\theta $ as (up to a constant term)
\begin{equation}
{\cal L} (\theta ) = {\cal L}_o(\theta )
-i{hS \over \lambda_{_A} } \dot{\theta }  \;,
\label{a.18}
\end{equation}
which results in a topological phase correction
\begin{equation}
i\pi S \to i\pi S +i \pi  { \gamma_e \delta {\hat {\vec H}}_y S \over
\Omega_o } \equiv  i\pi S + i { \pi \omega_o \over 2\Omega_o } {\hat \sigma}_y
\;.
\label{a.19}
\end{equation}
Thus we find one of the contributions to the coupling $\alpha_k$ to be
\begin{equation}
\alpha_k^{(top)}= { \pi \omega_o \over 2\Omega_o }\;; \;\;\;\;
{\vec n}_k^{(top)} = {\hat {\vec y }} \;,
\label{a.20}
\end{equation}

Now we turn to the last contribution to our instanton operator
(\ref{3.2}), which is due to the small, but unavoidable dynamics of $\{ {\vec
\sigma }_k \} $
during the transition. To define this we note that from the environmental spin
point of view the rotating central spin is seen as a time dependent external
magnetic field. Thus we calculate the evolution of the
microscopic spin wave-function according to
\begin{equation}
{\hat T}_k \mid \chi_{env} \rangle  = e^{i\int d\tau H_{int} (\tau ) }
\mid \chi_{env} \rangle \;,
\label{a.21}
\end{equation}
For the weak coupling case $\omega_k/ \Omega_o \ll 1$
we are considering here (sudden perturbation), this can be calculated as
\begin{equation}
{\hat T}_k \approx 1+ i\delta {\hat {\vec H}}
\int d\tau {\vec S}_o (\tau ) \approx  \exp \{ i \delta {\hat {\vec H}}
\int d\tau {\vec S}_o (\tau ) \} \;,
\label{a.22}
\end{equation}
Thus we find  exactly the same expression as in (\ref{a.16}) which is
hardly surprising since we are dealing with the same Schr{\H{o}}dinger
equation but now in the real time domain. From this we derive the second
contribution to the parameter $\alpha_k$. Again, for contact interactions
between the spins we have
\begin{equation}
\overline{\alpha}_k= { \pi \omega_o \over 2\Omega_o }\;; \;\;\;\;
\overline{{\vec n}}_k = {\hat {\vec x }} \;,
\label{a.23}
\end{equation}

These results for $\alpha_k$, $\xi_k$, $\phi_k$, and $\delta_k$ are all
for the realistic case where $\omega_k /\Omega_o \ll 1$. However in the
rare cases where $\omega_k /\Omega_o \ge 1$, the generalization of the
above method is easy. For example, in Fig.4 we show the variation of
$\alpha_k$, as a function of $\omega_k /\Omega_o$, for the $H_o(\vec{S})$
of (\ref{1}).  The actual form of $\alpha (\omega_o /\Omega_o )$ for
a contact hyperfine interaction is quite lengthy, but almost indistinguishable
from the simple form
\begin{equation}
\alpha (x) \sim (\pi /2 )\; \tanh x \;.
\label{a.78}
\end{equation}
Remarkably, this simple form is almost universal - it hardly depends on the
detailed form of $H_o(\vec{S})$. This point is dramatically illustrated if
we take the really pathological instanton
\begin{equation}
  \theta  (\tau ) = {\Omega_o \tau \over 2 } [\eta (\tau +\pi /\Omega_o)-\eta
(\tau -\pi /\Omega_o)]
 +{ \pi \over 2} [\eta (\tau +\pi /\Omega_o) + \eta (\tau -\pi /\Omega_o) ] \;.
\label{16}
\end{equation}
where $\eta (\tau )$ is the unit step function. Equation (\ref{16})
describes a "jerk start" and "jerk stop" instanton, with the polar angle
$\theta (\tau ) $ varying linearly in between; thus
\begin{equation}
\cos \alpha ={\Omega_o \over (\omega_o^2+\Omega_o^2)^{1/2}} \sin
[\pi /2 (1+\omega_o^2/\Omega_o^2)^{1/2} ]  \;,
\label{17}
\end{equation}
which is almost indistinguishable from Fig.4 except for some small
"wiggles" around $\alpha \approx \pi /2 $, when $\omega_o > \Omega_o$
(these arise from the infinite acceleration and deceleration at
$\tau = \pm \pi / \Omega_o$ respectively).

The saturation of $\alpha (x)$ at $\pi /2$ when $x>1$ is easy to understand -
it corresponds to the adiabatic (Berry phase) limit, when $\vec{\sigma}_k$
rotates adiabatically with $\vec{S}$. Similar calculations may be made
of $\xi_k$, $\phi_k$, $\delta_k$, etc., for the full range $0<x<\infty $ of
couplings.

This concludes our discussion of the instanton operator. Each term
in the exponent (\ref{3.2})  has its own physical meaning and may be
calculated from first principles once the form of $H_o ({\vec S} )$ and the
most relevant interactions are identified. Here we analyzed the most
important case of hyperfine interactions between the central spin and nuclear
spins when $H_o ({\vec S} )$ is given by (\ref{1}). We see that
the expression (\ref{3.2}) simplifies a lot for nonadiabatic
coupling ($\omega_k /\Omega_o \ll 1$):
\begin{equation}
\delta_k = \overline{\beta}_k = \phi_k =0 \;; \;\;\;\;
\alpha_k {\vec n}_k = {\pi \omega_k \over 2 \Omega_o} ({\hat {\vec x }},
{\hat {\vec y }})\;; \;\;
\xi_k {\vec v}_k = -{\pi \omega_k \over 2 \Omega_o} {\hat {\vec x }} \;.
\label{a.24}
\end{equation}
In the adiabatic limit ($\omega_k /\Omega_o \gg 1$) we have
\begin{equation}
 \overline{\beta}_k = \phi_k = \xi_k = 0 \;; \;\;\;\;
\delta_k = \lambda_{_A}^{-1/2} \;; \;\;
\alpha_k \to  \pi /2 \;.
\label{a.25}
\end{equation}


\section{Three important limiting cases}
\label{sec:4}

The effective Hamiltonian $H_{eff}$ in (\ref{3.4}) has a very rich
behaviour in general. To appreciate this behaviour it is best to lead up
to it by first studying some limiting cases, each  one of which brings out
an important aspect of the physics. This is done as follows. First we
assume that $\omega_k^{\parallel}$ and $\omega_k^{\perp}$ are zero -
although this is a rather artificial way of doing things, it enables us to see
"pure topological decoherence", without interference from anything else
\cite{1}.
Next we suppress topological decoherence, by making $\alpha_k$
very small, and we also make all the $\omega_k$ equal; however the ratio
$\omega_k^{\perp}/\omega_k^{\parallel} $ now becomes the crucial
parameter, governing the behaviour of "pure orthogonality blocking".
Finally we study "pure degeneracy blocking"; this is done by
making $\alpha_k \to 0$ (suppressing topological decoherence), and letting
$\omega_k^{\perp} =0$ (which suppresses orthogonality blocking) but now
allowing a {\it distribution} of different values of
$\omega_k=\omega_k^{\parallel} $. We shall see the precise meaning of these
appellations as we study each limiting case.

Naturally, the generic behaviour of $H_{eff}$, with completely arbitrary
values of $\alpha_k,\; \omega_k^{\perp},\;\omega_k^{\parallel}$ and $\xi_k$
involves all three mechanisms; it can also involve nuclear spin diffusion.
 We shall see how this works in section
\ref{sec:5}.

It is crucial in what follows ro remember that a necessary (but {\it not}
sufficient) condition for coherence is that the total effective bias
acting on $\vec{S}$ be very small - the initial and final states of
$\vec{S}$ must be within $\sim \Delta_o$ of each other, otherwise
tunneling cannot occur at all. Even if it can, it may be {\it incoherent}.

\subsection{Topological decoherence}

Formally the case of pure topological decoherence applies to the following
special case of $H_{eff}$:
\begin{equation}
H_{eff}^{top} = 2{\tilde \Delta}_o {\hat \tau }_x \cos \big[ \Phi   +
\sum_{k=1}^N
 \alpha_k {\vec n}_k \cdot {\hat {\vec \sigma }}_k  \big] \;,
\label{4.1}
\end{equation}
Notice here that (a) since $\omega_k^{\parallel}$ and
$\omega_k^{\perp}$ are zero, {\it all} environmental states are degenerate, and
so (b)  the initial and final states of the environmental spins,
and $\vec{S}$,
 are degenerate - there is no exchange of any
  energy between ${\vec S}$ and the $ \{ {\hat \sigma }_k \}$. The only thing
that is
exchanged is phase, i.e., the phase of ${\vec S}$ becomes entangled  with that
of the $ \{ {\hat \sigma }_k \}$, during the tunneling of ${\vec S}$. Thus the
initial
and final {\it states} of the spin environment are not the same.
The term proportional to $\xi_k$
which changes the real part of the instanton action and tends to choose
either clockwise or counter-clockwise trajectories will be included
in  section \ref{sec:5}.

The properties of $H_{eff}^{top}$ were entirely elucidated in refs.\cite{1,2}
(see section 2.2 (ii) of ref.\cite{2}); here we explain the derivation
and results in more detail. What we are interested in are the
correlation functions of ${\vec S} (t)$, most particularly in
$P_{\Uparrow \Uparrow} (t)$ and its Fourier transform
in the frequency domain. This function simply tells us
the  probability to find the system  in the same state  at time $t$
as  it was  at time $t=0$; its Fourier transform tells us about the
response of the system to a continuous perturbation at some
frequency $\omega$. In the case where all interactions are zero we
simply revert to the free central spin, for which
\begin{equation}
P_{\Uparrow \Uparrow}^{(0)} (t) = {1 \over 2} [1 + \cos (4\Delta_o t\cos \Phi )
]
\label{4.2}
\end{equation}
\begin{equation}
\chi_o^{\prime  \prime } (\omega )= \pi \delta
(\omega - \mid 4\Delta_o \cos \Phi \mid ) \; ; \;\;\;\; (for \;\omega >0) \;.
\label{4.3}
\end{equation}
Here $\chi_o^{\prime  \prime }$ is the imaginary part of $\chi (\omega )$,
the Fourier transform of $P_{\Uparrow \Uparrow} (t)$. In this non-interacting
case
$\Phi = \pi S$, and $\chi_o^{\prime  \prime }$ is just a sharp line at
the tunneling splitting energy $\mid 4\Delta_o \cos \pi S \mid$. The
spin oscillates coherently between
$\mid {\vec S}_1 \rangle $ and $\mid {\vec S}_2 \rangle $. The series
(\ref{4.2}) is
generated from (\ref{4.1}) (or alternatively, from the bare transition
amplitude (\ref{3.1})), in the absence of the $ \{ {\hat \sigma }_k \}$, by
the usual summation of all possible instanton flips in a time $t$ (see
refs.\cite{1,12,33,34,43}).

Returning now to $H_{eff}^{top}$, we see that we can formally write the
solution as
\begin{equation}
P_{\Uparrow \Uparrow} (t) = {1 \over 2} \left\{ 1 + \langle \cos \big[
4\Delta_o t \cos \big(
\Phi   + \sum_{k=1}^N
 \alpha_k {\vec n}_k \cdot {\hat {\vec \sigma }}_k \big)  \big]  \rangle
\right\}\;,
\label{4.4}
\end{equation}
simply treating the cosine function in $H_{eff}^{top}$ as a $c$-number,
because there are no other terms in the Hamiltonian, and it obviously
commutes with itself.
Or, if one prefers an expansion over all instanton trajectories,
we can  write
\begin{equation}
P_{\Uparrow \Uparrow}={1 \over 2} \left\{ 1+\sum_{s=0}^{\infty} (-1)^{s}
{ (2{\tilde \Delta}_o t)^{2s} \over (2s)! } \sum_{n=0}^{2s} {(2s)! \over
(2s-n)!n!}
e^{i\Phi (2s-2n)}\:  \langle \prod_{k=1}^{N}
e^{i\alpha_k {\vec n}_k \cdot {\hat {\vec \sigma }}_k (2s-2n)} \rangle \right\}
\;.
\label{4.5}
\end{equation}
 $\langle ... \rangle$ is a thermal average over
environmental states. Here
all these states are degenerate, the average is independent of $T$ and
 one finds
\begin{eqnarray}
F(n-m) &= &\langle \prod_{k=1}^{N}
e^{i\alpha_k {\vec n}_k \cdot {\hat {\vec \sigma }}_k (2s-2n)} \rangle
\nonumber \\
 & =& \prod_{k=1}^{N} \cos [ 2\alpha_k (n-m)] \;.
\label{4.6}
\end{eqnarray}
In the non-interacting case, $F(\nu )=1$, and we get  (\ref{4.2}).
For small $\alpha_k$ we may approximate the product in (\ref{4.6}) as
\begin{equation}
F_\lambda (\nu ) = e^{-4 \lambda \nu^2 }\;; \;\;\;
\lambda = \sum_{k=1}^N \alpha_k^2/2 \;.
\label{4.7}
\end{equation}

Notice that $\lambda$ is just the mean number of environmental spins
that are flipped.
We now employ a very useful identity
\begin{equation}
\int_{0}^{2 \pi} {d\varphi \over 2 \pi } e^{i(\Phi -\varphi )(2s-2n)}
\cos ^{2s} \varphi \equiv {(2s)! \over 2^{2s} (2s-n)!n!}
 e^{i\Phi (2s-2n)}
\label{4.88}
\end{equation}
to rewrite the expansion (\ref{4.5}) as a {\it weighted integration over
topological phase}:
\begin{eqnarray}
P_{\Uparrow \Uparrow} &= &
\sum_{m=-\infty }^{\infty} F_\lambda(m)
\int_{0}^{2 \pi} {d\varphi \over 2 \pi } e^{i2m(\Phi -\varphi )}
\left\{ {1 \over 2}+{1 \over 2} \cos (2 \Delta_o ( \varphi ) t) \right\}
\label{4.8a} \\
&= &{1 \over 2} \left\{ 1+\sum_{m=-\infty }^{\infty} F_\lambda(m)
e^{i2m\Phi } J_{2m} (4{\tilde \Delta}_o t ) \right\} \;,
\label{4.8}
\end{eqnarray}
where $\Delta_o(\varphi )= 2{\tilde \Delta}_o  \cos \varphi $, and
$J_{2m}(z) $ is the Bessel function of $m$-th order. For $F(m)=1$
this is just another representation of the coherent dynamics (\ref{4.2}).
On the other hand in the adiabatic strong-coupling limit, where all of the
$\alpha_k \to \pi /2$, we simply have $F(m) = (-1)^m$, so that
\begin{equation}
P_{\Uparrow \Uparrow} (t) \to {1 \over 2} [1 + \cos (4{\tilde \Delta}_o t\cos
\tilde{\Phi } )]\;;
\;\;\; ( \alpha_k \to \pi /2 )\;,
\label{4.9}
\end{equation}
where the renormalized phase is just $\Phi +N \pi /2 $, that is
\begin{equation}
\tilde{\Phi } = \pi S + \sum_{k=1}^N \phi_k^B
\label{4.10}
\end{equation}
where $\phi_k^B$ is the Berry phase, defined as before via
\begin{equation}
\phi_k^B = \phi_k -\pi /2  \;.
\label{4.11}
\end{equation}

Equation (\ref{4.9}) simply tells us that
in the strong coupling limit the total topological phase
now comes from the bare phase $\pi S$, plus the adiabatic Berry phase
accumulated by all the environmental spins rotating
adiabatically with it. If the
coupling Hamiltonian is such that, e.g., all the $\{ {\hat \sigma }_k \}$ are
forced
to be parallel to ${\vec S}$ in this limit, then this extra phase will just
be $N \pi /2$ (i.e., we have
effectively increased our spin quantum number from $S$ to $S+N/2$). However
this is by no means necessary  - if there is some other field acting on
${\vec \sigma }_k$ in addition to that due to ${\vec S}$ then the Berry phase
will
deviate from $\pi /2$.

By far the most important regime is the "intermediate coupling" regime,
in which $\alpha_k > N^{-1/2}$ (note that if $N$ is large, then
$\alpha_k$ may be very small here). This is typically going to be the case,
since usually $\alpha_k$ will be independent of $N$. In this case we have a
large parameter $\lambda$ in (\ref{4.7}) with
\begin{equation}
F_\lambda( \nu ) = \delta_{\nu ,0 } + \; small\; corrections  \;\;\;\;\;
(intermediate)\;.
\label{4.12}
\end{equation}
so that, very surprisingly, we get a {\it universal form} in the intermediate
coupling limit for $P_{\Uparrow \Uparrow} (t) $  (again, $\eta (x) $ is the
step function)
\begin{equation}
P_{\Uparrow \Uparrow}(t)  \longrightarrow  {1\over 2}\biggl[ 1+J_0(4{\tilde
\Delta}_o t) \biggr] \equiv
\int {d\varphi \over 2 \pi } P_{\Uparrow \Uparrow}^{(0)}(t,\Phi =\varphi)
\label{4.13a}
\end{equation}
(compare the angular average of the coherent series in (\ref{4.8})),
and $\chi^{\prime  \prime }(\omega )$
\begin{equation}
\chi ^{\prime \prime } (\omega ) \longrightarrow  {2 \over (16{\tilde
\Delta}_o^2 -\omega^2)^{1/2}}\:
\eta (4{\tilde \Delta}_o -\omega )\;\;\;\; (intermediate) \;.
\label{4.13b}
\end{equation}
We plot these universal forms  in Fig.5. The physics of this universal
form is simply one of phase cancellation in the expression
(\ref{4.6}) for $F(m)$. As explained in ref.\cite{1}, this phase cancellation
arises because successive flips of ${\vec S}$ cause, in general, a {\it
different}
topological phase to be accumulated by the spin environment, so that
when we sum over successive instantons for ${\vec S}$, we get phase
randomisation
and hence loss of coherence. This forces $2s-n=n$ in the sum in (\ref{4.5}),
i.e., the only paths that can contribute to $P_{\Uparrow \Uparrow} (t)$ are
those for
which the number of clockwise and anticlockwise flips of ${\vec S}$ are
{\it equal} - in this case the topological phase "eaten up" by the
environmental
spins is zero. By looking at expression (\ref{4.13a}) one sees another
explanation - the universal behaviour comes from complete phase
phase randomisation \cite{1},
so that all possible phases contribute equally to the
answer! The final form shows decaying oscillations , with an envelope
$\sim t^{-1/2}$ at long times. This decay can also be understood \cite{2}
by noting that the "zero  phase" trajectories that contribute to $P_{\Uparrow
\Uparrow}$
now constitute a fraction $(2s)!/(2^{s}s!)^2 \sim s^{-1/2}$ of the total
number of possible trajectories, where $s \sim {\tilde \Delta}_o t$. Because of
this decay,
the peak in the spectral function at $\omega = 4{\tilde \Delta}_o \cos \Phi $
is now
transformed to the spectral function of a 1-dimensional tight-binding model,
as shown in Fig.5b.

{}From the point of view of fundamental principles this result is really
rather interesting, for here we have a case of "{\it decoherence
without dissipation}"; the coherent oscillations of ${\vec S} (t)$ are
destroyed without any exchange of energy with the environment. This
mechanism would persist even if the bath were at zero temperature.
Whilst there is no reason why such "pure decoherence" cannot exist elsewhere
in nature, we are not aware of other examples \cite{44}.

It is actually very interesting to examine the intermediate coupling
regime a little bit more closely, especially for the $1/2$-integer spin,
when there is no tunneling for $\alpha_k=0$. In Fig.6 we see that
 weak coupling ($\alpha_k$ taking random values between
 $0$ and $0.05$) to 40 spins  completely changes this behaviour -
we now have an effective
splitting $\Delta_{eff} \sim {\tilde \Delta}_o /3$,
 but with strong damping .
Fig.7 shows rather stronger coupling ($\alpha_k$
having random values between $0$ and $1$). The result shown is for
$S=$integer, but in fact the result for $S=n+1/2$ is almost indistinguishable
from this. We see that the result is, within numerical accuracy, completely
indistinguishable from the universal behaviour in (\ref{4.13a}), for
only 20 environmental spins.

\subsection{Orthogonality blocking}

Orthogonality blocking arises when there is a mismatch between the initial and
final state wave-functions of the environment \cite{1,2,3}. Expressed in
this way it is an old idea,
as old as the Debye-Waller or Franck-Condon, or polaronic factors for
superohmic environments, or the "orthogonality catastrophe" of the $X$-ray
edge and Kondo problems \cite{24}, which reappears in the
discussion of Ohmic environments in the oscillator bath model of
the environment \cite{12}.

In the present case of the spin bath, the mismatch between
$\mid \sigma_{k}^{in} \rangle $ and $\mid \sigma_{k}^{fin} \rangle $
arises because the initial and final fields, ${\vec \gamma }_k^{(1)}$ and
${\vec \gamma }_k^{(2)}$, acting on ${\vec \sigma }_k$, are not exactly
parallel or antiparallel. Thus if ${\vec \sigma }_k$ is initially aligned
in ${\vec \gamma }_k^{(1)}$, it suddenly finds itself, after $\vec{S}$ flips,
to be misaligned with ${\vec \gamma }_k^{(2)}$. There is then an overlap with
a {\it flipped } state in the new basis, equivalent to an amplitude
$\beta_k$ for ${\vec \sigma }_k$ to have actually flipped (semiclassically,
$\mid \sigma_{k}^{fin} \rangle $ starts precessing in ${\vec \gamma
}_k^{(2)}$).
Depending on the readers' taste, one may think of this as a "quasi-flip"
generated by the basis change, or a real flip; however quantum-mechanically,
for all purposes the spin has actually flipped.

Thus superficially we are dealing with a standard orthogonality problem.
However as we now see the details are rather different.

We start by applying the following restrictions to $H_{eff}$ in (\ref{3.4}):

(i) We assume $\alpha_k^2N \ll 1$, i.e., extremely small $\alpha_k$; this
removes topological decoherence.

(ii) We assume that the effective fields ${\vec \gamma }_k^{(1)}$ and
${\vec \gamma }_k^{(2)}$ acting on ${\vec \sigma }_k$ at the beginning and end
of the
instanton are not either parallel or antiparallel; usually we will
assume however that they are {\it nearly} parallel or antiparallel, i.e., that
either  ${\vec \gamma }_k^{(1)} \approx {\vec \gamma }_k^{(2)}$
(i.e. that $\omega_k^{\perp} \gg \omega_k^{\parallel}$), or else
${\vec \gamma }_k^{(1)} \approx - {\vec \gamma }_k^{(2)}$
(so that $\omega_k^{\perp} \ll \omega_k^{\parallel}$). Both of these cases
are very common, indeed almost the rule - quite apart from any stray fields
that might disrupt perfect parallelism, there is also the contribution coming
from barrier fluctuations (cf. Appendix A) which will contribute to
orthogonality blocking even if parallelism is otherwise is perfect.
Stray fields will come from a variety of sources (weak external fields, or
strain fields in the sample, or
demagnetisation fields from the dipole forces, etc., etc.).

(iii) We assume that the couplings $\omega_k$ satisfy the inequalities
$\omega_k \gg {\tilde \Delta}_{eff} $, where ${\tilde \Delta}_{eff} $ will be
defined below.
The parameter ${\tilde \Delta}_{eff} < {\tilde \Delta}_o$ in general; often
${\tilde \Delta}_{eff} \ll {\tilde \Delta}_o$.
This condition  is virtually always obeyed for hyperfine interactions, since
we expect ${\tilde \Delta}_{eff} < 1MHz$.

(iv) We assume that all the $\omega_k$ are {\it equal}. This is of course
a very unrealistic assumption (even the very small spread in hyperfine
couplings will turn out to be important), but it is made to avoid
the introduction of degeneracy blocking, to which we will come presently.

To be specific let us consider the case where
${\vec \gamma }_k^{(1)} \approx - {\vec \gamma }_k^{(2)}$, which will arise
in many situations; the analysis for the other most common problem, where
${\vec \gamma }_k^{(1)} \approx  {\vec \gamma }_k^{(2)}$, goes through in
exactly the same way. Thus "pure orthogonality blocking" applies
to the effective Hamiltonian
\begin{equation}
H_{eff} = 2{\tilde \Delta}_\Phi {\hat \tau }_x  + {\hat \tau }_z
{\omega_o^{\parallel} \over 2}
\sum_{k=1}^N  \:
 {{\vec l}_k \cdot {\hat {\vec \sigma }}_k } +  {\omega_o^{\perp} \over 2}
 \sum_{k=1}^N
\: {{\vec m}_k \cdot {\hat {\vec \sigma }}_k } \;.
\label{4.14h}
\end{equation}
\begin{equation}
{\tilde \Delta}_\Phi = {\tilde \Delta}_o \cos \Phi  \;,
\label{4.14g}
\end{equation}
where $ \omega_o^{\parallel} \gg \omega_o^{\perp}$; in (\ref{4.14h}) we
use a "Kramers renormalised" splitting ${\tilde \Delta}_\Phi$.

We now make a number of observations, which are crucial to understanding
of orthogonality blocking:

(i) The spin bath spectrum is split by $\omega_o^{\parallel}$ into
"polarization groups" of degenerate lines, having polarisation
$\Delta \!N =
N^{\uparrow} - N^{\downarrow}$, and energy
$E_{\tau}( \Delta \!N )= \tau_z \omega_o^{\parallel} \Delta \!N/2$; that is,
we have $N+1$ polarisation groups each separated by energy
$\omega_o^{\parallel}$, whose energy {\it changes} (by
$\omega_o^{\parallel} \Delta \!N$) if $\vec{S}$ flips:
$E_{\Uparrow}( \Delta \!N )= -E_{\Downarrow}( \Delta \!N ) =
\omega_o^{\parallel} \Delta \!N/2$. There are $C^{(N+\Delta \!N )/2}_N$
degenerate states in polarisation group $\Delta \!N $.

(ii) For $\vec{S}$ to flip at all, we require near resonance between
initial and final states, i.e., they must be within
$\sim {\tilde \Delta}_\Phi $ in energy. Since $\omega_o^{\parallel} \gg
{\tilde \Delta}_\Phi $,
this means that if no nuclear spins are flipped during the
transition of $\vec{S}$, only states having $\Delta \!N =0$
(the "zero polarisation" states") can flip at all (since
$E_{\Uparrow}( \Delta \!N =0)= E_{\Downarrow}( \Delta \!N=0 ) = 0$).
At high $T$, only a fraction $f = \sqrt{2/\pi N}$ of grains in an ensemble
will have $\Delta \!N =0$.

(iii) However the presence of $ \omega_o^{\perp}$ means that some spins are
flipped (in general a different number of them in each transition of
$\vec{S}$).
Now consider a grain in polarisation state $\Delta \!N =M$. If when $\vec{S}$
tunnels, at the same time nuclear spins also flip so that
$\Delta \!N \to -\Delta \!N$, then since
$E_{\Uparrow}( M)= E_{\Downarrow}( -M)$, resonance is still preserved and the
transition is possible. Notice however that (a) for this change in polarisation
of $2M$, {\it at least} $M$ spins must flip, and (b) if we wish to preserve
resonance (i.e., $\vec{S}$ makes  back-and-forth
transitions) then the polarisation state must change by $\pm 2M$
{\it each time}. In fact $\vec{S}$ cannot flip at all {\it unless} there is
a change in polarisation state of magnitude $2M$.

These 3 observations will lead to a fourth, which is most easily given once we
have defined a function $P_M(t)$, which is simply the correlation function
$P_{\Uparrow \Uparrow}(t)$, now restricted only to those systems in an
ensemble for which $\Delta \!N =M$. Since, as remarked above, we can only have
transitions between $\pm\Delta \!N$ states, we can write the complete
correlation function for an ensemble of grains described by (\ref{4.14h}) as
\begin{equation}
P_{\Uparrow \Uparrow}(t;T)= \sum_{M=-N}^{N} w(T,M) P_{M} (t) \;,
\label{4.25x}
\end{equation}
\begin{equation}
w(T,M)= Z^{-1} C_N^{(N+M)/2} e^{-M\omega_o^{\parallel} /k_BT} \;,
\label{4.14x}
\end{equation}
where $Z$ is the partition function.

We now make the observation

(iv) If we are interested in the incoherent relaxation of $\vec{S}$,
particularly at long times, then obviously all the $P_{M} (t)$ in
(\ref{4.25x}) must be considered. However for {\it coherence} it is
obvious that the $P_{0} (t)$ must completely dominate, since it involves
transitions of $\vec{S}$ with possibly {\it no} flipping of the
$\{ \vec{\sigma}_k \}$, which are necessarily coherent. On the other hand
$P_{M\ne 0} (t)$ not only involves many spin flips each time (necessarily
causing loss of coherence), but it also involves only processes in which
some number $\ge M$ of spin flips leads to exactly a change $2M$ in
polarisation. From this argument we see that $P_{M\ne 0} (t)$ will
describe {\it slow} (i.e., for large $M$ on time scales
$\gg \tilde{\Delta}_\Phi ^{-1}$) and {\it incoherent} motions of $\vec{S}$.

These observations are borne out by the calculation of $P_{M} (t)$ in
Appendix A.2. Assuming that
$ \omega_o^{\parallel} \gg \omega_o^{\perp}$ as above, one easily sees that the
flip amplitude $\beta_k$ is just the small angle defined as shown in Fig.8, via
\begin{equation}
\cos 2\beta_k =  - {\hat \gamma }_k^{(1)} \cdot {\hat \gamma }_k^{(2)}  \;,
\label{4.15}
\end{equation}
where ${\hat \gamma }_k^{(i)}$ are unit vectors; it is then found that
\begin{eqnarray}
P_{M}(t) &= & \int_0^\infty d\!x x\:e^{-x^2} \bigg( 1+ \cos
[4{\tilde \Delta}_\Phi J_M(2\sqrt{\kappa } x)t]  \bigg)
\label{4.21o} \\
&= & 2\int_0^\infty d\!x x\:e^{-x^2}
P_{\Uparrow \Uparrow}^{(0)}(t,\Delta_M(x))\;,
\label{4.21a}
\end{eqnarray}
where $P_{\Uparrow \Uparrow}^{(0)}(t,\Delta_M)$ is just the {\it free}
function (Eq.(\ref{4.2})), but now a function of an $x$-dependent
splitting $\Delta_M(x) = 2{\tilde \Delta}_\Phi J_M(2\sqrt{\kappa } x)$;
and $\kappa$ is just the orhtogonality exponent, i.e.,
\begin{equation}
e^{-\kappa } = \prod_{k=1}^N \cos \beta_k  \;,
\label{4.20}
\end{equation}
so that $\kappa \approx 1/2 \sum_k \beta_k^2$. Thus the frequency
scale of $P_{M}(t)$ is set by $\Delta_M(x)$. The terms $P_{M}(t)$ are
easily verified to be incoherent, and so
\begin{equation}
P_{\Uparrow \Uparrow}(t) \sim f P_{0}(t) + {\rm incoherent} \;,
\label{4.20w}
\end{equation}
where $f=\sqrt{2/\pi N}$ as before; the exact answer is given still by
(\ref{4.25x}) above.

The frequency scale of oscillations in $P_{0}(t)$ is easily found; we write
it in terms of a renormalised splitting $\Delta_{eff}$:
\begin{equation}
\left.
\begin{array}{l}
P_{0}(t)=1/2 [ 1+ \cos (4{\tilde \Delta}_\Phi e^{-\kappa} t) ]\;\; \\
\Delta_{eff} = 2{\tilde \Delta}_\Phi e^{-\kappa}
\end{array} \right\} \;\; \kappa \ll 1
\label{4.22}
\end{equation}
for small $\kappa$, and
\begin{equation}
\left.
\begin{array}{l}
P_{0}(t)= 1  -  4({\tilde \Delta}_\Phi^2t^2 / (\pi \kappa )^{1/2})
+O({\tilde \Delta}_\Phi^4 t^4 ) \;\;  \\
{\tilde \Delta}_{eff} = 2{\tilde \Delta}_\Phi /(\pi \kappa)^{1/4}
\end{array} \right\} \;\; \kappa \gg 1
\label{4.23}
\end{equation}
for large $\kappa$. Thus the naive "orthogonality catastrophe
polaronic band narrowing" argument to derive
${\tilde \Delta}_{eff}$ only works for
$\kappa \ll 1$ (where it is superfluous!). In the interesting regime
$\kappa \gg 1$, we get a quite different answer (with a much smaller
suppression). This is understood as follows. In the usual oscillator
bath models, band narrowing comes essentially without any bath
transitions (most of the polaron "cloud" is in virtual high-frequency
modes) - it is adiabatic. Here, however, roughly $\kappa$ spins
flip each time $\vec{S}$ flips (the probability of $r$ flips is
$\kappa^re^{-\kappa}/r!$, which peaks at $r \sim \kappa$), even though
we only consider $P_0(t)$, i.e., even though $\Delta \!N=0$ and does not change
(the flips compensate - there are just as many one way as the other).

This inevitable flipping of the $\{ \vec{\sigma}_k \}$ means that even
$P_0(t)$ is incoherent if $\kappa \gg 1$. This is most easily seen by looking
at  $\chi_0 ^{\prime \prime} (\omega )$ which has extraordinary behaviour.
One finds
\begin{equation}
\chi_{M =0} ^{\prime \prime } (\omega ) =\sqrt{{2 \over \pi N}}\;
{\pi \over 4{\tilde \Delta}_\Phi \sqrt{\kappa } } \sum_j
{x_j e^{-x_j^2} \over \vert J_1(2\sqrt{ \kappa } x_j) \vert }\;\bigg|_{\it
J_0(2\sqrt{\kappa } x_j)=\pm \omega/ 4{\tilde \Delta}_\Phi } \;.
\label{4.25}
\end{equation}
 The first factor is just the statistical
fraction of zero polarization states in the spin environment.
It is the complicated structure of the set of roots $x_j$ to the equation
$J_0(2\sqrt{\kappa } x_j)=\pm \omega/ 4{\tilde \Delta}_\Phi $ that yields the
interesting
behaviour. For small $\kappa$ only the lowest square root Bessel function
singularity appears, coming from $x_1 \approx
[(1-\omega /4{\tilde \Delta}_\Phi )/\kappa ]^{1/2}$. This leads to
\begin{equation}
\chi_{M =0} ^{\prime \prime }(\omega ) =
{\pi \over 4{\tilde \Delta}_\Phi \kappa } \exp \left( {\omega -4{\tilde
\Delta}_\Phi \over 4{\tilde
\Delta}_\Phi \kappa} \right)
\:\eta (4{\tilde \Delta}_\Phi  - \omega )\;,
 \;\;\;\;( \kappa \ll 1)\;,
\label{4.26}
\end{equation}
and almost coherent dynamics for ${\vec S}$, as we see in Fig.9. However
as $\kappa$ increases, more  and more square root
singularities contribute to $\chi ^{\prime \prime }$ (which also gets pushed
to lower $\omega$, according to (\ref{4.23})), and it develops the
bizarre structure seen in Figs. 9$\&$ 10. This structure has
 quite counter-intuitive consequences for $P_{0}(t)$;
interference between the various peak structures can  conspire to give
the correlation function for ${\vec S}$ a multi-periodic behaviour. An example
is shown in Fig.11.

 Readers who  wish to more fully understand the mechanism of orthogonality
blocking are urged to study
Appendix A. Recall also that for a real system, the values of the $\beta_k$
will be renormalized by the $\overline{\beta}_k$,
 (Eq.(\ref{3.4}) {\it et seq.}),
which describe the effect of barrier fluctuations.
Thus the real value of $\kappa$, and the real values of
$\omega_k^{\perp}$ and $\omega_k^{\parallel}$, will already include
these barrier fluctuation effects.

\subsection{Degeneracy blocking}

The previous 2 limiting cases studied have imposed some rather artificial
restrictions in order to bring out 2 physical mechanisms involved in
decoherence from spin environments. We now study a third limiting case which
introduces a note of realism.
To do this we suppress topological decoherence (by letting
$\alpha_k^2N \ll 1$) and we suppress orthogonality blocking (by letting $
\omega_k^{\perp} =0$). On the other hand we now allow a {\it spread }
of coupling energies $\omega_k^{\parallel}$ around a mean value $\omega_o$.
Thus we consider the effective Hamiltonian
\begin{equation}
{\hat H}_{eff}=2{\tilde \Delta}_\Phi  {\hat \tau }_x +
{\hat \tau }_z \sum_{k=1}^N \omega_k^{\parallel} \:{\hat \sigma }_k^z \;;
\label{4.27}
\end{equation}
with a spread of values of $\omega_k^{\parallel}$ of
\begin{equation}
\sqrt{\sum_{k} (\omega_k^{\parallel}-\omega_o)^2 }
\equiv N^{1/2} \delta \omega_o \;.
\label{4.28}
\end{equation}
i.e., a distribution of width $\delta \omega_o$.
 Clearly, this Hamiltonian is identical to the standard biased
two-level system with the bias energy $\epsilon $ depending on the particular
environmental state; thus
 $\epsilon = \sum_{k=1}^N \omega_k^{\parallel} \sigma_k^z $.
The introduction of this spread is to destroy the exact degeneracy between
 states in the same polarisation group. For coherence to take place,
we require the initial and final states to be within roughly
${\tilde \Delta}_{eff}$ of each other,
otherwise Landau-Zener suppression of the
transition takes place (cf. Appendix A.1).
Since we assume a random distribution of
couplings around $\omega_o$, it is clear that there will be  a crossover
between unblocked behaviour for $\delta \omega_o \ll {\tilde \Delta}_o
/N^{1/2}$, and Landau-Zener
blocked behaviour for $\delta \omega_o \gg {\tilde \Delta}_o /N^{1/2}$. In
practice
once $\delta \omega_o \sim {\tilde \Delta}_o /N^{1/2}$,
the crossover is essentially complete,
and tunneling is blocked for almost all grains in an ensemble.
Thus for large $N$, degeneracy blocking is a very powerful mechanism
for suppressing coherent oscillations of ${\vec S}$, since  only a tiny spread
$\delta \omega_o$ will be required to do the job. Now we look at this
quantitatively.

The trivial case is of course where $\delta \omega_o =0$, and the spectrum of
 $\{ {\vec \sigma}_k \}$ is organized in highly degenerate lines corresponding
to
different polarizations $\epsilon = \omega_o \Delta \!N/2$.
Then, as we discussed in the last subsection  any  zero polarization
state of $\{ {\vec \sigma}_k \}$ gives an exact resonance for ${\vec S}$,
and the corresponding correlation function is an ideal one , i.e.,
we get $P^{(0)}_{\Uparrow \Uparrow}(t)$ as in (\ref{4.2});
however only a small fraction of possible environmental states have
$\Delta \!N =0$, and for large $N$ this fraction is defined by the
central limit theorem as $f=\sqrt{2/ \pi N}$. Thus the corresponding
spectral function is  like (\ref{4.3}) but with a total weight $f$, i.e.,
\begin{equation}
\chi_o^{\prime  \prime } (\omega )\to \sqrt{2\pi / N} \delta
(\omega - 4{\tilde \Delta}_\Phi  ) \; ; \;\;\;\; (\delta \omega_o =0) \;.
\label{4.29}
\end{equation}

We may simply disregard the contribution to $\chi^{\prime \prime }$
coming from states with nonzero $\Delta \!N$, because for
 $\omega_o \gg {\tilde \Delta}_{eff}$
these states are too far off the resonance. Thus, as we see from (\ref{4.29}),
in this special case the only effect of the coupling $\omega_o$ to the
$N$ spins is a reduction of the spectral weight - otherwise the
central spin behaviour is completely coherent. This is hardly
surprising, since we have removed all decoherence mechanisms - we have no
topological decoherence, no degeneracy blocking, and no orthogonality
blocking! This case was also studied very recently by Garg \cite{44} in an
attempt to discuss experiments in magnetic grains.

Going now to the case where we do have some degeneracy blocking, it is
useful to first study the regime where $\mu =\delta \omega_o N^{1/2} /\omega_o
> 1$ (in real
situations, this regime will almost always prevail in any case).
Under this regime the different groups of environmental states will be spread
out and overlap, and we will then get a continuous Gaussian distribution for
the bias energy $\epsilon$:
\begin{equation}
W(\epsilon )={1 \over \omega_o (2\pi N)^{1/2}} \exp \{ -\epsilon^2/
(2\omega_o^2 N) \}  \;;
\label{4.30}
\end{equation}
\begin{equation}
\epsilon = \sum_{k=1}^N \omega_k^{\parallel} \:{ \sigma }_k^z \;.
\label{4.31}
\end{equation}
This situation is depicted schematically in Figs.12a and 12b; Fig.12b also
depicts a case where $\mu <1$, i.e., with very small
degeneracy blocking. The correlation
function $P_{\Uparrow \Uparrow} (t)$ for an {\it ensemble} of central spins is
now obtained
by {\it averaging the known analytic behaviour for the simple biased
2-level system}. At high temperatures
\begin{eqnarray}
P_{\Uparrow \Uparrow}(t) &= &\int d\epsilon W(\epsilon )
{e^{-\beta \epsilon } \over Z(\beta )}
\bigg[ 1-
{2{\tilde \Delta}_\Phi ^2 \over \epsilon^2 + 4{\tilde \Delta}_\Phi^2 } \left(
1 -\cos (2t\sqrt{\epsilon^2+4{\tilde \Delta}_\Phi^2}  ) \right) \bigg] \\
\label{4.32}
&\longrightarrow & 1-2 A\sum_{k=0}^\infty J_{2k+1}(4{\tilde \Delta}_\Phi t) \;.
\label{4.33}
\end{eqnarray}
\begin{equation}
A=2\pi {\tilde \Delta}_\Phi /(2\omega_o \sqrt{2\pi N}) \equiv 2\pi f \cdot
{\tilde \Delta}_\Phi/4\omega_o \;.
\label{4.34}
\end{equation}
and a spectral function
\begin{equation}
\chi^{\prime \prime }(\omega ) =
A {8{\tilde \Delta}_\Phi \over \omega \sqrt{\omega^2-16{\tilde \Delta}_o^2} }
\eta (\omega -4{\tilde \Delta}_\Phi )\;,
\label{4.35}
\end{equation}
which is shown in Fig.13. (Integrating over $\epsilon$ we used the inequality
$W(0) {\tilde \Delta}_\Phi  \ll 1$).
 We see that not only is there line-broadening,
but also the amplitude of the oscillations,
for the ensemble of central spins, is further reduced by a factor
 ${\tilde \Delta}_\Phi /2\omega_o $, as compared with the
trivial  case in (\ref{4.29}), and by a total factor $A$ as compared with the
ideal free central spin in (\ref{4.2}). We should note  at this point
 that in almost any realistic situation, a random distribution of
values for $\omega_k^{\parallel}$ for a  {\it single} central spin will lead to
results identical to the above, i.e., the ensemble average here is
merely a device to define the distribution $W(\epsilon )$.
 The only difference is that for a single central spin one has to
consider (\ref{4.33}) as a statistical average over many experimental runs
during a time which is
longer then the NMR relaxation times in the nuclear subsystem.

Turning now to the general case, where the parameter $\mu = \delta \omega_o
N^{1/2}/
\omega_o $ can take arbitrary values, we notice that
the integral over $\epsilon$ in (\ref{4.32}) is convergent on a scale
defined by ${\tilde \Delta}_\Phi$, and this result will not change at all for
{\it any}
distribution which is flat around zero. Thus for
$\omega_o \gg \delta \omega_o N^{1/2} >{\tilde \Delta}_\Phi $, i.e., $\mu \ll
1$ we
have the same answer as before, but the amplitude $A$ is now
replaced by $A=f \cdot {\tilde \Delta}_\Phi /(2\sqrt{2\pi N} \delta \omega_o )
={\tilde \Delta}_\Phi /(2\pi N \delta \omega_o )$.
Thus we find that even a very small spread in the coupling
constants will be sufficient to
broaden the spectral line. Note that neither the energy dissipation nor
the phase randomisation are responsible for this behaviour, which is
totally due to the statistical average over the initial states of
$\{ {\vec \sigma}_k \}$.


\section{The generic case}
\label{sec:5}

As we shall see, if we combine all 3 mechanisms just studied,
and calculate $P_{M} (t)$ for the full $H_{eff}$, we get
a very complex (but still a closed form) answer. Thus in order to
make the derivations more transparent, we first combine the mechanisms
in different combinations. This tactic is not just pedagogically useful -
it also gives 3 more limiting cases which correspond to experimentally
realistic situations. All this is done in Sections
\ref{sec:5}.A - \ref{sec:5}.D. Finally, in \ref{sec:5}.E, we deal with the
residual inter-nuclear couplings mentioned in Section \ref{sec:3},
Eq.(\ref{3.112}). Within a particular polarisation group $\Delta \!N$, the
inter-nuclear dipolar couplings cause "diffusion in bias", simply by
flipping random pairs of nuclei. We study the role this has on the
general solution for the coherent part of
$P_{\Uparrow \Uparrow} (t)$ given in  \ref{sec:5}.D.

\subsection{Projected Topological decoherence}

 We start from the case of
"{\it projected topological decoherence}", defined by the following
special case of (\ref{3.4}):
\begin{equation}
 H_{eff}=2{\tilde \Delta}_o  {\hat \tau }_x \cos \big[ \Phi   + \sum_{k=1}^N
\alpha_k {\vec n}_k   \cdot {\hat {\vec \sigma }}_k  \big] +
{\hat \tau }_z \sum_{k=1}^N \omega_k {\hat \sigma }_k^z \;,
\label{5.3}
\end{equation}
with the restrictions:

(i) The effective fields ${\vec \gamma}_k^{(1)} = - {\vec \gamma}_k^{(2)}$ , so
$\omega_k^{\perp}=0$.

(ii) The couplings $\omega_k \gg {\tilde \Delta}_o$ , and all couplings are
equal (thereby
eliminating  degeneracy blocking).

The second condition forces the polarization of the environment {\it with
respect to the central spin orientation} to remain constant.
Now at
first glance we have no orthogonality blocking at all in (\ref{5.3}),
since $\omega_k^{\perp}=0$ (compare section \ref{sec:4}.B), and so it
looks as though {\it only} zero polarisation group states will have any
dynamics.
However this is incorrect; because now the {\it dynamic couplings}
$\alpha_k$ can flip spins when $\vec{S}$ flips, with amplitude $\alpha_k$;
they play exactly the same role as the $\beta_k$ in pure orthogonality
blocking (Section \ref{sec:4}.B), and so we can say that in (\ref{5.3}),
{\it orthogonality blocking is generated by the} $\alpha_k$. Thus, in
the same way as in Eqs.(\ref{4.25x}) and (\ref{4.14x}) we write
$P_{\Uparrow \Uparrow}(t)$ as a weighted sum over
$P_{M}(t)$, where now (cf. Appendix B.1):
\begin{eqnarray}
P_{M}(t) &=& \int dx x e^{-x^2} \sum_{m=-\infty }^{\infty}
F_{\lambda^\prime}(m) \int {d\varphi  \over 2\pi }
e^{i2m(\Phi -\varphi ) }  \left\{ 1+
\cos [4{\tilde \Delta}_o t J_M(2x\sqrt{\lambda -\lambda^\prime}) \cos \varphi ]
\right\}
\label{5.6a} \\
& =& \int dx x e^{-x^2}  \left\{ 1+ \sum_{m=-\infty }^{\infty}
F_{\lambda^\prime}(m)
e^{i2m\Phi }
J_{2m} [4{\tilde \Delta}_o t J_M(2x\sqrt{\lambda -\lambda^\prime})] \right\}
\;,
\label{5.6}
\end{eqnarray}
where $F_{\lambda^\prime}(m)$ is defined analogously to
$F_{\lambda}(m)$ in (\ref{4.7}), viz.,
\begin{equation}
F_{\lambda^\prime}(m) = e^{-4 \lambda^\prime m^2 }\;;
\label{5.7}
\end{equation}
\begin{equation}
\lambda^\prime = {1 \over 2} \sum_{k=1}^N \alpha_k^2(n_k^z)^2  \;,
\label{5.8}
\end{equation}
and we notice that $\lambda \ge \lambda^\prime$.
Eq.(\ref{5.6a}) can be interpreted in 2 ways, both of which are important.
First, we can interpret it as an orthogonality-blocked expression, now with
frequency scale
\begin{equation}
\Delta_M(\varphi ,x) =2{\tilde \Delta}_o \cos ( \varphi )
J_M(2x\sqrt{\lambda -\lambda^\prime})  \;,
\label{5.8a}
\end{equation}
which is then {\it averaged over $\varphi $}, to give phase randomisation
(compare Eq.(\ref{4.8a})). Alternatively we can regard
it is an integration $\int dx$ over an {\it already topologically decohered}
function $P_{M}^{(top)} (t, \Delta_M(x))$, with $\Delta_M (x)$ now given
by $\Delta_M (x) = {\tilde \Delta}_o J_M(2x\sqrt{\lambda -\lambda^\prime})$;
compare again with Eq.(\ref{4.21a}).

This is the result for $P_M(t)$; but now we can go through exactly the same
arguments as for pure degeneracy blocking, to show that only
$P_0(t)$ may behave coherently, with a weight $\sim \sqrt{2/\pi N}$. Again
the exact expression is given by the weighted sum  (\ref{4.25x}).

There are various interesting cases of (\ref{5.6}) for $M=0$.
If $\lambda =
\lambda^\prime $ (i.e., ${\vec n}_k $ is parallel to ${\hat {\vec z}}$), then
we go
back to pure topological decoherence - the projection operator then commutes
with the cosine operator. On the other hand if $\lambda^\prime = 0$, we
have pure orthogonality blocking as stated earlier, and in fact
when $\lambda^\prime = 0$, the parameter $\lambda $   plays
the role of $\kappa$ in (\ref{4.21o}). Notice that whereas the case $\lambda =
\lambda^\prime $ can only occur accidentally, $\lambda^\prime = 0$ is
quite common -  indeed it pertains to the model in Eqs. (\ref{6.1}) and
(\ref{6.2}) which we will examine below in connection with real experiments.

We really begin to see the analogy with degeneracy blocking when
 $\lambda ,\:\lambda^\prime \gg 1$;
just as with pure topological decoherence, $F_{\lambda^\prime}(m)$ collapses
to a Kronecker delta, and we get the {\it universal} projected
topological decoherence form:
\begin{equation}
P_{\Uparrow \Uparrow}(t)  \longrightarrow \int dx x e^{-x^2}
 \biggl[ 1+J_0[4{\tilde \Delta}_o tJ_0(2x\sqrt{\lambda -\lambda^\prime})]
\biggr]
\label{5.9}
\end{equation}
\begin{equation}
\chi ^{\prime \prime } (\omega ) \longrightarrow \sqrt{ {2 \over \pi N }}
\int dx x e^{-x^2}
  {4 \over [16{\tilde \Delta}_o^2 J_0^2(2x\sqrt{\lambda -\lambda^\prime})
-\omega^2]^{1/2}}\:
\eta (4{\tilde \Delta}_o \mid J_0(2x\sqrt{\lambda -\lambda^\prime}) \mid
-\omega ) \;,
\label{5.10}
\end{equation}
which generalizes the result of (\ref{4.13b}) for pure topological
decoherence. Eq.(\ref{5.9}) should be compared to (\ref{4.21o}).

We show in Figs.14 and 15 some results for
$\chi ^{\prime \prime } (\omega )$ for selected values of $\lambda
-\lambda^\prime$. The results are startling; even a very small value of
$(\lambda -\lambda^\prime )$ significantly  washes out pure topological
decoherence; but for any large value of $\lambda^\prime $, we never get
back the pure orthogonality blocking spectrum.

\subsection{Orthogonality blocking plus degeneracy blocking}

We now look at what might be called "biased orthogonality blocking";
as explained in section \ref{sec:4}.C, if we introduce a spread
$\delta \omega_k$ about some $\omega_o$, then in effect we introduce
a bias $\epsilon$ which destroys the degeneracy of the environmental states.
Thus we now study the Hamiltonian
\begin{equation}
 H_{eff}=2{\tilde \Delta}_\Phi  {\hat \tau }_x   +
{\hat \tau }_z \sum_{k=1}^N \omega_k^{\parallel} \:
 {{\vec l}_k \cdot {\hat {\vec \sigma }}_k } + \sum_{k=1}^N
\omega_k^{\perp}\: {{\vec m}_k \cdot {\hat {\vec \sigma }}_k } \;.
\label{5.13}
\end{equation}
with the spread in frequencies  $\omega_k =[(\omega_k^{\parallel})^2+
(\omega_k^{\perp})^2]^{1/2} $ defined by Eq.(\ref{4.28}). The effective bias
$\epsilon =\sum_{k=1}^N \omega_k^{\parallel} {\hat \sigma }_k^z $ again, and
now we would like to take account of the effect of this bias on the
orthogonality blocking. Again, one starts by calculating a function $P_M$
referring to
polarisation subgroup $M$, but this time {\it in a bias} $\epsilon$. One finds
(Appendix B.2):
\begin{eqnarray}
P_{M} (t,\epsilon ) &= &2 \int dx xe^{-x^2}  \left\{
 1-{\Delta_M^2 ( x  )\over 2 E^2( x ) }
\bigg(
1 -\cos \big[ 2tE( x ) \bigg) \right\} \nonumber \\
 &= &2 \int dx xe^{-x^2} P_{\Uparrow \Uparrow}^{(0)} (t,\Delta_M ( x  ),
\epsilon ) \;.
\label{5.16}
\end{eqnarray}
\begin{equation}
E^2(x) = \epsilon^2 + \Delta_M^2(x) \;.
\label{5.16b}
\end{equation}

This expression could have been guessed without any calculation - it is just
the usual orthogonality blocking average $\int dx$, but now for a biased system
(instead of, as in pure orthogonality blocking, an unbiased expression -
cf. (\ref{4.21o}).

To get $P_{\Uparrow \Uparrow} (t)$, we must now both integrate over bias
{\it and} sum over $M$. Thus the generalisation of our previous weighted
average (\ref{4.25x}) is just
\begin{equation}
 P(t;T ) = \int d\epsilon W(\epsilon ) { e^{-\beta \epsilon }
\over Z(\beta )} \sum_{M=-N}^N
P_M(t,\epsilon -M\omega_o )\;.
\label{5.25x}
\end{equation}

However, yet again, we use the usual argument that only $P_0$ may produce
{\it coherent} response.
 Assuming again that the spread in couplings
satisfies $\mu =N^{1/2}\delta \omega_o /\omega_o >1$ (a trivial  modification
is necessary
for $\mu \omega_o/{\tilde \Delta}_o \gg 1$ but $\mu <1$, as discussed in
section
\ref{sec:4}.C), we then get for $M=0$
\begin{eqnarray}
P_{0} (t) &=& 1- 4 A  \int dx xe^{-x^2}
\mid J_0(2\sqrt{\kappa } x ) \mid  \sum_{k=0}^{\infty} J_{2k+1}
\big[ 4{\tilde \Delta}_\Phi \mid J_0(2\sqrt{\kappa } x ) \mid t  \big]
\label{5.18} \\
&=& 1-  2 \int dx xe^{-x^2}  2A(x)  \sum_{k=0}^{\infty} J_{2k+1}
\big[ 2 \Delta_0 ( x ) t  \big] \;,
\label{5.18b}
\end{eqnarray}
where the $x$-dependent spectral weight is
\begin{equation}
A(x) =A J_0(2\sqrt{\kappa } x )\;.
\label{5.18c}
\end{equation}

Yet again, by comparing with the previous results for pure orthogonality
blocking and pure degeneracy blocking, we could have guessed this (compare
Eq.(\ref{4.33}); we have just done an "orthogonality average" over a
"biased averaged" expression for the free system with $x$-dependent
tunneling frequency $\Delta_0(x)$.

The expression for $\chi_{M=0}^{\prime \prime }(\omega )$ follows
immediately; we have
\begin{equation}
\chi_0^{\prime \prime }(\omega ) =
{ 1 \over \omega } \int dx xe^{-x^2} 4A(x)
{ \Delta_0^x ) \over
[ \omega^2 - 4 \Delta_0 ^2 (x)  ]^{1/2} }
\eta ( \omega - 2 \mid \Delta_0  (x) \mid ) \;,
\label{5.19}
\end{equation}

 Figs.16 \& 17  show some representative plots for this coherent part
of $\chi^{\prime \prime }(\omega )$; it is in fact almost completely
incoherent, with total spectral weight

\begin{eqnarray}
\int_{-\infty}^{\infty} (d\omega /2\pi )
\chi^{\prime \prime }(\omega )
 & =& 2A  \int dx xe^{-x^2} \mid J_0(2\sqrt{\kappa } x)\mid
\nonumber \\
&=& {2\Gamma(3/4) \over \pi^{3/2} }{A \over \kappa^{-1/4} } \;;
\label{5.21}
\end{eqnarray}
a result which is very accurate even for $\kappa \sim 0.02 $. Of course,
once we include all other polarization sectors $M \ne 0$, the total
power increases - in fact it now becomes $\sim A \kappa^{1/4} $ for
large $\kappa$, since $\sim \kappa^{1/2}$ different polarization sectors
contribute. However none of this is coherent either.

\subsection{Pure topological decoherence plus degeneracy blocking}

Here we discuss the case when the central spin dynamics is governed
by pure topological decoherence is a biased system. With what we have
calculated already the answer is readily obtained as follows. We note that
the correlation function  in Eq.(\ref{4.8}) is clearly given as
a {\it coherent} correlation function $P_{\Uparrow \Uparrow}^{(0)} (t) $ (see
(\ref{b1.3}))
integrated over $\varphi $ and summed over $m$. Now, if we add the bias,
all we need to do is to substitute for the unbiased  correlation
function with the {\it non-interacting  biased} one from Eq.(\ref{b1.10}). Thus
we have
for the biased topological decoherence
\begin{equation}
P_{\Uparrow \Uparrow}(t) =
\sum_{m=-\infty }^{\infty} F_\lambda(m)
\int {d\varphi \over 2 \pi } e^{i2m(\Phi -\varphi )}
\left\{
1-{\Delta_0 ^2 (\varphi ) \over \epsilon^2 +  \Delta_0(\varphi )}
\bigg(
1 -\cos \big[ 2t\sqrt{\epsilon^2+ \Delta_0^2(\varphi )} \big] \bigg) \right\}
+ {\rm inc.} \;,
\label{u.1}
\end{equation}
Integrating over the bias with the usual assumption about the distribution
function we immediately find
\begin{equation}
P_{\Uparrow \Uparrow} (t) = 1-    \sum_{m=-\infty }^{\infty} F_\lambda(m)
\int {d\varphi \over 2 \pi } e^{i2m(\Phi -\varphi )}
2 A( \varphi )  \sum_{k=0}^{\infty}
J_{2k+1} [ 2 \Delta_0 ( \varphi )t ]  + {\rm inc.} \;,
\label{u.2}
\end{equation}
for the correlation function, with $\Delta_0 ( \varphi ) = 2{\tilde
\Delta}_\Phi
\cos \varphi $ as before, and $A( \varphi ) =A \cos \varphi $; and
\begin{equation}
\chi^{\prime \prime }(\omega ) =
{ 2A \over \omega } \sum_{m=-\infty }^{\infty} F_\lambda(m)
\int {d\varphi \over 2 \pi } e^{i2m(\Phi -\varphi )}
 { \cos^2 \varphi \over
[ (\omega /4{\tilde \Delta}_o )^2 - \cos^2 \varphi ]^{1/2} }
\eta ( \omega /4{\tilde \Delta}_o -
\mid  \cos \varphi \mid ) \;,
\label{u.3}
\end{equation}
for the spectral function.

One might be confused whether this calculation applies to any real
system, because on one hand the pure topological decoherence
implies  no (or at least very weak $\omega_k \ll {\tilde \Delta}_o $)
interaction between the
central spin and its environment; on the other hand including degeneracy
blocking we assume that the coupling is strong enough to move the system
out of the resonance. How one can reconcile this? Below we demonstrate that
this situation is described as a special limiting case of  our effective
Hamiltonian.

Suppose we have a very weak interaction between the central spin
and its environment so that $\omega_k \ll {\tilde \Delta}_o $. Since ${\tilde
\Delta}_o $ is exponentially
small as compared to $\Omega_o$ we are dealing with the case of
extremely small  $\alpha_k \sim \omega_k/\Omega_o \ll {\tilde
\Delta}_o/\Omega_o $.
 Imagine now that we have a really macroscopic number of environmental
spins contributing to the decoherence so that the small value of $\alpha_k$ is
compensated or even superseded by $N$ so that $\lambda \sim N \alpha_k^2$ is
of order unity or even large. One immediately understands that in this case
it is impossible for the central spin to tunnel and {\it not to flip} some
environmental spins, but since $\omega_k$ is much smaller than ${\tilde
\Delta}_o$ we are
still in resonance (if $\omega_k \lambda^{1/2} < {\tilde \Delta}_o$) and may
neglect energy dissipation.

It is quite obvious, however, that for such a large
number of microscopic spins both the parameter
$\mu = \delta \omega_k N^{1/2} /\omega_o \gg 1$,
and the level distribution width $N^{1/2} \omega_o \gg {\tilde \Delta}_o$ are
very large,
and we do have to consider tunneling in a biased system. The reason why the two
mechanisms are combined so easily is that the bias is produced by the total
effect of {\it all} spins, while only an extremely small fraction of them
flips when the central spin tunnels. As described elsewhere the numbers are
such that this special case has direct application to the environmental spin
decoherence in SQUIDs. For large parameter $\lambda $ Eq.(\ref{u.3})
simplifies to
\begin{equation}
\chi^{\prime \prime }(\omega ) =
{ 2A \over \omega }  \int {d\varphi \over 2 \pi }
 { \cos^2 \varphi \over
[ (\omega /4{\tilde \Delta}_o )^2 - \cos^2 \varphi ]^{1/2} } \eta ( \omega/
4{\tilde \Delta}_o -
\mid  \cos \varphi \mid ) \;,
\label{u.5}
\end{equation}
which can be expressed in terms of Elliptic functions.

\subsection{Combining the 3 mechanisms: the generic case}

With the preceding studies we hope that the following basic message has come
through, viz., that combining combinations of the 3 mechanisms simply
boils down to combining 3 weighted averages. These averages are
\begin{equation}
\hbox{ (a)  A "topological phase average" given by } \;\;\;\;\sum_{\nu =-\infty
}^{\infty} F_\lambda^\prime (\nu )
\int {d\varphi \over 2 \pi } e^{i2\nu (\Phi -\varphi )} \;;
\label{q.5}
\end{equation}
\begin{equation}
\hbox{ (b) An "orthogonality average" given by  } \;\;\;\;
2 \int_0^\infty  dx x e^{-x^2} \;;
\label{q.6}
\end{equation}
\begin{equation}
\hbox{ (c) A "bias average" } \;\;\;\;\int d\epsilon W(\epsilon ) { e^{-\beta
\epsilon }
\over Z(\beta )}\;.
\label{q.7}
\end{equation}

These averages apply to any polarisation group $M$; the final expression for
$P_{\Uparrow \Uparrow}(t)$ is then given by averaging the quantity
\begin{equation}
P_{\Uparrow \Uparrow}^{(0)} (t; \Delta_M(\varphi ,x); \epsilon )=
1-{ \Delta_M^2(\varphi ,x) \over E_M^2(\varphi ,x) }
\sin ^2 ( E_M(\varphi ,x) t ) \;,
\label{q.8}
\end{equation}
i.e., the {\it free} biased correlator for a splitting $\Delta_M $:
\begin{equation}
 \Delta_M(\varphi ,x) =  2{\tilde \Delta}_o \cos \varphi
\: J_M(2x\sqrt{\lambda -\lambda^\prime}) \;,
\label{q.9}
\end{equation}
\begin{equation}
 E_M^2(\varphi ,x) = \epsilon^2 + \Delta_M^2(\varphi ,x) \;.
\label{q.10}
\end{equation}

Thus, using this prescription, we can immediately write down {\it almost}
the final generic case; we have, assuming that in the effective Hamiltonian
both $\omega_k^{\perp} =0$ and $\xi_k =0$
\begin{equation}
 H_{eff}=2{\tilde \Delta}_o  {\hat \tau }_x
\cos \big[ \Phi   + \sum_{k=1}^N
 \alpha_k {\vec n}_k  \cdot {\hat {\vec \sigma }}_k  \big]
+ {\hat \tau }_z
\sum_{k=1}^N  { \omega_k \over 2} {\hat \sigma }_k^z\: \;.
\label{q.11}
\end{equation}
the full correlator is given by
\begin{equation}
 P_{\Uparrow \Uparrow }(t;T ) =
\int d\epsilon W(\epsilon ) { e^{-\beta \epsilon }
\over Z(\beta )} \sum_{M=-N}^N
P_M(t,\epsilon -M\omega_o ) \;;
\label{6.25x}
\end{equation}
\begin{equation}
 P_M(t;\epsilon ) = 2 \int_0^\infty  dx x e^{-x^2}
 \sum_{\nu =-\infty }^{\infty} F_\lambda^\prime (\nu )
\int {d\varphi \over 2 \pi } e^{i2\nu (\Phi -\varphi )}
 \bigg[ 1-{ \Delta_M^2(\varphi ,x) \over E_M^2(\varphi ,x) }
\sin ^2 ( E_M(\varphi ,x) t ) \bigg]               \;,
\label{6.25y}
\end{equation}
at any temperature. It should be emphasized that this expression is exact
provided the $\alpha_k$ are small
 and $\omega_k \gg \Delta_\Phi$ (which is virtually always the case in nature).

If we are only  interested in possible coherence, then it suffices only to
look at $M=0$ term in this expression, for reasons we have previously
discussed. Carrying out the bias average then just gives the obvious
generalization of Eq.(\ref{5.18b}), i.e.,
\begin{equation}
P_{\Uparrow \Uparrow}(t) = 1 -2 \int_0^\infty  dx x e^{-x^2} 2
\sum_{\nu =-\infty }^{\infty} F_\lambda^\prime (\nu )
\int {d\varphi \over 2 \pi } e^{i2\nu (\Phi -\varphi )}
A(\varphi ,x) \sum_{k=0}^{\infty} J_{2k+1}
\big[ 2 \Delta_0 (\varphi , x ) t  \big] + \hbox{inc.} \;;
\label{6.25z}
\end{equation}
\begin{equation}
A(\varphi ,x) =A \cos \varphi \: J_0(2x\sqrt{\lambda -\lambda^\prime}) \;;
\label{6.26z}
\end{equation}
and an equally obvious
generalization for $\chi^{\prime \prime }(\omega )$
\begin{eqnarray}
\chi_0^{\prime \prime }(\omega ) = & &
{ 2 \over \omega } \int dx xe^{-x^2}
\sum_{\nu =-\infty }^{\infty} F_\lambda^\prime (\nu )
\int {d\varphi \over 2 \pi } e^{i2\nu (\Phi -\varphi )} \nonumber \\
 &\times  & { A(\varphi ,x) \Delta_0 (\varphi , x ) \over
[\omega^2 - 4 \Delta_0^2 (\varphi , x )]^{1/2} }  \;,
\label{6.27z}
\end{eqnarray}

These results for the effective Hamiltonian in (\ref{3.4}) are all we will
derive here. We now argue that these are all we will ever really need for
qualitative understanding.
We hope it is clear that if we put the $\omega_k^{\perp}$ term back into
$H_{eff}$, it would be equally straightforward to calculate $P_{\Uparrow
\Uparrow}(t)$, since the role of the $\beta_k$ in the calculations is
analogous to that of the $\alpha_k$, but even simpler to deal with.
The only term we have still not dealt with is the term $\xi_k \vec{v}_k \cdot
\hat{\vec{\sigma}}_k$ which also appears in the cosine term in $H_{eff}$.
In Appendix C.3 we show how this can be included - the final expression
is a monster, but can easily be put onto a
computer for comparison with real systems (as in the next Section).
We also show plots here of $\chi_0^{\prime \prime }(\omega )$
in Fig.18 and Fig.19 for a few representative values of $\lambda -
\lambda^\prime$; one should again notice how rapidly decoherence sets in,
and how large is the power reduction factor.

\subsection{Nuclear Spin Diffusion}

 Finally we comment on the role  of the role of the very weak
dipolar/Nakamura-Suhl couplings between the nuclei (Eq.(\ref{3.112}).
There  is potentially very important decoherence effect that can come
from these interactions, because the dipolar coupling can cause pairwise
flipping ($ \mid \uparrow \downarrow \rangle \to \mid \downarrow  \uparrow
\rangle $) of the nuclei. Even though this interaction is weak (it is of
strength $T_2^{-1}$, and will be in the range $10^3-10^6\: Hz$ - see
Table \ref{t2}) it is important because it allows both real space spin
diffusion
{\it and} diffusion of the system in a bias energy $\epsilon$, within a given
polarisation group. For a set of $N$ nuclei the pairwise flips occur at a rate
$\sim N T_2^{-1} $; each flip will change the bias energy by $\sim \delta
\omega_o$ in energy, and the bias will random walk throughout the energy
range $N^{1/2} \delta \omega_o =\mu \omega_o $ of the polarisation group.
Now suppose $\Delta \!N =M$, so that transitions of $\vec{S}$ can only proceed
if the time it takes $\epsilon (t)$ to diffuse out of the energy window
around resonance, of width $\Delta_M$, exceeds the "waiting time"
$\Delta_M^{-1}$ necessary for the transition to proceed.
Since in a time $\Delta_M^{-1}$, roughly $N/T_2\Delta_M$ pairwise flips occur,
$\epsilon (t)$ will change by roughly $\delta \omega_o (N/T_2\Delta_M)^{1/2}$
in this time, and so coherence requires
\begin{equation}
 \Delta_M^3 \gg {N \over T_2 } (\delta \omega_o)^2 \;.
\label{8.1}
\end{equation}
If the spread $\delta \omega_k$ arises itself from Nakamura-Suhl interactions
between the nuclei of the same kind, $ T_2^{-1} \sim \delta \omega_k$, and this
criterion becomes $ \Delta_M \gg T_2^{-1} N^{1/3}$.
These criteria are very hard to fulfill unless $M$ is small,
i.e., for the zero polarisation group. One obvious way to avoid this
nuclear spin diffusion decoherence mechanism is to choose
a system like $Fe$ or $Ni$ in which nuclear spin isotopes are rare (cf. Table
\ref{t1}). Then $T_2^{-1}$ will be very small, and coherence may be visible
for quite large $N$ (and even larger $S$, of course).

In the opposite case, when $\epsilon (t)$ fluctuates too rapidly for tunneling,
we will simply get incoherent relaxation of $P_M(t)$, with a correlation
time $\tau_M = (2\Delta_M^2/\mu \omega_o)^{-1}$; or, defining a
correlation time $\tau_M(x)$ via
\begin{equation}
\tau_M^{-1}(x) = {2\Delta_M^2(x) \over \mu \omega_o } \;,
\label{8.2}
\end{equation}
we would find a completely incoherent form for
$P_{\Uparrow \Uparrow}(t) $:
\begin{equation}
P_{\Uparrow \Uparrow}(t) = 2\sum_M w(M;T) \int dx x e^{-x^2}
\exp \{ -t/\tau_M(x) \} \;,
\label{8.3}
\end{equation}
(and obvious generalisation, if necessary,
to include integration over $\varphi$).

This concludes our analysis of the central spin model. We should emphasize
that whereas in this paper we concentrated on the coherence problem, the
general
formalism we have constructed, and the results we established, can also be used
in analysing any of the low-$T$ relaxation properties of $\vec{S}$. The
generalisation to a finite applied field, for example, can be done
straightforwardly,starting from our generic expression.

\section{A realistic example}
\label{sec:6}

Readers (particularly experimentalists) may feel that in spite of these
analytic results, there is still a very long path to follow from the bare,
untruncated  Hamiltonian $H_o({\vec S} )$, with some set of couplings to
nuclear spins, paramagnetic spins, etc., to our final answers. We therefore,
in this section take a specific bare Hamiltonian, viz.,
\begin{equation}
H({\vec S} , \{ {\vec I}_k \} ) = H_0( {\vec S} ) + {1 \over 2S}
\sum_{k} \omega_k {\vec S} \cdot {\vec I}_k \;,
\label{6.1}
\end{equation}
\begin{equation}
H_o( {\vec S} ) = -K_{\parallel}\:s_z^2 + K_{\perp}s_y^2 \;,
\label{6.2}
\end{equation}
with a range of couplings $\omega_k$ around a central hyperfine coupling
$\omega_o$ between ${\vec S} $ and the nuclear spins ${\vec I}_k $. We then
solve this problem almost completely, by giving the analytic forms for
$P_{\Uparrow \Uparrow} (t) $ and $\chi^{\prime \prime} (\omega )$. We hope that
this will make it clear how one may deal with whatever realistic central
spin Hamiltonian that Nature may care to throw at us.

We observe that this model Hamiltonian actually provides a very good
description
of magnetic grains (compare the examples in section \ref{sec:3}.B.) provided we
ignore:

(a) The coupling of the grain order parameter to both nuclei and electron spins
{\it outside } the grain;

(b) The coupling to "loose spins" on the grain surface;

(c) The breaking of the degeneracy between $\mid {\vec S}_1 \rangle = {\hat
{\vec z}}
\mid S \rangle $ and $\mid {\vec S}_2 \rangle = -{\hat {\vec z}}
\mid S \rangle $ in (\ref{6.2}), caused by strain fields, etc., coming from
non-uniformity in the shape of the grain, surface defects, etc. - this will
lead to a finite "bare" orthogonality blocking term, in which $\beta_k$ now
is given by $\cos 2\beta =- {\vec S}_1 \cdot {\vec S}_2 /S^2 $.

The neglect of these terms may, for a real experiment, lead to much
greater decoherence than a calculation based solely on (\ref{6.1}) and
(\ref{6.2})
can give. Thus the results we will give here will describe what one would see
in an experiment  with the {\it minimum possible decoherence}. However we shall
see that even this minimum is very large - and therefore our
prognosis for observation of any coherence in magnetic grains is very poor (as
already noted in refs.\cite{1,2,3}).

The decoherence that {\it does} come into (\ref{6.1}) and  (\ref{6.2}) is as
follows:

(i) There is the inevitable  topological decoherence, and also the effect of
asymmetric barrier fluctuations.

(ii) As discussed in section \ref{sec:5}.A, even if there is no bare
orthogonality
blocking in (\ref{6.2}), it will be {\it generated} by the solution, via
projected
topological decoherence.

(iii) finally we note that there will be a {\it spread} $\delta \omega_k$ in
the
couplings $\{ \omega_k \}$ in (\ref{6.1}). This is an "NMR linewidth"
effect, and is accessible experimentally. Thus if one was able experimentally
to eliminate the effects (a), (b), and (c) described above (using, e.g., a
perfect grain or magnetic macromolecule, in a matrix of ultrapure $^4$He-II),
then our theory would be
testable by direct experiment: all parameters in (\ref{6.1}) and  (\ref{6.2})
would be
known from the start.

The spread will in reality arise from various sources. First, there are the
Suhl-Nakamura
interactions, which can give a spread $\delta \omega_k \sim 10-1000\:kHz$; in
most magnetic materials, this spread is at least $100\:kHz$ for the magnetic
nuclei;
some examples are shown in Table \ref{t2}. Second, there will be a spreading
caused
by the non-uniform dipolar field from the {\it electronic} spins which make
up the giant spin (the nuclear dipole-dipole interaction will give a spreading
$\sim 1\:kHz$, and can be completely ignored). This spreading will depend on
the size
and shape of the grain (it is {\it not} present in a bulk sample , and so
NMR linewidth measurements in the bulk will miss this effect). Its size can be
estimated  from simple magnetostatics to be $\sim 100\:kHz$ for small grain
($S < 10^3$) to above $1\:MHz$ for large grains. Thus to find $\delta \omega_k$
in a set of real grains, NMR linewidth measurements should be done
{\it on the same grains}.

 The solution to the general problem with the Hamiltonian (\ref{6.1}) and
(\ref{6.2})
has already been given; the relevant effective Hamiltonian is
\begin{equation}
 H_{eff}=2{\tilde \Delta}_o \big\{ {\hat \tau }_+\cos \big[ \Phi   +
\sum_{k=1}^N
( \alpha_k {\vec n}_k -i \xi_k {\vec v}_k ) \cdot {\hat {\vec \sigma }}_k
\big] +H.c. \big\} +
{\hat \tau }_z \sum_{k=1}^N \omega_k^{\parallel} \:
 {{\vec l}_k \cdot {\hat {\vec \sigma }}_k } \;.
\label{6.3}
\end{equation}
in which the parameters are given by
\begin{eqnarray}
 {\tilde \Delta}_o &=& \Delta_o \nonumber \\
\Phi &=& \pi S \nonumber \\
\alpha_k {\vec n}_k &=& {\pi \omega_k \over 2 \Omega_o} ({\hat {\vec x }},
{\hat {\vec y }})
\nonumber \\
\xi_k {\vec v}_k &=& -{\pi \omega_k \over 2 \Omega_o} {\hat {\vec x }} \;,
\label{6.4}
\end{eqnarray}
and the unit vector ${\vec l}_k = {\hat {\vec z }}$. As just discussed, the
parameter
$\omega_k$ (and its spread) is accessible experimentally using NMR, as is
$\Delta_o$ (which is unrenormalised here, for weak interactions). If we can
determine
experimentally the values of $K_{\parallel}$ and $K_{\perp}$, we can determine
$\Omega_o$ via \cite{4,20}
\begin{equation}
\Omega_o = {2 \over S} (K_{\parallel}K_{\perp})^{1/2}  \;.
\label{6.5}
\end{equation}
or, alternatively, $\Omega_o$ can be determined directly from microwave
absorption
experiments (which "rock" the central spin back and forth in one of
the degenerate wells).

Equation (\ref{6.3}) corresponds to the case of projected topological
decoherence
with no orthogonality blocking, but degeneracy blocking  coming from the spread
$\delta \omega_k$. This case was completely solved in section \ref{sec:5};
these equations are valid for all the examples we will discuss, since they
assume
$ \mu \omega_o /{\tilde \Delta}_o =N^{1/2} \delta \omega_k /{\tilde \Delta}_o
>1 $. To see what we get, let us look at two examples:

\underline{Example A: $TbFe_3$ grains}: We take this example as a
representative
case in which topological decoherence is very important; as noted in
section \ref{sec:3}.B, one has for this system $\Omega_o \sim 60\:GHz$, and
the quadrupolar split lines have $\alpha_k \sim 0.08,\:0.07,\:$and $0.05$
respectively. Now since no experiments to look for
coherence have been done on this system, we will simply imagine one in which we
choose
$S=10^4$, and $\Delta_o=1\:MHz$ (this  value of $\Delta_o$ may be optimistic,
but is not
in conflict with equation (\ref{3.02}), given the uncertainty in the ratio of
$K_{\parallel}/K_{\perp}$ for $TbFe_3$). We also note that the spread in
$\alpha_k$
caused by the $Fe^{57}$ nuclei will give $\delta \omega_k \sim 50\:MHz$, so
that
with $N_{Tb} =2.5\times 10^3 $ we have $\mu \sim 1$. Thus we have the following
parameters
\begin{eqnarray}
&&\alpha =\sqrt{2} \xi = 0.05,\:0.07,\:0.08 \nonumber \\
&&\mu \sim 1 \nonumber \\
&&(\omega_o /\Delta_o )\mu \sim  3\times 10^3 \;\;\;\;\;\;\;\;(TbFe_3) \;,
\label{6.6}
\end{eqnarray}
leading to the values
\begin{equation}
\lambda \sim 10 \;;\;\;\;\;\;\;(\lambda^\prime =0 )\;.
\label{6.7}
\end{equation}
The form of
$\chi^{\prime \prime} (\omega )$ for this case is shown in Fig.20
(again only the $M=0$ contribution is shown). We see that
there is complete loss of coherence, and extremely strong suppression of the
absorption power ($A \sim 10^{-5}-10^{-6}$). As noted previously
in Sec. \ref{sec:5}.B adding in the $M\ne 0$ contributions will increase the
total spectral weight, but this contribution will be even more incoherent and
centered around $\omega =0$. Nuclear spin diffusion effects will of course
change this somewhat.

\underline{Example B: Ferritin grains}:
This is a more interesting example, since topological decoherence is
unimportant here.
We shall first analyze the experiments of Awschalom et al. \cite{17}
whilst {\it ignoring}
the interaction between the ferritin molecules.
As we saw in section \ref{sec:3}.B, ferritin is an
antiferromagnetic macromolecule with $4500\;Fe$ ions. It also has an excess
ferromagnetic moment with a somewhat uncertain value (estimates range between
$217$ and $640 \mu_B$). In reality, because of this moment, ferritin molecules
will interact. There will also be a coupling to nuclear and electronic moments
outside the ferritin, which we ignore here.

We are then left with an $Fe$ hyperfine coupling $\omega_k \sim
64\:MHz$; there will also be other hyperfine couplings to the $H^1$ protons in
the
$H_2O$ in ferritin; we also ignore this for the moment (but see our discussion
below).
We assume  a $\delta \omega_k \sim 200\:kHz$ arising from a combination of
Suhl-Nakamura
and non-uniform dipolar interactions - this is probably an underestimate, but
the quantity
has not yet been measured, for ferritin molecules, as far as we know.

According to Awschalom et al., the anisotropy  constants $K_{\parallel}$ and
$K_{\perp}$
obey $(K_{\parallel}K_{\perp})^{1/2} \sim 1.72\: Tesla$, so we shall assume a
value
for $\Omega_o$ of $40\:GHz$, as a rough estimate \cite{45}. Finally, the
experiments of Awschalom et al. claim
that a resonance they see of frequency just under $1\:MHz$ corresponds to
coherent tunneling; therefore we assume $\Delta_o \sim 1\:MHz$. We thereby
arrive at the following
numbers, using (\ref{6.4}), (\ref{6.5}), and the definition of $\mu$:
\begin{eqnarray}
& &\alpha = \sqrt{2} \xi \sim 2\times 10^{-3}   \nonumber \\
& &\mu = N^{1/2} \delta \omega_k /\omega_o \sim 2\times 10^{-2}  \nonumber \\
& & (\omega_o /\Delta_o )\mu \sim 2  \;,
\label{6.8}
\end{eqnarray}
where we note that there will be $\sim 100\:Fe^{57}$ nuclei in the molecule,
so $N^{1/2} =10$. Now from (6.8) we find
\begin{equation}
\lambda \sim 4\times 10^{-5} \;;\;\;\;\;(\lambda^\prime
=0)\;;\;\;\;\;(Ferritin)\;,
\label{6.9}
\end{equation}
which rules out topological decoherence; since we assume $\beta_k =0$ (perfect
sample)
this leaves only degeneracy blocking, and since $(\omega_o /\Delta_o )\mu >1$
we
can immediately apply the result (\ref{4.35}) in the text, which is depicted in
Fig.13.
If one compares the curve in Fig.13 with the data of Awschalom et al.
\cite{17},
one is immediately surprised that the lineshape in the experiments is not at
all
that different from the calculation. Now it is important to understand that the
reason
for the sharp theoretical lineshape in Fig.13 is precisely because topological
decoherence
and orthogonality blocking have been dropped. It will be immediately obvious,
on referring to Table I, that the value of the dimensionless topological
decoherence
parameter $\lambda $ is extremely low for ferritin; the value of $10$ for
$TbFe_3$ is
more normal. If we can ignore any orthogonality blocking, then it rather seems
as
though the experimentalists have hit upon an almost ideal system for the
investigation of MQC in magnets (it could be made even better if one could
isotopically purify \cite{2} the $Fe$ in ferritin, to get rid of all $Fe^{57}$
nuclei).
Do our calculations then support the thesis of Awschalom et al., to have
seen MQC in ferritin? At this point we feel that one must be cautious. First,
we
notice there is a strong reduction, by a factor $A={\tilde \Delta}_o
/(2\omega_o \sqrt{2\pi N}) \sim
3.5 \times 10^{-4} $, in the total spectral weight of this peak, as compared
to the situation where we ignore the nuclear spins (notice that the
question of power absorption in these experiments has already caused some
controversy \cite{18,46}). This absorption is a consequence of the restriction
to the $M=0$ polarization sector, and the requirement of near resonance.

Perhaps more seriously, it is not obvious that there really is no
orthogonality blocking in the ferritin. The {\it intrinsic} blocking will be
very
small - the only obvious other sources are either the dipolar interactions
between
the ferritin molecules, "loose spins" on the surface of the ferritin, or
strain fields in the sample. The effect of dipolar fields is probably small,
but the other
effects are difficult to quantify - we simply call attention to the severe
effect
on the lineshape that will be wrought by even a fairly small amount of
orthogonality blocking (see Fig.16, for $\kappa =2$). Orthogonality blocking
will also further reduce the spectral weight (Eq.(\ref{5.21})).
On the other hand nuclear spin diffusion between the protons may be important.

Thus there is apparently a disagreement between the theory and experiment for
ferritin.
Nevertheless it may not be insurmountable, provided the discrepancy concerning
power absorption can be dealt with \cite{18,46}, and provided one can rule out
any
significant orthogonality blocking, and we certainly do not think that one
can yet rule out MQC as a possible explanation of the experimental results.
Another
question that needs further investigation is the role of the dipole interaction
between
the ferritin molecules
 - this may well lead to some
kind of cooperative behaviour between the ferritin molecules. Elsewhere we have
also
suggested \cite{3} a search for a "spin echo" which would help to reveal the
source of the experimental resonance at $\sim 1\:MHz$.

It should also be noted that very recently Garg has given an analysis
\cite{46} of the effects of the nuclear spins on coherence in ferritin
molecules.
Garg concludes (in agreement with our earlier papers \cite{1,2,3}),
that the experiments do not agree
with the theory. However his reasoning and detailed results are different, and
are based upon the reduction in the total spectral
weight of the signal (the factor $f$ in Eq.(\ref{4.34}), where $f=(2/\pi
N)^{1/2}$;
recall this factor comes solely from the assumption of zero nuclear spin
polarization).
In effect the analysis of Garg ignores topological decoherence
$(\alpha_k =0)$, ignores orthogonality blocking ($\kappa =0$) and it ignores
degeneracy blocking ($ \delta \omega_k =0$). He therefore arrives at a set of
sharp
lines for the spectrum of $\chi^{\prime \prime} (\omega )$.

However, as we have just seen this is not  realistic. Dropping $\alpha_k$ is
justified for ferritin (but not for most other systems - compare the values for
the examples in section \ref{sec:3}.B), but   we see that the
spectrum in Fig.13 is not a sharp Lorentzian line.
 Thus the analysis of Garg misses the crucial physical mechanism (of
decoherence);
his lineshape is one of completely coherent motion for the ferritin N\'{e}el
vector
(albeit with a reduced total signal amplitude). We note moreover
 that the amplitude reduction factor found by Garg is just the
 "bare" value $f$, which is incorrect; the correct reduction factor
is that given in Eq.(\ref{4.34}), for pure degeneracy blocking (or
(\ref{5.21}),
if orthogonality blocking is also present).

A further point, upon which we will elaborate in detail in another paper
dealing with finite field effects, is that the field dependence of
$\chi^{\prime \prime} (\omega )$ will be quite different from that of
a 2-level system, even one coupled to a bath of oscillators. This is
because degeneracy blocking causes $\chi^{\prime \prime} (\omega )$ to
vary on an energy scale determined by $W(\epsilon = \gamma_e SH_o)$,, which
has nothing to do with $\tilde{ \Delta}_o$ and is usually much greater.

We hope that from our analysis of these 2 examples the reader will now have a
fairly
concrete idea of how our effective Hamiltonian is used in
the analysis of a real
central spin; in fact with the analytic results we have given, it should be
possible for
experimentalists to produce plots of $\chi^{\prime \prime} (\omega )$ for
any given system, provided the relevant parameters are known.

\section{Conclusions}
\label{sec:7}

In this paper we have given an analysis of the behaviour of a central spin
${\vec S}$, coupled in an essentially arbitrary way
 to a spin environment. That one  can do this, and moreover recover
closed analytic forms for the correlation functions $P_{\Uparrow \Uparrow} (t)$
and $\chi^{\prime \prime} (\omega )$, is perhaps surprising. It was certainly
not obvious to us when this work was initiated (cf. ref.\cite{1}); although the
3 mechanisms
involved in the decoherence of $P_{\Uparrow \Uparrow} (t)$
 are already contained in the analysis of ref.\cite{2}, it has taken some time
to find closed analytic forms
for all cases. It is perhaps even more surprising when one examines
the equation
produced by the only other method we could think of using for this problem,
viz.,
the Bethe ansatz (we are dealing with a set of $N$ coupled 1-dimensional
problems here
\cite{47},
with in general random couplings). However examination of the relevant Bethe
ansatz equations reveals them to be apparently intractable.

Our solution relies on our novel technique of introducing "operator instantons"
in the Hilbert space of the environmental spins, and then separating out the
relevant energy scales. This then allowed identification of 3 important
limiting
cases of the general problem, each of which involved a single specific
physical mechanism. These mechanisms were "topological decoherence"
(already completely solved in ref.\cite{2}), and "degeneracy blocking" and
"orthogonality blocking", whose essential physics was also described in
ref.\cite{2}.
The solution is then achieved by combining these 3 mechanisms; this allows
full coverage of the parameter space  when there is decoherence without energy
dissipation. In this paper we have used these methods to examine specifically
the {\it coherence properties} of ${\vec S}$ in zero external field.
This allows us to deal only with the zero polarization sector of the
environment. In fact a fairly trivial extension of these methods can be used
to deal with the finite field case, and to include all polarisation sectors.
This then yields a theory of the relaxation of ${\vec S} (t)$
(including tunneling of ${\vec S} (t)$) in an applied field, or in its absence.
This will be discussed in detail in another paper.

As a result of these investigations we find that the spin environment acts very
differently from the usual "oscillator bath" environments, and has a far more
powerful suppressive effect on quantum coherence at low temperature. One can
also
analyze the effect on coherence, of spin environments, for other systems. For
example, an analysis of possible macroscopic quantum coherence for SQUIDs
yields quite enormous
values of our parameters $\lambda ,\: \lambda^\prime$, etc. (typically $\lambda
>
10^{5}$). This implies  that the previous analyses of SQUIDs, with a view to
observing such "MQC", are far too optimistic, and that in fact it is
unlikely that it will be seen in any superconductor (although the prognosis is
more
hopeful if the superconductor lacks nuclear spins). This work will be reported
elsewhere.

These results lead us to suspect  that in nature the overwhelmingly dominant
mechanism of decoherence on the
mesoscopic or macroscopic scales is likely to come from environmental spins,
particularly nuclear spins. Whilst this may not change the fundamental
attitude of most physicists towards the foundations of quantum mechanics, or
the "measurement problem", it is certainly an interesting and non-trivial
discovery about the physics of decoherence in
the real world. It is also of fundamental importance for efforts to either
observe MQC,
or to at least build devices which operate coherently on more than just the 1-
or
2-particle level. As noted in the introduction, it very severely restricts the
possible physical systems in which such coherence might be seen -
the only really
clear example that avoids the problem discussed here is superfluid
$^4$He - again, the details of this example will be discussed elsewhere. Our
prognosis for attempts to see MQC in magnetic systems is somewhat pessimistic,
although if a system can be found with $\lambda \ll 1$ {\it and} $\kappa \ll
1$,
we may be in business again.

Finally, let us note again that once an applied field starts to noticeably {\it
bias}
the potential $H_o({\vec S} )$, things change a great deal. Our problem then
becomes effectively one of the effect of nuclear and other spins on the
crossover
to {\it dissipative tunneling} of ${\vec S} (t)$. This is a quite different
problem
from the one we have just analyzed, and we will return to it in another paper.

\section{Acknowledgement}
This work was supported by NSERC in Canada,
by the International Science Foundation (MAA300), and by the
Russian Foundation for
Basic Research (95-02-06191a). Some
of it was carried out at the Aspen Center for Physics. We would like to thank
I. Affleck, B.G. Turrell, and W.G. Unruh for some very helpful discussions.

\appendix

\section{ Derivations for three limiting cases}

We describe here the derivations of the key formulae in section \ref{sec:4}. We
begin
with a few remarks on the free central spin, and the derivation of $P_{\Uparrow
\Uparrow} (t)$ for it;
it will then be seen how the expressions (\ref{4.4}), (\ref{4.5}) involve
obvious generalizations of the free spin algebra. Then we give an account
of the lengthy derivation required to obtain the expression (\ref{4.21o}) for
pure
orthogonality blocking.

\subsection{Free central spin, plus topological and bias effects}

Once $H_o({\vec S} )$ has been truncated, an expression for  $P_{\Uparrow
\Uparrow} (t)$ can be derived using standard instanton techniques \cite{12,42},
employing the free amplitude $K_o^\pm =
\Delta_o {\hat \tau }_x \exp \{ \pm i\pi S \} $. In the absence of the phase
$\pi S$, one has a
"2-level system" correlation function $P_{TLS} (t) $
\begin{equation}
 P_{TLS} (t) = {1 \over 2} [1+\cos (2\Delta_ot )]\;,
\label{b1.1}
\end{equation}
which is derived in the instanton framework by summing over all paths involving
an
{\it even} number of flips, with an amplitude $i\Delta_o dt$ to flip in a time
$dt$,
giving
 \begin{equation}
 P_{TLS} (t) = {1 \over 2} \left\{ 1+ \sum_{s=0}^{\infty}
{ (2i\Delta_o t)^{2s} \over (2s)! } \right\} \;.
\label{b1.2}
\end{equation}

Adding the phase $\pi S$ clearly changes  the flip amplitude in time $dt$ to
$2i\Delta_o \exp \{ \pm i\pi S \} $, and adding the $\pm $ processes then gives
\begin{equation}
 P_{\Uparrow \Uparrow}^{(0)} (t) = {1 \over 2} [1+\cos (4\Delta_o \cos \pi S)t
]\;,
\label{b1.3}
\end{equation}
which can, if one wishes, be derived \cite{1,33} by summing over all paths with
an even
number of flips, and summing over all combinations of clockwise and
counterclockwise
flips:
\begin{equation}
 P_{\Uparrow \Uparrow}^{(0)}(t) = {1 \over 2} \left\{ 1+ \sum_{s=0}^{\infty}
{ (2i\Delta_o t)^{2s} \over (2s)! } \sum_{n=0}^{2s}
{(2s)! \over (2s-n)!n! } e^{i\Phi_o(2s-2n) } \right\} \;,
\label{b1.4}
\end{equation}
where $\Phi_o =\pi S$ for the free spin. This is equation (\ref{4.2}) of the
text.

The generalization to topological decoherence is now easy. We either observe
that the
interference factor $(e^{i\pi S} +e^{-i\pi S})=2 \cos \pi S$ in (\ref{b1.3})
will now
be generalized via
\begin{equation}
 \pi S \to (\Phi +\sum_{k=1}^{N}\alpha_k {\vec n}_k \cdot {\hat {\vec \sigma
}}_k )\;; \;\;\;
\mbox{(pure topological)}  \;,
\label{b1.5}
\end{equation}
which gives Eq.(\ref{4.4}) (including the trace over spin states); or we simply
include the extra phase  (\ref{b1.5}) into (\ref{b1.4}), to give Eq.(\ref{4.5})
of the text.

Now consider the modification introduced by a bias $\epsilon$. For the 2-level
system this is discussed very thoroughly in one way by Leggett et al.
(ref.\cite{12},
Appendix A). We use a different representation here. Consider first the
amplitude
$A_{\Uparrow \Uparrow}(t)$ for a return to an initial state in the biased case;
this is clearly
\begin{equation}
  A_{\Uparrow \Uparrow}^{^{TLS}}(t, \epsilon ) = \sum_{n=0}^{\infty}
(-i\Delta_o )^{2n}
\int_0^t dt_{2n} \dots \int_0^{t_2} dt_1 e^{i(\epsilon (t-t_{2n})-\epsilon
(t_{2n} -
 t_{2n-1}) + \dots \epsilon t_1 )} \;,
\label{b1.6}
\end{equation}
since the action coming from a pause of length $dt$ is $e^{\pm i\epsilon dt }$.
We
now write this in Laplace transform form as
\begin{equation}
 A_{\Uparrow \Uparrow}^{^{TLS}}(t, \epsilon ) = \int_{-i\infty}^{i\infty}
e^{pt}
A_{\Uparrow \Uparrow}^{^{TLS}}(p, \epsilon ) \;.
\label{b1.7}
\end{equation}
\begin{equation}
 A_{\Uparrow \Uparrow}^{^{TLS}}(p, \epsilon ) = {1 \over p-i\epsilon }
\sum_{n=0}^{\infty} \left(
{ (-i\Delta_o )^2 \over p^2+\epsilon^2 } \right)^{n} =
 {1 \over p-i\epsilon } {   p^2+\epsilon^2 \over p^2+E^2 }\;.
\label{b1.8}
\end{equation}
\begin{equation}
E = [\epsilon^2 +\Delta_o^2 ]^{1/2} \;,
\label{b1.9}
\end{equation}
which gives the standard answer for $P_{\Uparrow \Uparrow}^{(0)}(t, \epsilon
)$:
\begin{eqnarray}
P_{\Uparrow \Uparrow}^{^{TLS}} (t,\epsilon )&=& \int_{-i\infty }^{i\infty} dp_1
dp_2 { e^{(p_1+p_2)t}  \over (p_1 -i
\epsilon ) ( p_2 -i\epsilon )} \sum_{n=0}^{\infty}\sum_{m=0}^{\infty}
\left( { (-i\Delta_o )^2 \over p_1^2+\epsilon^2 } \right)^{n}
\left( { (-i\Delta_o )^2 \over p_2^2+\epsilon^2 } \right)^{m} \nonumber \\
&=& 1-{\Delta_o^2 \over 2E^2 } (1- \cos 2Et) = 1- {\Delta_o^2 \over E^2 }\sin^2
Et
\label{b1.10}
\end{eqnarray}

It is then obvious that in the case of the free spin, for a  bias that does not
distinguish
between the clockwise and counterclockwise  flips,
$P_{\Uparrow \Uparrow} (t,\epsilon )$ will look exactly like (\ref{b1.10}),
except that $\Delta_o \to 2\Delta_o \cos \pi S$.
This representation avoids the difficulty in the usual representation which
comes essentially
from the square root $E = [\epsilon^2 +\Delta_o^2 ]^{1/2}$ in the cosine.

\subsection{Orthogonality blocking}

We consider the situation  described in section \ref{sec:4}.B. For the spin
${\vec \sigma }_k$, the "initial" and "final" local fields are ${\vec
\gamma}_k^{(1)} $ and
${\vec \gamma}_k^{(2)} $. We will begin, to make the derivations easier, by
assuming that the angle relating ${\vec \gamma}_k^{(1)} $ and
${\vec \gamma}_k^{(2)} $ is close to $\pi $
 for all the ${\vec \sigma }_k$. This angle $\beta_k$ is defined as
\begin{equation}
\cos 2\beta_k = - \gamma _k^{(1)} \cdot  \gamma _k^{(2)} \;.
\label{b.1}
\end{equation}

We now choose axes such that the final state wave-function of ${\vec \sigma
}_k$ is
given in terms of the initial state wave-function by
\begin{equation}
\mid {\vec \sigma }_k^f \rangle = {\hat U}_k
\mid {\vec \sigma }_k^{in} \rangle =
 e^{ -i\beta_k {\hat \sigma }_k^x }  \mid {\vec \sigma }_k^{in} \rangle \;.
\label{b.2}
\end{equation}
\begin{equation}
\mid \{ {\vec \sigma }_k^f \} \rangle = \prod_{k=1}^N {\hat U}_k
\mid \{ {\vec \sigma }_k^{in} \} \rangle = {\hat U}  \mid \{ {\vec \sigma
}_k^{in} \} \rangle \;.
\label{b.3}
\end{equation}

In general the initial state of the spin bath will belong to the
polarisation group $\Delta \!N$ (not necessarily $\Delta \!N =0$),
where the polarisation is defined along the $\vec{S}_1$-direction. As
explained in the text, when $\vec{S}$ rotates from $\vec{S}_1$ to
$\vec{S}_2 \approx -\vec{S}_1 $, the energy conservation requires to
change the polarisation from $\Delta \!N$ to $-\Delta \!N$. In future
discussion of degeneracy blocking effects we will have to consider
more general tunneling processes, when the nuclear polarisation is
changed from $\Delta \!N$ to $\Delta \!N -2M$ and back.
In what follows we calculate the correlation function $P_{\Delta \!N,M} (t)$,
which gives the dynamics of $\vec{S}$ when the spin bath transitions
are restricted to be between the $\langle \hat{\cal P} \rangle =\Delta \!N$ and
 $\langle \hat{\cal P} \rangle =\Delta \!N -2M$ subspaces, which are
supposed to be in resonance (for pure orthogonality blocking $M=\Delta \!N$).
The statistical weight
of states with $\Delta \!N \gg N^{1/2}$ is negligible, so we will
assume in our calculation that $\Delta \!N,M \ll N$.

The above restriction is enforced by
 the projection operator
\begin{equation}
{\hat \Pi}_{\Delta \!N}
= \delta (\sum_{k=1}^N {\hat \sigma }_k^z -\Delta \!N) =
\int_0^{2\pi} {d\xi \over 2\pi} e^{ i\xi (
\sum_{k=1}^N {\hat \sigma }_k^z -\Delta \!N ) } \;.
\label{b.4}
\end{equation}
We can now, including this restriction, write down an expression for the
amplitude for ${\vec S}$ to go from $\mid {\vec S}_1 (t=0) \rangle$ to
$\mid {\vec S}_1 (t) \rangle$ starting from some
$\langle \hat{\cal P} \rangle =\Delta \!N$ environmental state;
\begin{equation}
A_{\Delta \!N,M} (t) =  \left\{
\sum_{n=0}^\infty {(i{\tilde \Delta}_o t)^{2n} \over (2n)!}
\prod_{i=1}^{2n}  \int {d\xi_i \over 2\pi }
e^{-i\Delta \!N (\xi_{2n}+\xi_{2n-1}+\dots +\xi_1 )}
e^{2iM(\xi_{2n-1}+\xi_{2n-3}+\dots +\xi_1 )}
{\hat T}_{2n} \right\}
\mid \{ {\vec \sigma }_k^{in} \} \rangle \;,
\label{b.5}
\end{equation}
where  ${\hat T}_{2n}$ is
\begin{equation}
 {\hat T}_{2n} =\bigg[ e^{i\xi_{2n}\sum_{k=1}^N {\hat \sigma }_k^z}
{\hat U}^{\dag }
e^{i\xi_{2n-1}\sum_{k=1}^N {\hat \sigma }_k^z}
{\hat U} \dots {\hat U}^{\dag }
e^{i\xi_{1} \sum_{k=1}^N {\hat \sigma }_k^z}
{\hat U} \bigg] \;.
\label{b.6}
\end{equation}
{}From (\ref{b.5}) we can now
write the correlation function  $P_{\Delta \!N,M} (t)$ as
\begin{eqnarray}
&P_{\Delta \!N,M}(t) & \equiv  \langle A^{*}_{\Delta \!N,M} (t) A_{\Delta
\!N,M} (t) \rangle \nonumber \\
 &= &
\sum_{n=0}^\infty \sum_{m=0}^\infty {(i{\tilde \Delta}_o t)^{2(n+m)} \over
(2n)!(2m)!}
\prod_{i=1}^{2n} \prod_{j=1}^{2m} \int {d\xi_i \over 2\pi }
\int {d\xi_j^{\prime} \over 2\pi }
e^{-i\Delta \!N (\sum_i^{2n}\xi_{i} - \sum_j^{2m}\xi_{j}^\prime )}
e^{2iM( \sum_{i=odd}^{2n-1} \xi_{i} - \sum_{j=odd}^{2m-1} \xi_{j}^\prime )}
\langle
{\hat T}_{2m}^{\dag} {\hat T}_{2n} \rangle \;.
\label{b.7}
\end{eqnarray}
The  restrictions required to arrive at this
formula are discussed in the text.

We now return to the assumption  that the $\beta_k$ are small;
specifically we will assume that the orthogonality exponent $\kappa $,
defined by
\begin{equation}
e^{-\kappa } = \prod_{k=1}^N \cos \beta_k  \;,
\label{b.8}
\end{equation}
can be approximated by the perturbative expansion
\begin{equation}
\kappa  \approx  { 1\over 2}  \sum_{k=1}^N  \beta_k^2 \;.
\label{b.9}
\end{equation}
This assumption makes it much easier to calculate the average
in (\ref{b.7}). Consider first the problem with only one environmental spin
${\vec \sigma }_k$, and calculate the average
$\langle {\hat T}_{2m}^{\dag} {\hat T}_{2n} \rangle_k $ in this case;
since ${\hat T}_{2m}^{\dag} {\hat T}_{2n}$ is a product of operators acting
separately on each $\vec{\sigma}_k$, the average over all spins is also
the product of single spin results.

Because of (\ref{b.9}) we need only to consider processes with
$0,\; 1$, or $2$ flips of the environmental spin, i.e., we will make an
expansion in powers of $\beta_k$ for ${\vec \sigma }_k$, and stop at
$\beta_k^2$.
Then it is clear that, if the initial state of ${\vec \sigma }_k$ is
$\mid \uparrow_k \rangle $
\begin{eqnarray}
{\hat T}_{2n}^{(k)} \mid \uparrow_k \rangle & = &
e^{i\xi_{2n} {\hat \sigma }_k^z} e^{ -i\beta_k {\hat \sigma }_k^x } \dots
e^{ -i\beta_k {\hat \sigma }_k^x } e^{i\xi_{1} {\hat \sigma }_k^z} e^{ i\beta_k
{\hat \sigma }_k^x }
\mid \uparrow_k \rangle \nonumber \\
 & = & e^{i\sum_{i=1}^{2n}\xi_i} \bigg[
(1-n\beta_k^2)\mid \uparrow_k \rangle +i\beta_k \mid \downarrow_k \rangle
\sum_{l=1}^{2n} (-1)^{l+1} e^{ -2i \sum_{i=l}^{2n} \xi_i } \nonumber \\
 & \; & \;\;\;\;\;\;\;\;\;\;\; -\beta_k^2 \mid \uparrow_k \rangle
\sum_{l^\prime =l+1}^{2n} \sum_{l=1}^{2n-1} (-1)^{l^\prime -l}
e^{-2i \sum_{i=l}^{l^\prime -1} \xi_i } + O(\beta_k^3) \bigg] \;,
\label{b.10}
\end{eqnarray}
where the first term arises from the sequence $[ \uparrow \uparrow \uparrow
\dots \uparrow \uparrow ]$, the second from the sequence
$[ \uparrow \uparrow \uparrow \dots \uparrow \downarrow \downarrow
\downarrow \dots \downarrow \downarrow ]$, with a flip when $j=l$; and so on.
In the same way we find
\begin{eqnarray}
\langle \uparrow_k \mid ({\hat T}_{2m}^{(k)})^{\dag}
{\hat T}_{2n}^{(k)} \mid \uparrow_k \rangle  &= &
e^{i(\sum_{i=1}^{2n} \xi_i - \sum_{j=1}^{2m} \xi_j^\prime )}
\bigg[ 1 - \beta_k^2 \big[ (n+m) + \sum_{l^\prime =l+1}^{2n} \sum_{l=1}^{2n-1}
(-1)^{l^\prime -l}  e^{-2i \sum_{i=l}^{l^\prime -1} \xi_i } \nonumber \\
&+ &
\sum_{p^\prime =p+1}^{2m} \sum_{p=1}^{2m-1}
(-1)^{p^\prime -p}  e^{2i \sum_{j=p}^{p^\prime -1} \xi_j^\prime } \nonumber \\
 &- &
\sum_{p =1}^{2m} \sum_{l=1}^{2n}
(-1)^{l+p}  e^{-2i (\sum_{i=l}^{2n} \xi_i - \sum_{j=1}^{p-1} \xi_j^\prime )}
\big] \bigg] \;,
\label{b.11}
\end{eqnarray}
to order $\beta_k^2$. The sequence
$\langle \downarrow_k \mid ({\hat T}_{2m}^{(k)})^{\dag}
{\hat T}_{2n}^{(k)} \mid \downarrow_k \rangle$
will have a similar expression, but with reversed signs coming from the
$e^{i\xi_j {\hat \sigma }_k^z}$ factors.

We now observe that the state with polarisation $\Delta \!N $ consists of
$(N+\Delta \!N)/2$ spins up and $(N-\Delta \!N)/2$ spins down.
Consequently, for each ${\vec \sigma }_k$, we add $\uparrow $ or $\downarrow$
averages like (\ref{b.11}), and then take  the product
\begin{equation}
\langle {\hat T}_{2m}^{\dag} {\hat T}_{2n} \rangle =\prod_{k=1}^{N_{\uparrow}}
\langle ({\hat T}_{2m}^{(k)})^{\dag} {\hat T}_{2n}^{(k)} \rangle\;
\prod_{k^\prime =1}^{N_{\downarrow}}
\langle ({\hat T}_{2m}^{(k)})^{\dag} {\hat T}_{2n}^{(k)} \rangle
\label{b.12}
\end{equation}
Substituting (\ref{b.11}) into this expression we get
\begin{equation}
\langle {\hat T}_{2m}^{\dag} {\hat T}_{2n} \rangle =
e^{i\Delta \!N (\sum_i^{2n}\xi_{i} - \sum_j^{2m}\xi_{j}^\prime )}
\exp \big\{ -K^{eff}_{nm} (\xi_i ,\xi_j^\prime , \Delta \!N ) \big\}
\;,
\label{b.12b}
\end{equation}
where the "effective action" $K^{eff}_{nm} (\xi_i ,\xi_j, \Delta \!N )$
has two contributions $K^{eff} = K_1+K_2$:
\begin{eqnarray}
K_1= 2\kappa (1-{\Delta \!N \over N })
 \bigg\{ (n+m) & + & \sum_{l^\prime > l}(-1)^{l^\prime -l}
\cos [2\sum_{i=l}^{l^\prime -1}\xi_i ] + \sum_{p^\prime > p}(-1)^{p^\prime -p}
\cos [2\sum_{j=p}^{p^\prime -1}\xi_j^\prime ] \nonumber \\
 &-& \sum_{p =1}^{2m} \sum_{l=1}^{2n} (-1)^{l+p}
\cos [2\sum_{i=l}^{2n} \xi_i - \sum_{j=1}^{p-1} \xi_j^\prime ] \bigg\}
\;,
\label{b.12c}
\end{eqnarray}
\begin{eqnarray}
K_2= 2\kappa {\Delta \!N \over N }
 \bigg\{ & & \sum_{l^\prime > l}(-1)^{l^\prime -l}
\exp [-2i \sum_{i=l}^{l^\prime -1}\xi_i ] +
\sum_{p^\prime > p}(-1)^{p^\prime -p}
\exp [2i \sum_{j=p}^{p^\prime -1}\xi_j^\prime ] \nonumber \\
 &-& \sum_{p =1}^{2m} \sum_{l=1}^{2n} (-1)^{l+p}
\exp [2i \sum_{j=1}^{p-1} \xi_j^\prime -2i \sum_{i=l}^{2n} \xi_i ] \bigg\}
\;,
\label{b.12d}
\end{eqnarray}
We recall now that $\Delta \!N \le N^{1/2} \ll N$, which allows to neglect
the contribution due to $K_2$ and  drop the correction  $\Delta \!N / N$ to
the coefficient $\kappa $ in $K_1$.
 Notice also that the phase factor in front of
$\exp \{ -K^{eff} \} $ in (\ref{b.12}) cancels exactly the phase
proportional to $\Delta \!N$ in the formular (\ref{b.7}) for $P_{\Delta
\!N,M}(t)$. Thus, quite surprisingly, we find the correlation function to
be independent of $\Delta \!N$ in this limit:
\begin{equation}
P_{M}(t)  =\sum_{n=0}^\infty \sum_{m=0}^\infty {(i{\tilde \Delta}_o t)^{2(n+m)}
\over (2n)!(2m)!}
\prod_{i=1}^{2n} \prod_{j=1}^{2m} \int {d\xi_i \over 2\pi }
\int {d\xi_j^{\prime} \over 2\pi }
\exp \big\{2iM(\xi_{2n-1}+\xi_{2n-3}+\dots +\xi_1 )
 -K^{eff}_{nm} (\xi_i ,\xi_j^\prime ) \big\}
\;,
\label{b.13}
\end{equation}

We can render this expression more useful by changing variables; first
we consider the whole sequence $\xi_\alpha = (\xi_1, \dots , \xi_{2n},
-\xi_1^\prime, \dots , -\xi_{2m}^\prime)$ together, and then define new
angular variables
\begin{equation}
\chi_\alpha = \sum_{\alpha^\prime =
\alpha }^{2(n+m)} 2\xi_\alpha +\pi \alpha \;,
\label{b.15}
\end{equation}
so that now
\begin{eqnarray}
P_{M}(t)  = & &\sum_{n=0}^\infty \sum_{m=0}^\infty {(i{\tilde \Delta}_o
t)^{2(n+m)} \over (2n)!(2m)!}
\left( \prod_{\alpha =1}^{2(n+m)} \int {d\chi_\alpha \over 2\pi } \right)
\nonumber \\
& \times &
\exp \bigg\{iM \sum_{\alpha} (-1)^{\alpha +1}\chi_{\alpha}
-2\kappa \big[ (n+m) +\sum_{\alpha^\prime > \alpha }
\cos (\chi_\alpha - \chi_{\alpha^\prime} ) \big]   \bigg\} \;.
\label{b.16}
\end{eqnarray}
Thus we have mapped our problem onto the partition function of a rather
peculiar system of spins, interacting via infinite range forces, with
interaction strength $2\kappa$.

To deal with this partition function , we define "pseudo-spins"
${\vec s}_\alpha $ and ${\vec {\cal S}}$, via
\begin{eqnarray}
{\vec s}_\alpha &=& (\cos \chi_\alpha , \sin \chi_\alpha ) \nonumber \\
{\vec {\cal S}} &= &\sum_{\alpha =1}^{2(n+m)} {\vec s}_\alpha
\;,
\label{b.17}
\end{eqnarray}
so that
\begin{eqnarray}
& &{\vec s}_\alpha \cdot {\vec s}_{\alpha^\prime} =
\cos (\chi_\alpha  -\chi_{\alpha^\prime} ) \nonumber \\
& & \sum_{\alpha^\prime , \alpha }
\cos (\chi_\alpha - \chi_{\alpha^\prime} ) = {\vec {\cal S}}^2 \;,
\label{b.18}
\end{eqnarray}
We can think of ${\vec s}_\alpha $ as rotating in our fictitious angular space
defined by the projection operator (\ref{b.4}). Now consider the term
$G({\vec {\cal S}} )$ in (\ref{b.16}) defined by
\begin{eqnarray}
G({\vec {\cal S}} ) & = & \left(
\prod_{\alpha =1}^{2(n+m)} \int {d\chi_\alpha \over 2\pi }
e^{iM(-1)^{\alpha +1}\chi_{\alpha} } \right) \:
\exp \big\{ -\kappa \sum_{\alpha^\prime , \alpha }
\cos (\chi_\alpha - \chi_{\alpha^\prime} ) \big\} \nonumber \\
& =&  \left(
\prod_{\alpha =1}^{2(n+m)} \int {d\chi_\alpha \over 2\pi }
e^{iM(-1)^{\alpha +1}\chi_{\alpha} } \right)  \:
e^{ -\kappa {\vec {\cal S}}^2} \;.
\label{b.19}
\end{eqnarray}
This is easily calculated, viz.,
\begin{eqnarray}
G({\vec {\cal S}} ) &=& \int d{\vec {\cal S}} e^{ -\kappa {\vec {\cal S}}^2}
\prod_{\alpha =1}^{2(n+m)} \int {d\chi_\alpha \over 2\pi }
e^{iM(-1)^{\alpha +1}\chi_{\alpha} }
\delta ({\vec {\cal S}} - \sum_\alpha {\vec s}_\alpha ) \nonumber \\
&=& \int {d{\vec  z } \over 2\pi }
\int d{\vec {\cal S}} e^{ -\kappa {\vec {\cal S}}^2 +i{\vec  z } \cdot
{\vec {\cal S}}}\: \left( \int_0^{2\pi } {d\chi_\alpha \over 2\pi }
e^{-i{\vec  z} {\vec s}_\alpha +i M \chi_{\alpha }} \right) ^{2(n+m)}
\nonumber \\
 &=& {1 \over 2\kappa } \int d z  z e^{- z^2/4\kappa }
J_M^{2(n+m)}( z ) \;,
\label{b.20}
\end{eqnarray}
where $J_M(\lambda )$ is the zeroth-order Bessel function. Using
\begin{equation}
\sum_{l=0}^{2(n+m)} {(2(n+m))! \over (2m)! (2n)! } =
\delta_{n+m,0} +2^{2(n+m)-1}
\;,
\label{b.21}
\end{equation}
to reorganize the sum over $n$ and $m$ in (\ref{b.16}) and changing
the integration variable
$z \to 2x \sqrt{\kappa }$, we then find
\begin{eqnarray}
P_{\Uparrow \Uparrow}(t)&=& 2\int_0^\infty dx x\:e^{-x^2} \; {1 \over 2}
\left( 1+ \sum_{s=0}^{\infty} {
[2it{\tilde \Delta}_o J_M(2\sqrt{\kappa } x)]^{2s} \over (2s)! } \right)
\nonumber \\
&\equiv & 2\int_0^\infty dx x\:e^{-x^2} P^{(0)}_{\Uparrow
\Uparrow}(t,\Delta_M(x)) \;.
\label{b.21b}
\end{eqnarray}
\begin{equation}
\Delta_M(x) = {\tilde \Delta}_o J_M(2\sqrt{\kappa } x)
\;.
\label{b.21c}
\end{equation}

Here we come to the crucial point in our derivation. Eq.({b.21b}) gives us
the final answer as a {\it superposition  of
 non-interacting} correlation functions for
effective tunneling rates $\Delta_M(x)$  with the proper weighting
\begin{equation}
P_{\Uparrow \Uparrow}(t)= \int_0^\infty dx x\:e^{-x^2} \big( 1+ \cos
[2{\tilde \Delta}_o J_M(2\sqrt{\kappa } x)t]  \big) \;,
\label{b.22}
\end{equation}
For $M=0$ this is the form quoted in Eq.(\ref{4.21o}) of the text.

It is worth noting that non-zero $M$ enters  this calculation as the overall
phase factor starting from (\ref{b.7}) and crown it in (\ref{b.20}) while
finally intergating over $\{ \chi_\alpha \} $ to produce the Bessel function
of order $M$. This observation allows to generalise any calculation done
for $M=0$ to finite $M$ by simply replacing $J_0 \to J_M$ in the final answer -
the  prescription which we make use in other Appendices.

\section{ Derivations for the generic case}

We outline here the derivations for section \ref{sec:5}, incorporating
all three mechanisms.

\subsection{Orthogonality blocked topological decoherence}

 As discussed in the text, a formal expression for this case can be easily
written down, by including the induced topological phase $\pm [\Phi +
\sum_{k=1}^N (\alpha_k {\vec n}_k -i\xi_k {\vec v}_k )\cdot  {\hat {\vec \sigma
}}_k ]$
which accumulates during each instanton, and then averaging over the
trajectories
of the $\{ {\vec \sigma }_k \}$. Here we evaluate the $M=0$ contribution;
The result for $P_{M}(t)$ then follows from obvious generalisation. We have
\begin{equation}
P_{0}(t) =\sum_{n=0}^\infty \sum_{m=0}^\infty {(i{\tilde \Delta}_o t)^{2(n+m)}
\over (2n)!(2m)!}
\sum_{\{ g_l=\pm \} } e^{i\Phi \sum_{l=1}^{2(n+m)}g_l }
\prod_{i=1}^n \prod_{j=1}^m \int {d\xi_i \over 2\pi }
\int {d\xi_j^{\prime} \over 2\pi } \langle
{\hat T}_{2m}^{\dag}(g_l) {\hat T}_{2n}(g_l) \rangle \;,
\label{c.1}
\end{equation}
where the transition matrices now include these phases, i.e.,
\begin{equation}
 {\hat T}_{2n}(g_l) = e^{i\xi_{2n}{\hat {\cal P}}} {\hat U}^{\dag }_{top}
e^{i\xi_{2n-1}{\hat {\cal P}}} \dots
e^{i\xi_{1} {\hat {\cal P}}} {\hat U}_{top}  \;.
\label{c.2}
\end{equation}
and ${\hat U}_{top}$ now generalizes the ${\hat U}$ involved in orthogonality
blocking, viz.,
\begin{equation}
{\hat U}_{top} =\prod_{k=1}^N \exp \left\{
ig_l[(\alpha_k {\vec n}_k -i\xi_k {\vec v}_k )]\cdot  {\hat {\vec \sigma }}_k
-i\beta_k {\hat \sigma }_k^x \right\}  \;.
\label{c.3}
\end{equation}
To evaluate (\ref{c.1}), we rewrite it as
\begin{equation}
P_{0}(t) =\sum_{n=0}^\infty \sum_{m=0}^\infty {(i{\tilde \Delta}_o t)^{2(n+m)}
\over (2n)!(2m)!}
\sum_{\{ g_l=\pm \} } e^{i\Phi \sum_{l=1}^{2(n+m)}g_l }
\prod_{\rho=1}^{2(n+m)} \int {d\xi_\rho \over 2\pi } \exp \left\{ -
K^{eff}_{nm}( \{ g_l \} , \{ \xi_\rho \})  \right\} \;,
\label{c.4}
\end{equation}
and calculate the $K^{eff}_{nm}(\{ g_l \} ,  \{ \xi_\rho \})$ up to order in
$\alpha_k^2,\;\beta_k^2,\;\xi_k^2$  (see Appendix A.2).

Consider first the case where $\beta_k$ and $\xi_k$ are zero. Then, with the
usual
assumption that the individual $\alpha_k$ are small, but not necessarily
$\lambda =1/2 \sum_k  \alpha_k^2$, the "effective action" $K^{eff}_{nm}$
simplifies to
(compare Eq.(\ref{b.16}))
\begin{equation}
K^{eff}_{nm}( \{ \xi_\rho \}) =  \lambda^\prime
 \sum_{\rho^\prime , \rho }^{2(n+m)} g_\rho g_{\rho^\prime}+
(\lambda - \lambda^\prime ) \sum_{\rho^\prime ,\rho }^{2(n+m)}
\cos (\chi_\rho - \chi_{\rho^\prime} ) g_\rho g_{\rho^\prime} \;,
\label{c.5}
\end{equation}
which generalizes from orthogonality blocking; the $\chi_\rho $ are defined as
in
 (\ref{b.15}), and
\begin{eqnarray}
\lambda &= &{1 \over 2} \sum_{k=1}^N \alpha_k^2 \nonumber \\
\lambda^\prime &=& {1 \over 2} \sum_{k=1}^N \alpha_k^2(n_k^z)^2  \;.
\label{c.6}
\end{eqnarray}

The term in $\lambda^\prime $ comes from the expansion to $O(\alpha_k^2)$
(analogous to
the expansion in $\beta_k^2$ in Appendix B), which produces the average
\begin{equation}
\langle \sigma_k \mid {\hat {\vec \sigma }}_k \cdot {\vec n}_k e^{-i
\sum_{j=\rho}^{\rho^\prime -1} \xi_j
({\hat \sigma }_k^z -\sigma_k) } {\hat {\vec \sigma }}_k \cdot {\vec n}_k \mid
\sigma_k \rangle =
[(n_k^z)^2 +(1-(n_k^z)^2)e^{-2i \sigma_k \sum_{j=\rho}^{\rho^\prime -1} \xi_j }
]\;,
\label{c.7}
\end{equation}
(compare (\ref{b.10}); again, we assume equal statistical weightings for
$\uparrow $ and $\downarrow $ spins).

We now use the same trick of (\ref{b.17}) and (\ref{b.18}), to write
\begin{equation}
{\vec {\cal S}}^2 =  \left( \sum_{\rho =1}^{2(n+m)} g_\rho {\vec s}_\rho
\right) ^2
= \sum_{\rho^\prime , \rho } g_\rho g_{\rho^\prime}
 \cos (\chi_\rho - \chi_{\rho^\prime} ) \;,
\label{c.8}
\end{equation}
with the ${\vec s}_\rho =(\cos \chi_\rho ,\sin \chi_\rho )$ as before; making
the
change of variables $\chi_\rho =\chi_\rho +\pi g_\rho /2 $, we get
the generalization of (\ref{b.21b}), viz.,
\begin{eqnarray}
P_{0}(t) &=& 2\int dx x e^{-x^2} {1 \over 2} \bigg[ 1 +
\sum_{s=0}^{\infty} {(2i{\tilde \Delta}_o t)^{2s} \over (2s)!}
 J_0^{2s}(2x\sqrt{\lambda - \lambda^\prime})  \nonumber \\
&\times &
\sum_{\{ g_l=\pm \} } e^{i\Phi \sum_{l=1}^{2(n+m)}g_l - \lambda^\prime
 (\sum_{l=1}^{2(n+m)}g_l)^2}  \bigg] \nonumber \\
&=& 2\int dx x e^{-x^2} {1 \over 2}
 \bigg[ 1+ \sum_{s=0}^{\infty} {[2i{\tilde \Delta}_o J_0(2x\sqrt{\lambda -
\lambda^\prime})t]^{2s} \over (2s)!}   \nonumber \\
& \times &
\sum_{\{ g_l=\pm \} } e^{i\Phi \sum_{l=1}^{2(n+m)}g_l - \lambda^\prime
 (\sum_{l=1}^{2(n+m)}g_l)^2}  \bigg] \;,
\label{c.9}
\end{eqnarray}
Writing the sum over $\{ g_l \} $ in the form
\begin{equation}
\sum_{\{ g_l=\pm \} } e^{i\Phi \sum_{l=1}^{2(n+m)}g_l - \lambda^\prime
 (\sum_{l=1}^{2(n+m)}g_l)^2} = \sum_{m=0}^{2s} {(2s)! \over (2s-m)! m!}
e^{i\Phi (2s-2m) - \lambda^\prime (2s-2m)^2 }
\label{c.10}
\end{equation}
we now use the identity in (\ref{4.88}) to eliminate this term in favour of a
phase
integration over $\varphi $; defining $ F_{\lambda^\prime}(\nu ) =
e^{-4 \lambda^\prime \nu^2 }$ as
in the text, we get $P_{\Uparrow \Uparrow} (t)$ in the form
\begin{equation}
P_{0} (t) = 2\int dx x e^{-x^2} \int {d\varphi \over 2\pi }
\sum_{m=-\infty}^{\infty} F_{\lambda^\prime}(m)  e^{i2m(\Phi -\varphi )}
P_{\Uparrow \Uparrow}^{(0)} (t, {\tilde \Delta}_o (\varphi ,x ))
\label{c.11}
\end{equation}
where
\begin{equation}
{\tilde \Delta}_o (\varphi ,x ) = 2 {\tilde \Delta}_o \cos \varphi
J_0(2x\sqrt{\lambda - \lambda^\prime}) \;.
\label{c.12}
\end{equation}

If we do the integration over $\varphi $, this gives
\begin{equation}
P_{0}(t) = \int dx x e^{-x^2}  \left\{ 1+ \sum_{m=-\infty }^{\infty}
F_{\lambda^\prime}(m) e^{i2m\Phi }
J_{2m} [4{\tilde \Delta}_o t J_0(2x\sqrt{\lambda -\lambda^\prime})] \right\}
\;,
\label{c.13}
\end{equation}
which is Eq.(\ref{5.6}) in the text.

The above calculation is generalized for non-zero $M$ by noting that
the only modification will be in a phase factor
$iM\sum_{\rho }(-1)^{\rho +1}\chi_{\rho} $ throughout the whole
derivation, which will finally result in  the replacement
$J_0(2x\sqrt{\lambda - \lambda^\prime}) \to J_M(2x\sqrt{\lambda -
\lambda^\prime})$  in
Eq.(\ref{c.12}) for the effective tunneling rate.

\subsection{Orthogonality blocking plus degeneracy blocking}

We wish to evaluate the "biased orthogonality blocking" problem described in
section \ref{sec:5}.B. We generalize the expression (\ref{b1.8}) for a biased
2-level system (topological phase is not yet included here) to include
orthogonality blocking via our projection operators (cf. Eq.(\ref{b.5}); this
gives
\begin{equation}
 A_{M=0}(p, \epsilon ) = {1 \over p-i\epsilon }
\sum_{n=0}^{\infty} \left(
{ (-i\Delta_\Phi )^2 \over p^2+\epsilon^2 } \right)^{n}
\prod_{i=1}^{2n} \int {d\xi_i \over 2\pi } {\hat T}_{2n} \mid \{ \sigma_k^{in}
\} \rangle \;,
\label{c.16}
\end{equation}
with $A_{M=0}(t, \epsilon )$ given by the Laplace transform (\ref{b1.7}) of
(\ref{c.16}). It then follows that
\begin{equation}
P_{\Uparrow \Uparrow} (t, \epsilon ) = \int \! \int_{-i\infty }^{i\infty} dp_1
dp_2 e^{(p_1+p_2)t} {1 \over p_1 -i
\epsilon } {1 \over p_2 -i\epsilon } \sum_{n=0}^{\infty}\sum_{m=0}^{\infty}
B_{nm}(p_1,p_2,\epsilon )
\label{5.14}
\end{equation}
\begin{equation}
B_{nm}(p_1,p_2,\epsilon ) =
\left( { (i{\tilde \Delta}_\Phi )^2 \over p_1^2+\epsilon^2 } \right)^{n}
\left( { (i{\tilde \Delta}_\Phi )^2 \over p_2^2+\epsilon^2 } \right)^{m}
\prod_{i=1}^n \prod_{j=1}^m \int {d\xi_i \over 2\pi }
\int {d\xi_j^{\prime} \over 2\pi } \langle
{\hat T}_{2m}^{\dag} {\hat T}_{2n} \rangle \;,
\label{5.15}
\end{equation}

Surprisingly, we find  that the problem of calculating
the correlation function in a bias  (after writing it as a series
expansion in the tunneling rate) can be again solved in terms of
non-interacting $P_{\Uparrow \Uparrow}^{(0)} (t, \epsilon )$, because the
 averages in (\ref{5.15}) are identical to those evaluated in Appendix A,
Eq.(\ref{b.20}), to give
\begin{equation}
\prod_{i=1}^{2n} \int {d\xi_i \over 2\pi }
\prod_{j=1}^{2m} \int {d\xi_j^\prime \over 2\pi }
\langle {\hat T}_{2m}^{\dag} {\hat T}_{2n} \rangle =
2\int dx x e^{-x^2} J_0^{2(n+m)}(2x\sqrt{\kappa}) \;.
\label{c.17}
\end{equation}
This can be absorbed into defining an effective $\Delta_o(x)
=2{\tilde \Delta}_\Phi J_0(2x\sqrt{\kappa})$, just as for pure orthogonality
blocking, and the series
become identical to Eq.(\ref{b1.10}) giving
the result (\ref{5.16}) in the text. The generalisation to $M\ne 0$
just consists
in the replacement $\Delta_o(x) \to \Delta_M(x)$.

\subsection{The generic case}

Here we calculate $P_{\Uparrow \Uparrow} (t)$
with all terms ($\alpha_k$, $\omega_k^{parallel}$, $\Phi$, and $\xi_k$)
{\it except} $\omega_k^{perp}$ included from the effective Hamiltonian.
Thus we return to the expression (\ref{c.3}) and (\ref{c.4}) and calculate the
"effective action" $K^{eff}_{nm}$ with nonzero $\xi_k$. Now, evaluating the
contribution from individual spins  from the expansion to
$O(\alpha_k^2,\:\xi_k^2)$,
we find for  the average   (cf. (\ref{c.7}))
\begin{eqnarray}
\langle \sigma_k \mid (i\alpha_k  {\vec n}_k + (-1)^\rho \xi_k {\vec v}_k )
\cdot {\hat {\vec \sigma }}_k & &
e^{-i \sum_{j=\rho}^{\rho^\prime -1} \xi_j ({\hat \sigma }_k^z -\sigma_k) }
\nonumber \\
 & \times &
(i\alpha_k  {\vec n}_k +(-1)^{\rho^\prime}
 \xi_k {\vec v}_k ) \cdot {\hat {\vec \sigma }}_k   \mid \sigma_k \rangle \;,
\label{c.22}
\end{eqnarray}
the following expression
\begin{eqnarray}
& & (i\alpha_k  n_k^z +(-1)^\rho \xi_k v_k^z )\:
(i\alpha_k  n_k^z +(-1)^{\rho^\prime} \xi_k v_k^z )
  \nonumber \\
& +& e^{-2i \sum_{j=\rho}^{\rho^\prime -1} \xi_j } \nonumber \\
& & \times
 [ i\alpha_k (n_k^x+in_k^y) +(-1)^\rho \xi_k (v_k^x+iv_k^y) ]
\:[ i\alpha_k (n_k^x-in_k^y) +(-1)^{\rho^\prime} \xi_k (v_k^x-iv_k^y) ] \;,
\label{c.23}
\end{eqnarray}
Thus, in full analogy to the derivation done in Appendix B (Eq.(\ref{b.11})),
we find the
contribution of an individual spin to the action to be
\begin{eqnarray}
& &{1 \over 2} \sum_{\rho ,\rho^\prime =1}^{2(n+m)} g_\rho g_{\rho^\prime}
\bigg\{
\big[ \alpha_k^2(n_k^z)^2 -(-1)^{\rho +\rho^\prime}\xi_k^2(v_k^z)^2
-i\alpha_k\xi_k n_k^z v_k^z ((-1)^\rho +(-1)^{\rho^\prime} ) \big] \nonumber \\
& + & \cos (\chi_\rho - \chi_{\rho^\prime }) \big[
\alpha_k^2(1-(n_k^z)^2) -(-1)^{\rho +\rho^\prime}\xi_k^2(1-(v_k^z)^2) \nonumber
\\
&-& i\alpha_k\xi_k ({\vec n}_k \cdot {\vec v}_k -n_k^z v_k^z )((-1)^\rho
+(-1)^{\rho^\prime} ) \big]
\bigg\} \;,
\label{c.24}
\end{eqnarray}
Summing up the contributions from $N$ spins we generate the "effective action",
which now takes the form
\begin{eqnarray}
K^{eff}_{nm} &= & \sum_{\rho ,\rho^\prime =1}^{2(n+m)}
g_\rho g_{\rho^\prime}  \bigg\{
\big[ \lambda^\prime -\eta^\prime (-1)^{\rho +\rho^\prime}
-i\gamma^\prime ((-1)^\rho +(-1)^{\rho^\prime} ) \big] \nonumber \\
& + & \cos (\chi_\rho - \chi_{\rho^\prime }) \big[
(\lambda - \lambda^\prime ) -(-1)^{\rho +\rho^\prime} (\eta - \eta^\prime )
-i(\gamma - \gamma^\prime )((-1)^\rho +(-1)^{\rho^\prime} ) \big] \bigg\} \;,
\label{c.25}
\end{eqnarray}
where the constants are  defined by:
\begin{equation}
\lambda ={1 \over 2} \sum_{k=1}^N \alpha_k^2 \;; \;\;\;\;
\lambda^\prime ={1 \over 2} \sum_{k=1}^N \alpha_k^2(n_k^z)^2 \;;
\label{c.26}
\end{equation}
\begin{equation}
\eta ={1 \over 2} \sum_{k=1}^N \xi_k^2 \;; \;\;\;\;
\eta^\prime ={1 \over 2} \sum_{k=1}^N \xi_k^2(v_k^z)^2 \;;
\label{c.27}
\end{equation}
\begin{equation}
\gamma ={1 \over 2} \sum_{k=1}^N \alpha_k\xi_k {\vec n}_k \cdot {\vec v}_k \;;
\;\;\;\;
\gamma^\prime ={1 \over 2} \sum_{k=1}^N \alpha_k\xi_k n_k^z v_k^z \;;
\label{c.28}
\end{equation}

As before we change variables according to $\chi_\rho =\chi_\rho +\pi $ when
$g_\rho =-1$, to introduce the odd and even spin fields
\begin{equation}
{\vec {\cal S}}_o =\sum_{\rho =odd}^{2(n+m)-1} {\vec s}(\chi_\rho ) \;;
\;\;\;\;\;
{\vec {\cal S}}_e =\sum_{\rho =even}^{2(n+m)} {\vec s}(\chi_\rho ) \;.
\label{c.29}
\end{equation}
Now we notice that the effective action is a {\it quadratic} form in
$({\vec {\cal S}}_o,{\vec {\cal S}}_e )$:
\begin{eqnarray}
K^{eff}_{nm} &= &  \lambda^\prime \big( \sum_{\rho } g_\rho \big)^2 -
\eta^\prime \big( \sum_{\rho } (-1)^{\rho } g_\rho \big)^2 -
2i\gamma^\prime \big( \sum_{\rho } g_\rho \big)
\big( \sum_{\rho } (-1)^{\rho } g_\rho \big) \nonumber \\
& & + (\lambda - \lambda^\prime )({\vec {\cal S}}_e + {\vec {\cal S}}_o)^2 -
(\eta - \eta^\prime )({\vec {\cal S}}_e - {\vec {\cal S}}_o)^2 -
2i(\gamma - \gamma^\prime )( {\cal S}_e^2 -{\cal S}_o^2)  \;,
\label{c.300}
\end{eqnarray}

The rest is simple now. First we employ spectral representations for the
$\delta$-functions
$\delta ({\vec {\cal S}}_o -\sum_{\rho =odd}^{2(n+m)-1} {\vec s}_\rho )$ and
$\delta ({\vec {\cal S}}_e -\sum_{\rho =even}^{2(n+m)} {\vec s}_\rho )$ to
integrate
over the angle variables $\chi_\rho$ (compare Eq.(\ref{b.20})) to give
($s=n+m$)
\begin{eqnarray}
& &\prod_{\rho=1}^{2s} \int {d\xi_\rho \over 2\pi } \exp \left\{ -
K^{eff}_{nm}( \{ g_l \} , \{ \xi_\rho \})  \right\} =
e^{-\lambda^\prime ( \sum_{\rho } g_\rho )^2 +
\eta^\prime ( \sum_{\rho } (-1)^{\rho } g_\rho )^2 -
2i\gamma^\prime ( \sum_{\rho } g_\rho )
( \sum_{\rho } (-1)^{\rho } g_\rho ) } \nonumber \\
& & \times  \int {d{\vec x}_1d{\vec x}_2 \over (2\pi )^4 }
J_0^{s}(x_1)J_0^{s}(x_2)  \int d{\vec {\cal S}}_o d{\vec {\cal S}}_e \nonumber
\\
& & \times \exp
\bigg\{ -a({\vec {\cal S}}_e + {\vec {\cal S}}_o)^2+
b({\vec {\cal S}}_e - {\vec {\cal S}}_o)^2 +2ic
( {\cal S}_e^2 -{\cal S}_o^2) +i{\vec x}_1\cdot {\vec {\cal S}}_e+
i{\vec x}_2\cdot {\vec {\cal S}}_o \bigg\} \;.
\label{c.30}
\end{eqnarray}
with obvious definition
\begin{equation}
a=\lambda - \lambda^\prime \;;\;\;\;\;
b=\eta - \eta^\prime \;;\;\;\;\;\;
c=\gamma - \gamma^\prime \;.
\label{c.31}
\end{equation}
The Gaussian integration over $d{\vec {\cal S}}_o$, $d{\vec {\cal S}}_e$, and
then
the integration over the angle between the ${\vec x}_1 $ and ${\vec x}_2 $
gives for the
 expression (\ref{c.30}) the formula
\begin{eqnarray}
& & e^{-\lambda^\prime ( \sum_{\rho } g_\rho )^2 +
\eta^\prime ( \sum_{\rho } (-1)^{\rho } g_\rho )^2 -
2i\gamma^\prime ( \sum_{\rho } g_\rho )
( \sum_{\rho } (-1)^{\rho } g_\rho ) } \int dx_1 \int dx_2 x_1 x_2
J_0^{s}(x_1)J_0^{s}(x_2) \nonumber \\
 &\times & {1 \over 8(ab-c^2)} I_0\left( {(a+b) x_1x_2 \over 8(ab-c^2) }
\right)
\exp \left\{ { (a-b+2ic)x_1^2 + (a-b-2ic)x_2^2 \over 16(ab-c^2) } \right\} \;.
\label{c.32}
\end{eqnarray}

Finally we note that the sum over $\{ g_\rho \}$ can be written as
\begin{eqnarray}
\sum_{\{ g_\rho =\pm \} } & \equiv & \sum_{m_1,m_2=-\infty}^{\infty}
\sum_{\{ g_\rho =\pm \} }
\delta (2m_1 - \sum_{\rho } g_\rho)\: \delta (2m_2 - \sum_{\rho } (-1)^{\rho }
g_\rho )
\nonumber \\
& =& \int \! \int {d\varphi_1 d\varphi_2 \over (2\pi )^2 }
\sum_{m_1,m_2=-\infty}^{\infty}
e^{i2m_1\varphi_1 +i2m_2\varphi_2} \sum_{\{ g_\rho =\pm \} }
e^{-i\varphi_1 \sum_{\rho } g_\rho -i\varphi_2 \sum_{\rho }  (-1)^{\rho }
g_\rho }
\nonumber \\
& =& \int \! \int {d\varphi_1 d\varphi_2 \over (2\pi )^2 }
\sum_{m_1,m_2=-\infty}^{\infty}
e^{i2m_1\varphi_1 +i2m_2\varphi_2} \big[
2\cos (\varphi_1 +\varphi_2)\:2\cos (\varphi_1 -\varphi_2) \big] \;.
\label{c.33}
\end{eqnarray}
Combining now (\ref{c.32}), (\ref{c.33}), and (\ref{c.4}) together we observe
that the sum over $n$ and $m$ is nothing but the series for the coherent
dynamics expansion in powers of the renormalized tunneling splitting
\begin{equation}
{\tilde \Delta}_o^2 (x_1, x_2, \varphi_1, \varphi_2 ) = 4{\tilde \Delta}_o^2
\cos (\varphi_1 + \varphi_2)
\cos (\varphi_1 - \varphi_2 ) J_0(x_1)J_0(x_2) \;,
\label{c.34}
\end{equation}
which is then integrated over the variables $x_1,x_2,\varphi_1,\varphi_2$
and summed
over $m_1,m_2$ with the weight given by
\begin{eqnarray}
{\cal Z} &= & e^{2i[ m_1(\Phi -\varphi_1) -m_2
\varphi_2 +4m_1m_2\gamma^\prime ]} e^{4(\eta^\prime m_2^2 - \lambda^\prime
m_1^2)}
\nonumber \\
 &\times & {x_1x_2 \over 8(ab-c^2)} I_0\left( {(a+b) x_1x_2 \over 8(ab-c^2) }
\right)
\exp \left\{ { (a-b+2ic)x_1^2 + (a-b-2ic)x_2^2 \over 16(ab-c^2) } \right\} \;.
\label{5.34}
\end{eqnarray}
One then finds
\begin{eqnarray}
P_{\Uparrow \Uparrow} (t) = \int {d\varphi_1  \over 2\pi }\int {d\varphi_2
\over 2\pi } & &
\sum_{m_1 =-\infty }^{\infty}\sum_{m_2 =-\infty }^{\infty}
\int dx_1 \int dx_2  \nonumber \\
& \times &
{\cal Z}(\varphi_1,\varphi_2,x_1,x_2,m_1,m_2) P_{\Uparrow \Uparrow}^{(0)}
[t, {\tilde \Delta}_o (x_1,x_2,\varphi_1,\varphi_2)] \;,
\label{5.29}
\end{eqnarray}
and (\ref{5.29}) has an obvious generalisation to include the bias
integration $\int d \epsilon$.

\begin{center}
{\bf FIGURE CAPTIONS }
\end{center}

\figure{{\bf Figure 1 }Schematic depiction of the Central Spin Model}

\figure{{\bf Figure 2 }The range of coupling energies between a typical central
spin and its
spin environment. The hyperfine contact interaction is shown for a variety
of nuclear spins in various hosts.}

\figure{{\bf Figure 3 }A typical trajectory for ${\vec S} (\tau )$,
contributing to the path
integral.}

\figure{{\bf Figure 4 }The function $\alpha_o (\omega_o )$ for the instanton of
the toy model
Hamiltonian. This form is typical of virtually any instanton describing the
tunneling of our central spin.}

\figure{{\bf Figure 5 }Behaviour of the central spin in the topological
decoherence limit.
In (a) is shown $P_{\Uparrow \Uparrow} (t)$ for the intermediate  coupling
case; and (b)
shows the spectral function $\chi^{\prime \prime} (\omega )$ deriving
from this. These forms are the {\it universal forms} arising for topological
decoherence (see text).}

\figure{{\bf Figure 6 }$P_{\Uparrow \Uparrow} (t)$ for an $S=1/2$-integer spin
coupled to $40$
environmental spins with random  $\alpha_k$ between $0$ and $0.05$.}

\figure{{\bf Figure 7 }$P_{\Uparrow \Uparrow} (t)$ for an integer $S$, and
coupling to $20$
environmental spins with random  $\alpha_k$ between $0$ and $1$.}

\figure{{\bf Figure 8 }Definition of the orthogonality angle $\beta_k$, in
terms of the
unit vectors ${\vec \gamma}_k^{(1)}$ and ${\vec \gamma}_k^{(2)}$. This also
defines the
unit vectors ${\vec l}_k$ and ${\vec m}_k$ appearing in the effective
Hamiltonian (\ref{3.4}).}

\figure{{\bf Figure 9 }The effect of pure orthogonality blocking when
$\kappa$ is small $( \kappa \le O(1) )$.}

\figure{{\bf Figure 10 }The effect of pure orthogonality blocking when $\kappa
\gg 1 $.
As well as being pushed to lower frequencies  in accordance with
Eq.(\ref{4.23}), $\chi^{\prime \prime} (\omega )$ develops a complicated
peak structure.}

\figure{{\bf Figure 11 }$P_{\Uparrow \Uparrow} (t)$, in the orthogonality
blocking limit, for
$\kappa = 1000$. Notice the small resurgence of oscillation for
${\tilde \Delta}_o t \sim 300 $, coming from constructive interference  between
the
environment-induced peaks in $\chi^{\prime \prime} (\omega )$.}

\figure{{\bf Figure 12 }The effect of a finite but small spread in the
couplings
$\{ \omega_k \} $ on the energy levels of the system. In (a) the set of levels
for the
total "central spin + environment" is shown as we add environmental spins one
by one.
In (b) the resulting distribution $W(\epsilon )$ is shown.  The parameter
$\mu =N^{1/2} \delta \omega_k /\omega_o$. The separation between peaks for
$\mu \ll 1$ is $\omega_o$. The tunneling splitting is far smaller, indeed
invisible
on this scale for typical values of $\omega_o$. Usually $\mu >1$, giving
the smooth Gaussian; if $\mu \ll 1$ a peak structure appears in $W(\epsilon
)$.}

\figure{{\bf Figure 13 }The spectral weight $\chi^{\prime \prime} (\omega )$
for the case
of pure degeneracy blocking.

\figure{{\bf Figure 14 } A plot of the result for projected topological
decoherence,
in Eq.(\ref{5.10}), for small values of $(\lambda -\lambda^\prime )$; notice
that
even if $(\lambda -\lambda^\prime )$ is only $0.1$, the result is very
noticeably
different from pure topological decoherence (for which $\lambda -\lambda^\prime
=0$),
shown in Fig.5b.}

\figure{{\bf Figure 15 } Graphs of $\chi^{\prime \prime} (\omega )$  for
projected
topological decoherence, but now for $\lambda -\lambda^\prime \gg 1$.}

\figure{{\bf Figure 16 } The spectral function $\chi^{\prime \prime} (\omega )$
for degeneracy blocked orthogonality blocking (eq.(\ref{5.19})). Here we see
the
results for small $\kappa$ .}

\figure{{\bf Figure 17 } $\chi^{\prime \prime} (\omega )$
for degeneracy blocked orthogonality blocking, for $\kappa \gg 1$.}

\figure{{\bf Figure 18 } $\chi^{\prime \prime} (\omega )$ for the case where
all
3 decoherence mechanisms enter, and we assume that $\lambda ,\: \lambda^\prime
>1 $
(so there is topological decoherence). Here we show $\chi^{\prime \prime}
(\omega )$
for small values of $\lambda - \lambda^\prime $. }

\figure{{\bf Figure 19 } $\chi^{\prime \prime} (\omega )$  when all
3  mechanisms operate, and $\lambda ,\: \lambda^\prime >1 $; here we show
results
for large $(\lambda -\lambda^\prime )$.}

\figure{{\bf Figure 20 } $\chi^{\prime \prime} (\omega )$ for $TbFe_3$ grains,
calculated using the numbers in the text.}

\newpage

\begin{table}
\setdec 0.000
\caption{SELECTED HYPERFINE COUPLINGS}
\begin{tabular}{ccccc}
Nucleus & Host & State &$\omega_k$ ($MHz$) &Abundance (\%) \\
\tableline
$H^1$ & $CuCl_2 \cdot H_2O$ &(AFM) & $3.5$ & $99.99$ \\
$F^{19}$ & $MnF_2  $ & (AFM) & $160$ & $100$ \\
$Mn^{55}$ & $MnF_2$ &(AFM) & $680$ & $100$ \\
 & $MnB$ & (FM) & $217.7$ & " \\
$Fe^{57}$ & $Fe$ &(FM) & $46.65$ & $2.19$ \\
 & $\alpha - Fe_2O_3$ &(FM) & $71.5$ & " \\
 & $ Fe_3O_4$ &(Ferrite) & $63.55$ & " \\
 & $ Y_3Fe_5O_{12}$ &(YIG-FM) & $64.9/76.05$ & " \\
 & $ FeCo_5;$ &(FM) & $44.8$ & " \\
$Co^{59}$ & $ FeCo_5$ &(FM) & $289.2$ & $100$ \\
 & $ Co$ &(FM) & $228$ & " \\
$Ni^{61}$ & $ Ni$ &(FM) & $28.35$ & $1.19$ \\
$Tb^{159}$ & $ Tb$& (FM) & $3776,\:3108,\:2439$ & $100$ \\
$Dy^{161}$ & $ Dy$& (FM) & $1603,\:1219,\:830,\:445$ & $18.88$ \\
$Dy^{163}$ &  " & " & $1984,\:1574,\:1163$ & $24.97$ \\
    &   & &$755,\:347$ &  \\
\end{tabular}
\label{t1}
\end{table}

\begin{table}
\caption{SOME TYPICAL  NMR  LINEWIDTHS IN ANTIFERROMAGNETS COMING FROM
THE SUHL - NAKAMURA INTERACTION.
For $Mn^{55}$, a $1\:Oe $ linewidth roughly corresponds to a
 $\delta \omega_k \sim 1\: kHz$.}
\begin{tabular}{cccc}
Nucleus & Host &  & Linewidth (Oe) \\
\tableline
$F^{19}$  & $MnF_2  $ & (AFM) & $14$ \\
$Mn^{55}$ & "         &(AFM)  &   $1000$ \\
$Co^{59}$ & $ CoF_2$  &(AFM)  & $268$ ($1/2 \leftarrow \!\! \rightarrow -1/2$)
\\
        " & " & " & $258$ ($\pm 3/2 \leftarrow \!\! \rightarrow \pm 1/2$) \\
        " & " & " & $222$ ($5/2   \leftarrow \!\! \rightarrow 3/2$) \\
        " & " & " & $165$ ($7/2 \leftarrow \!\! \rightarrow  5/2$) \\
\end{tabular}
\label{t2}
\end{table}


\begin{thebibliography}{99}
\bibitem{44}
      One other example we have considered is the Aharonov-Bohm
    effect in a metallic ring. Here one needs a coupling which is sensitive to
    the direction of electron velocity in order to randomize the phase
    interference between the clockwise and anticlockwise trajectories of the
electron.
 We found that the coupling between the nuclear spins and the
single electron current is very weak to be observed experimentally. Note
however that
the coherence in this example is only 1-particle (individual electron)
coherence.
\end{thebibliography}
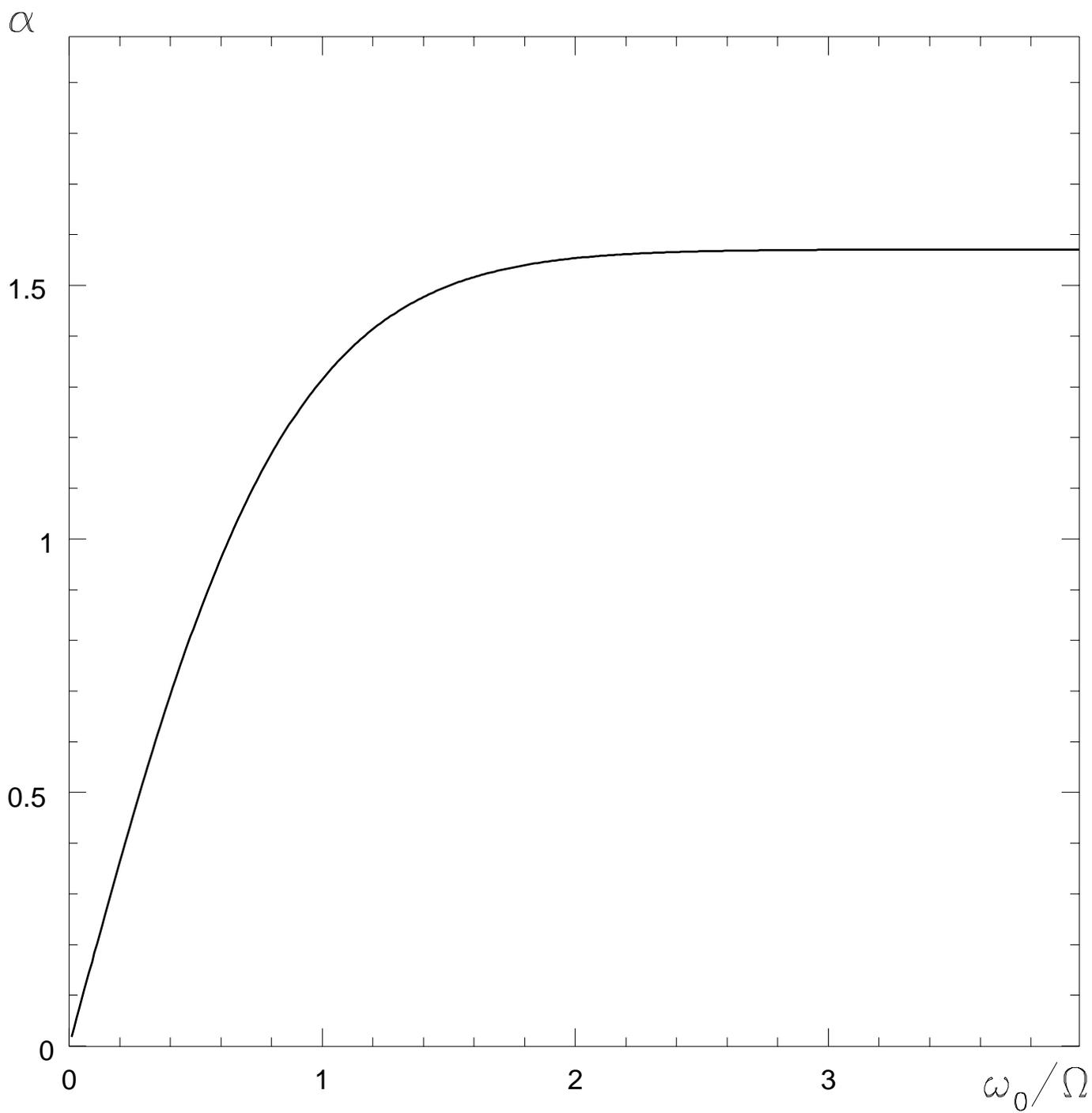
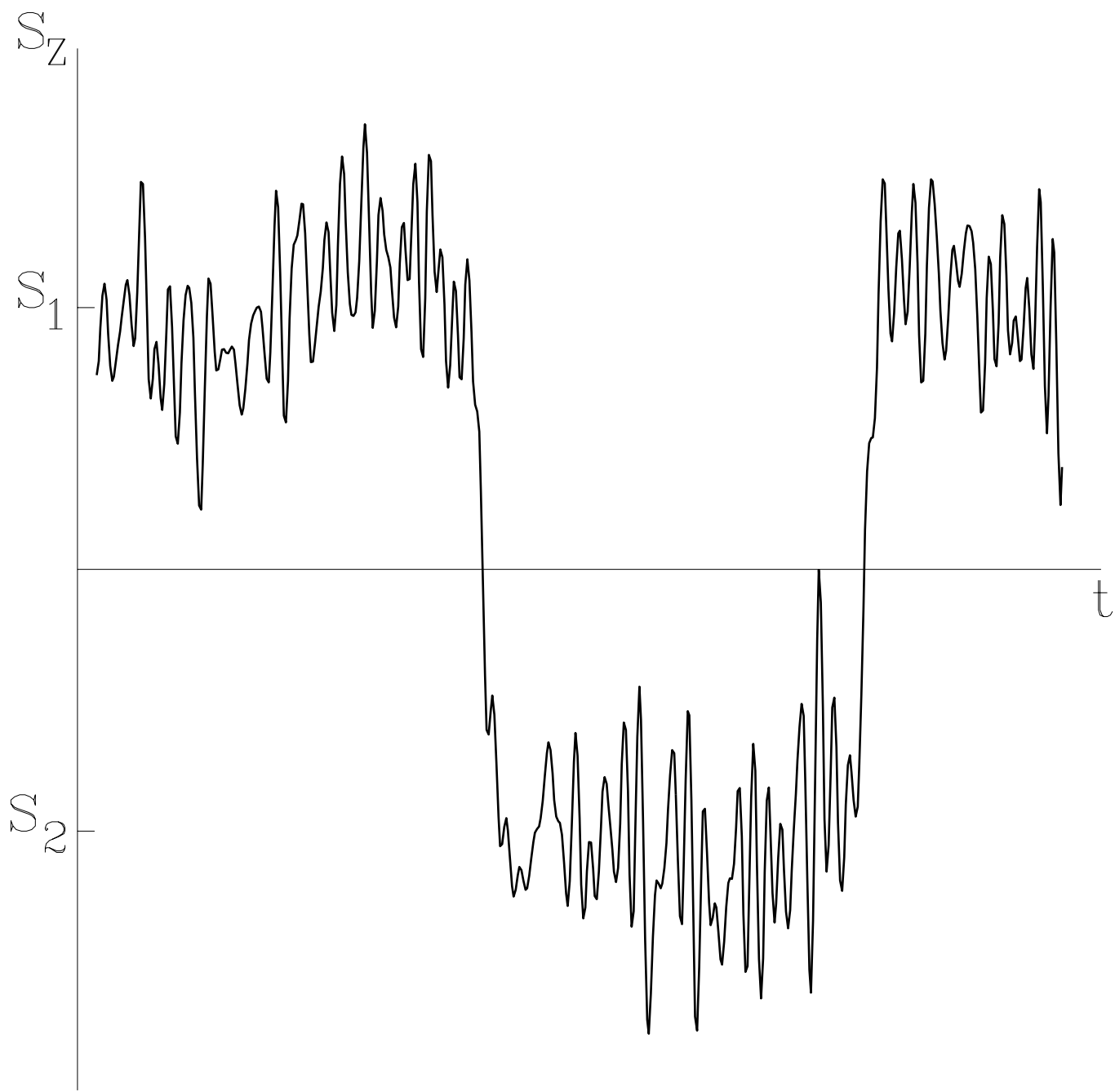
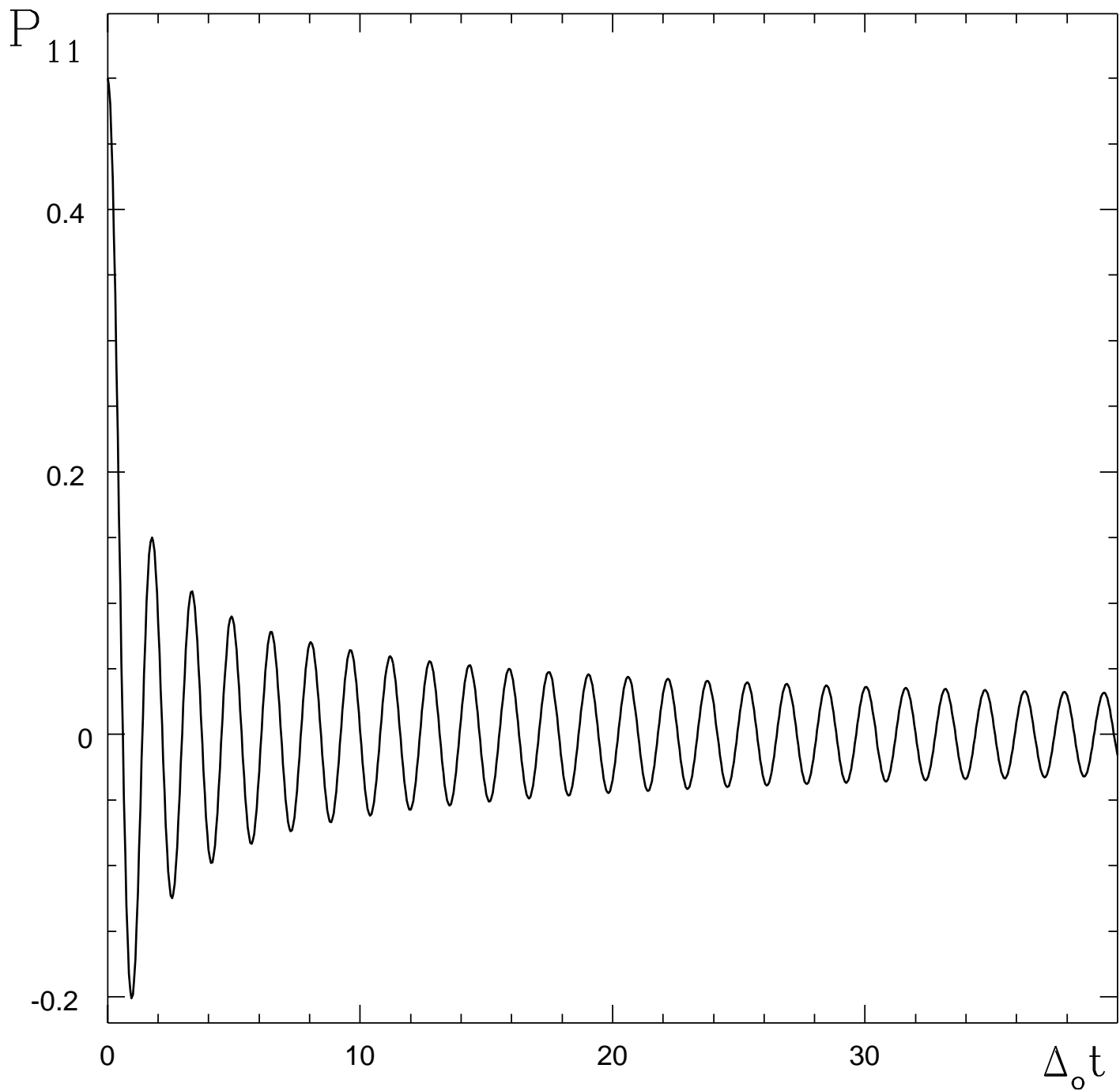
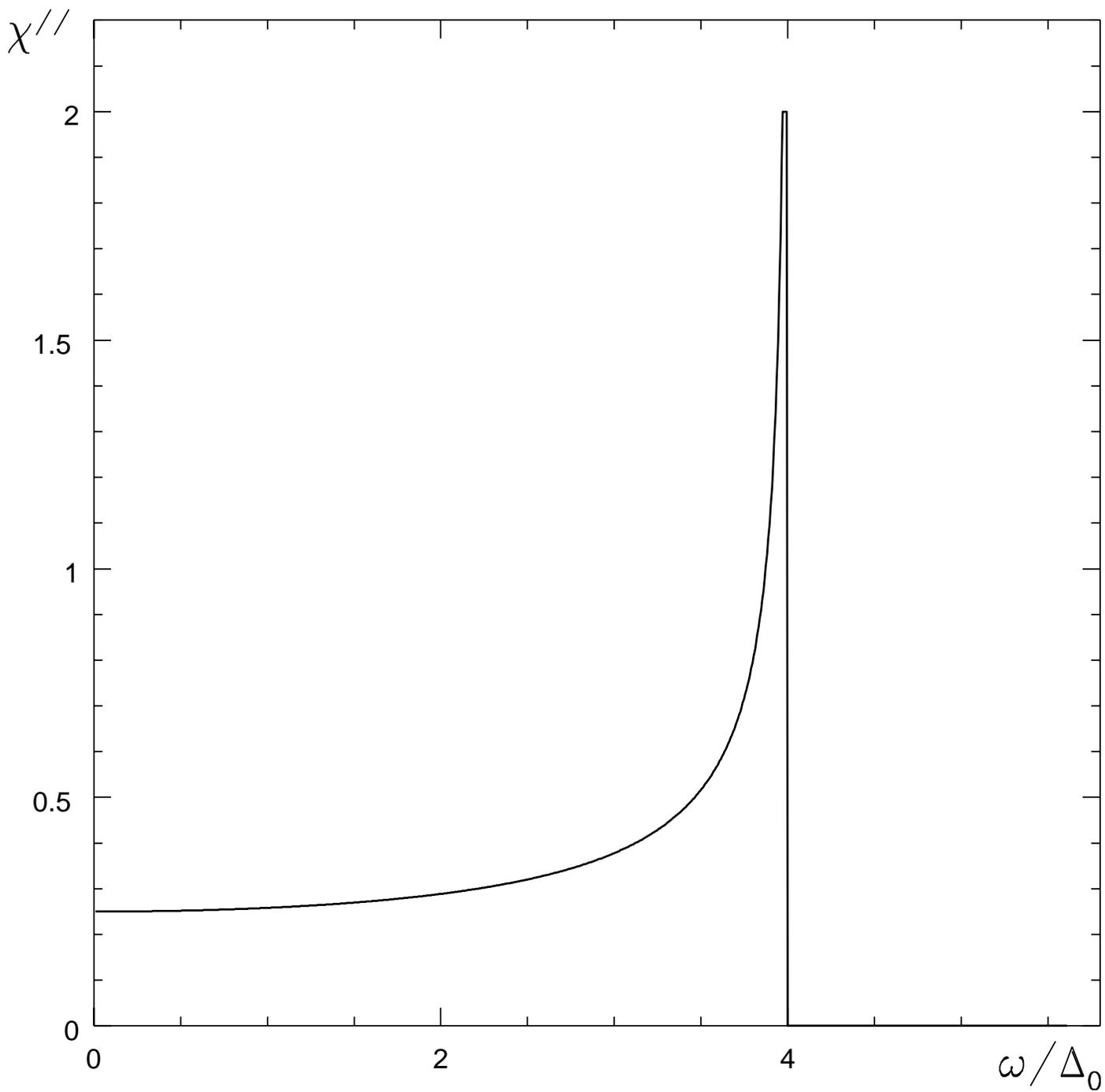
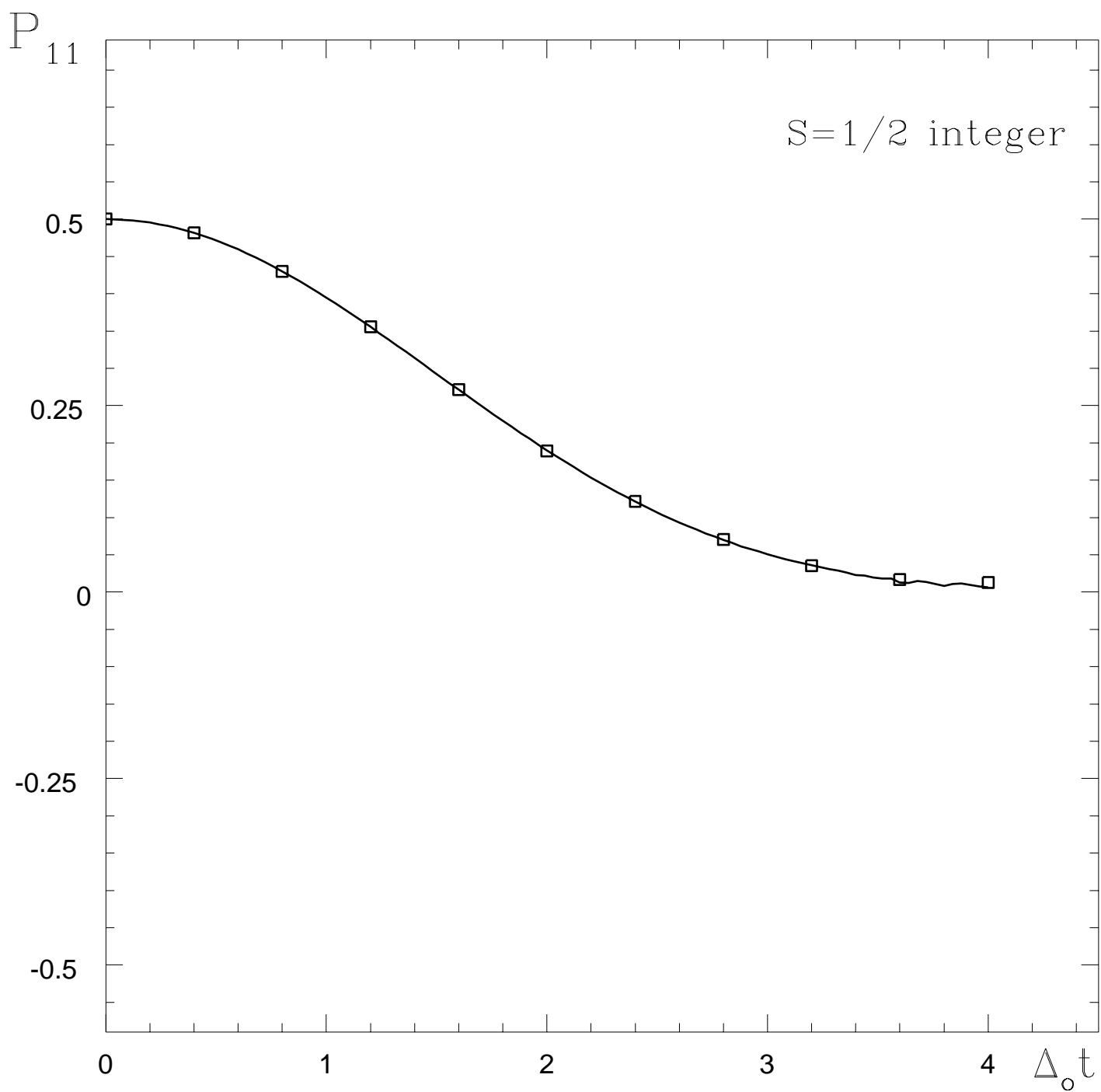
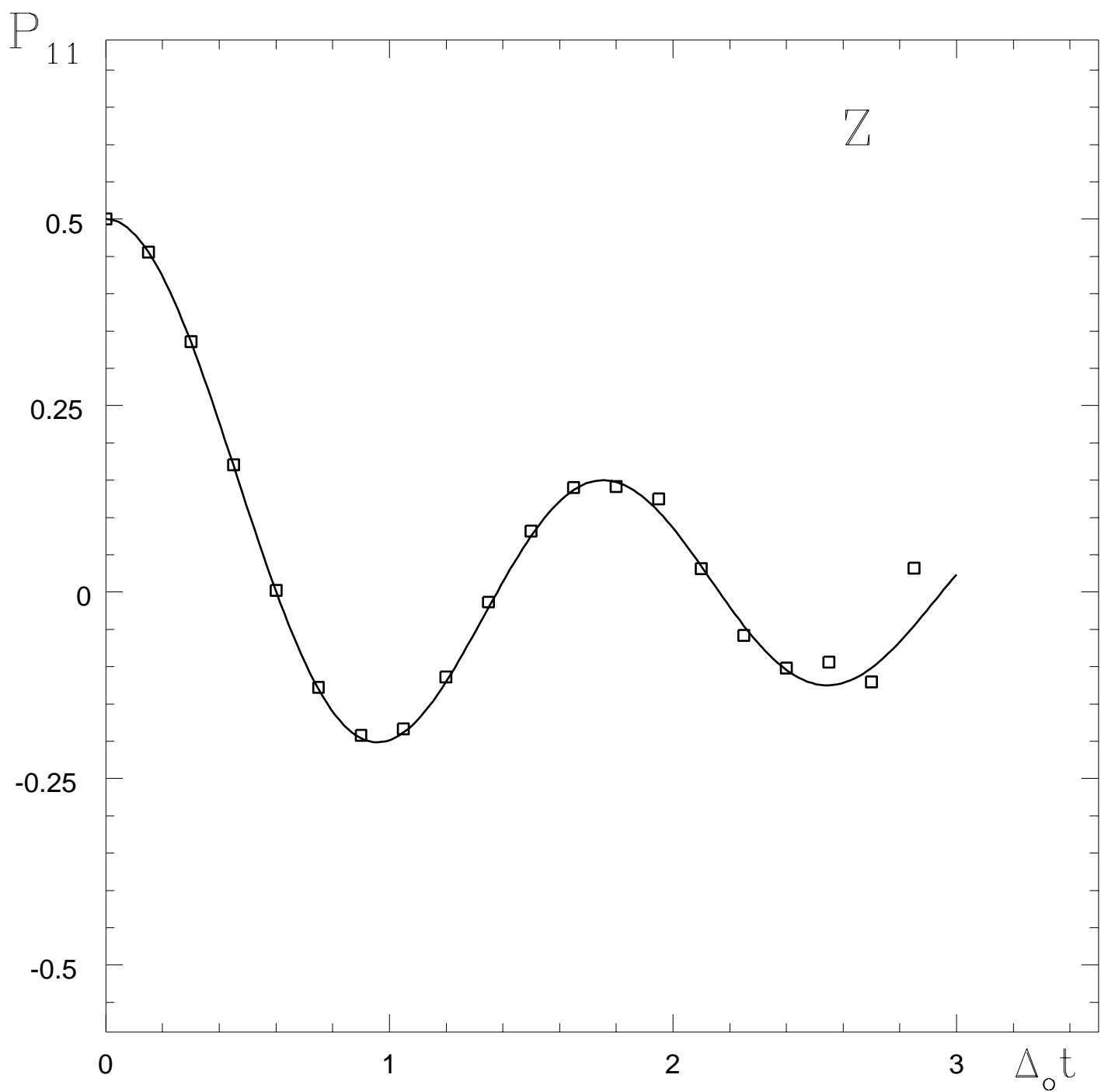
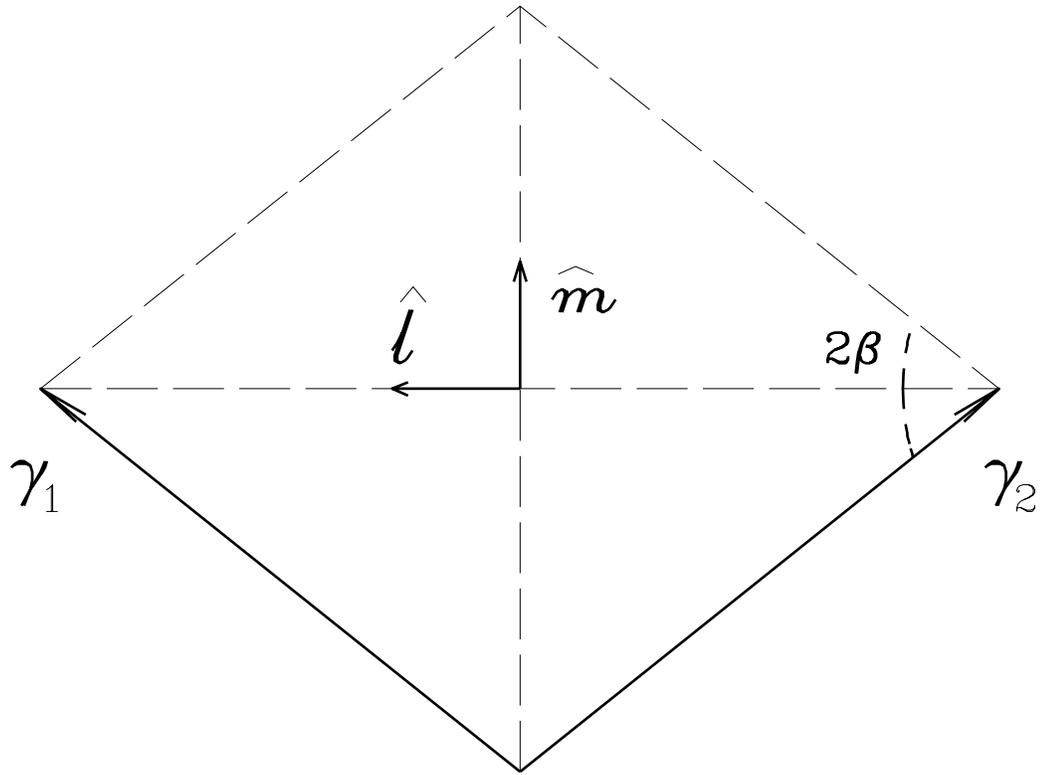
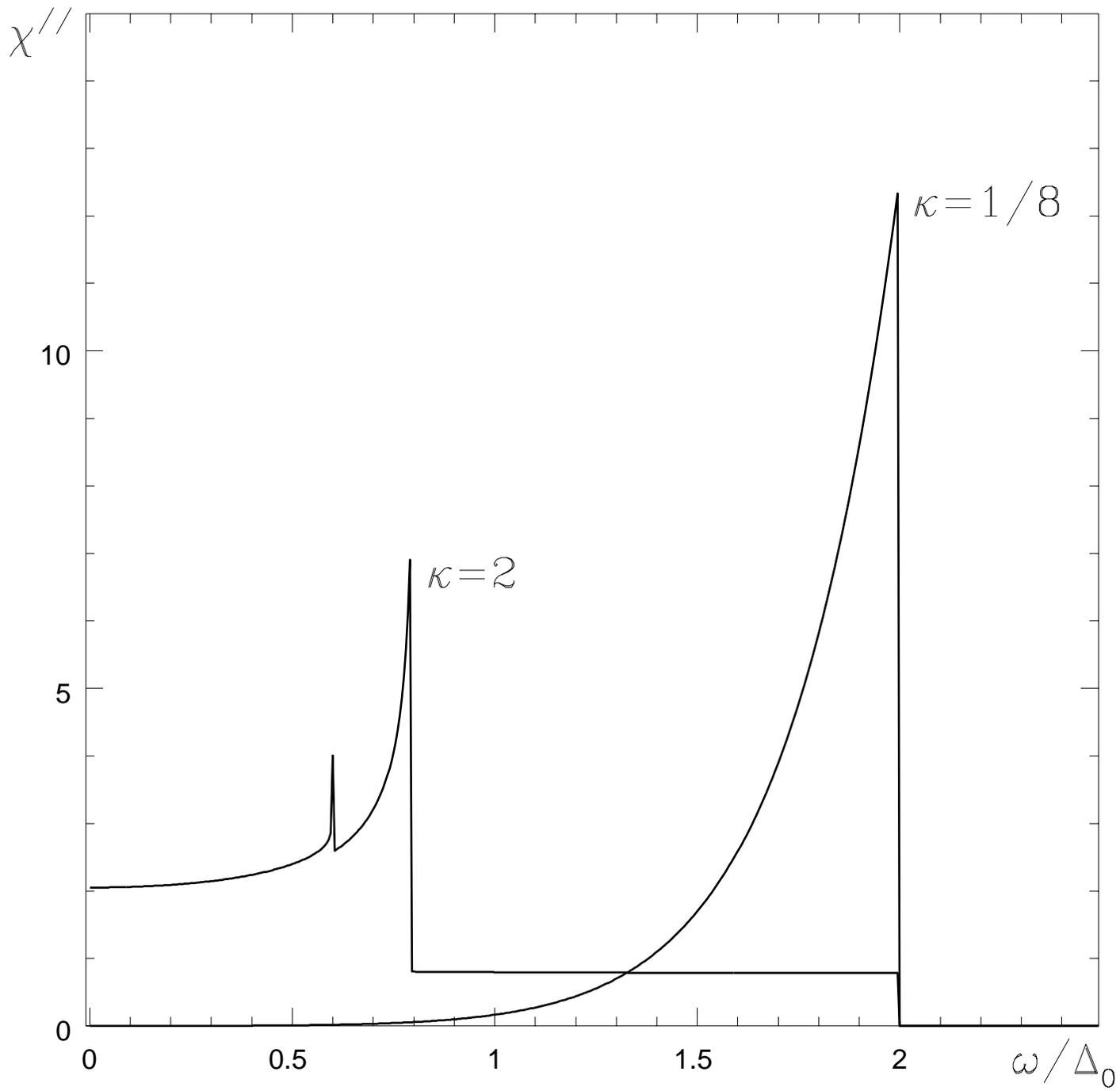
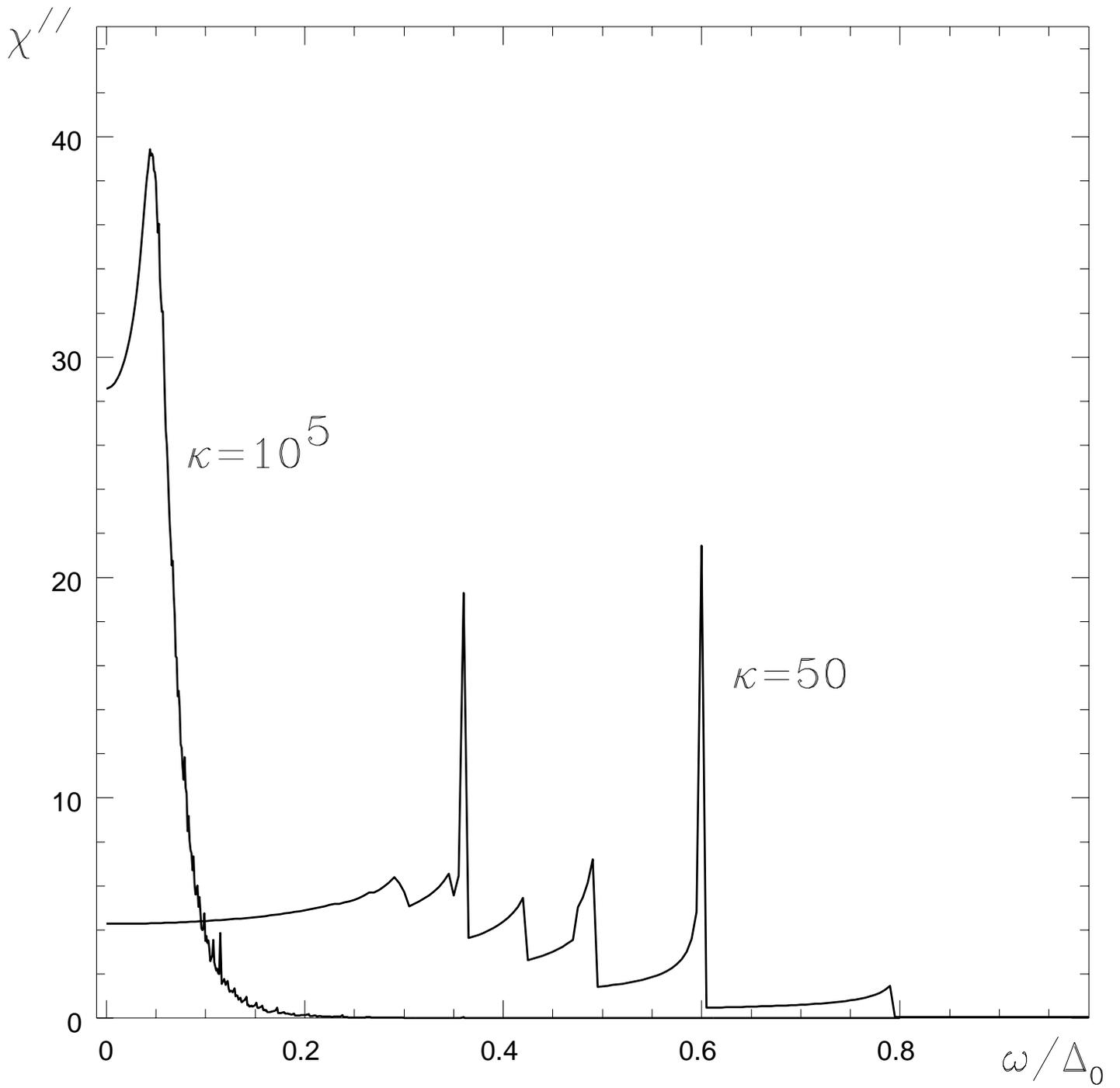
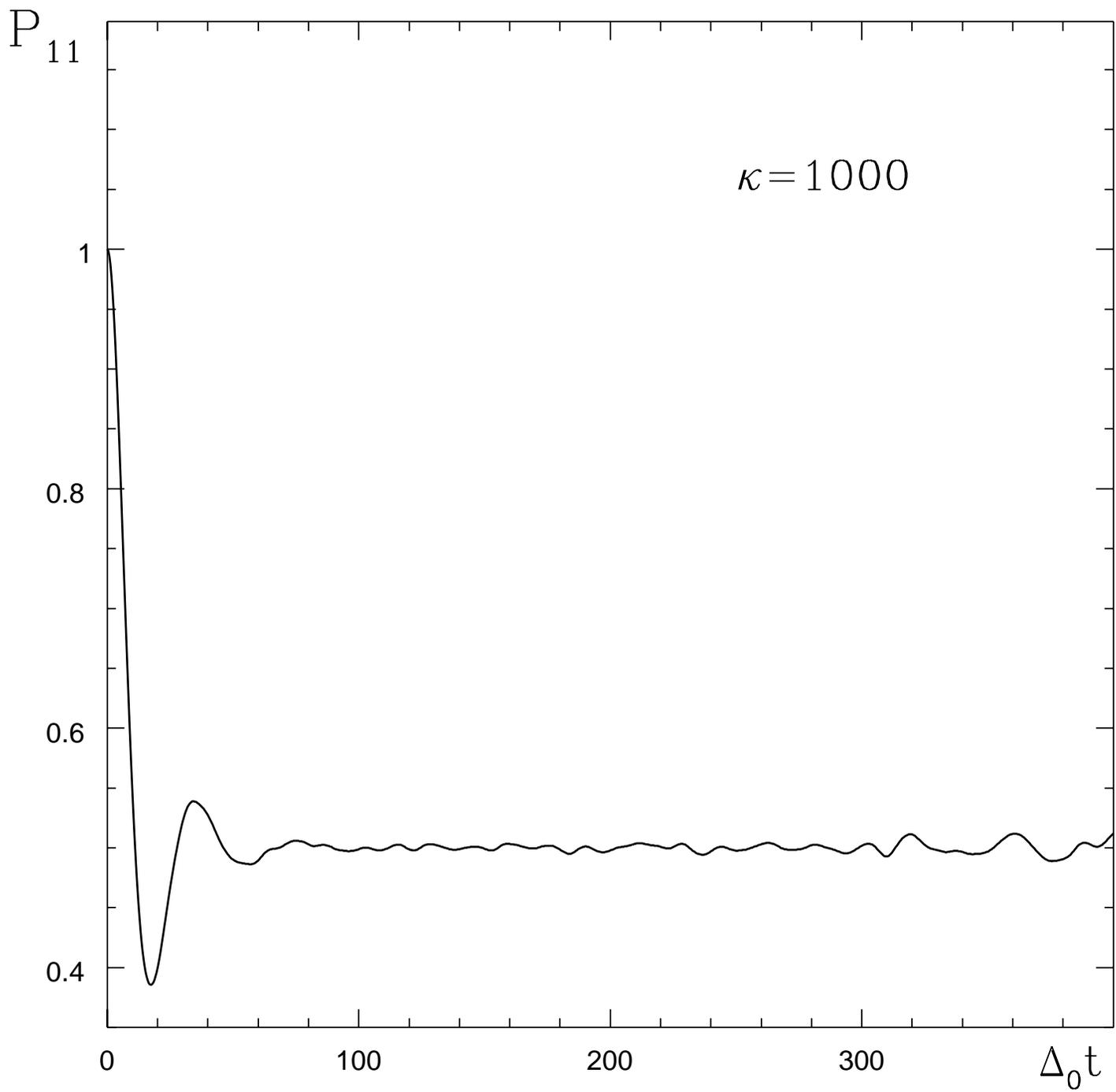
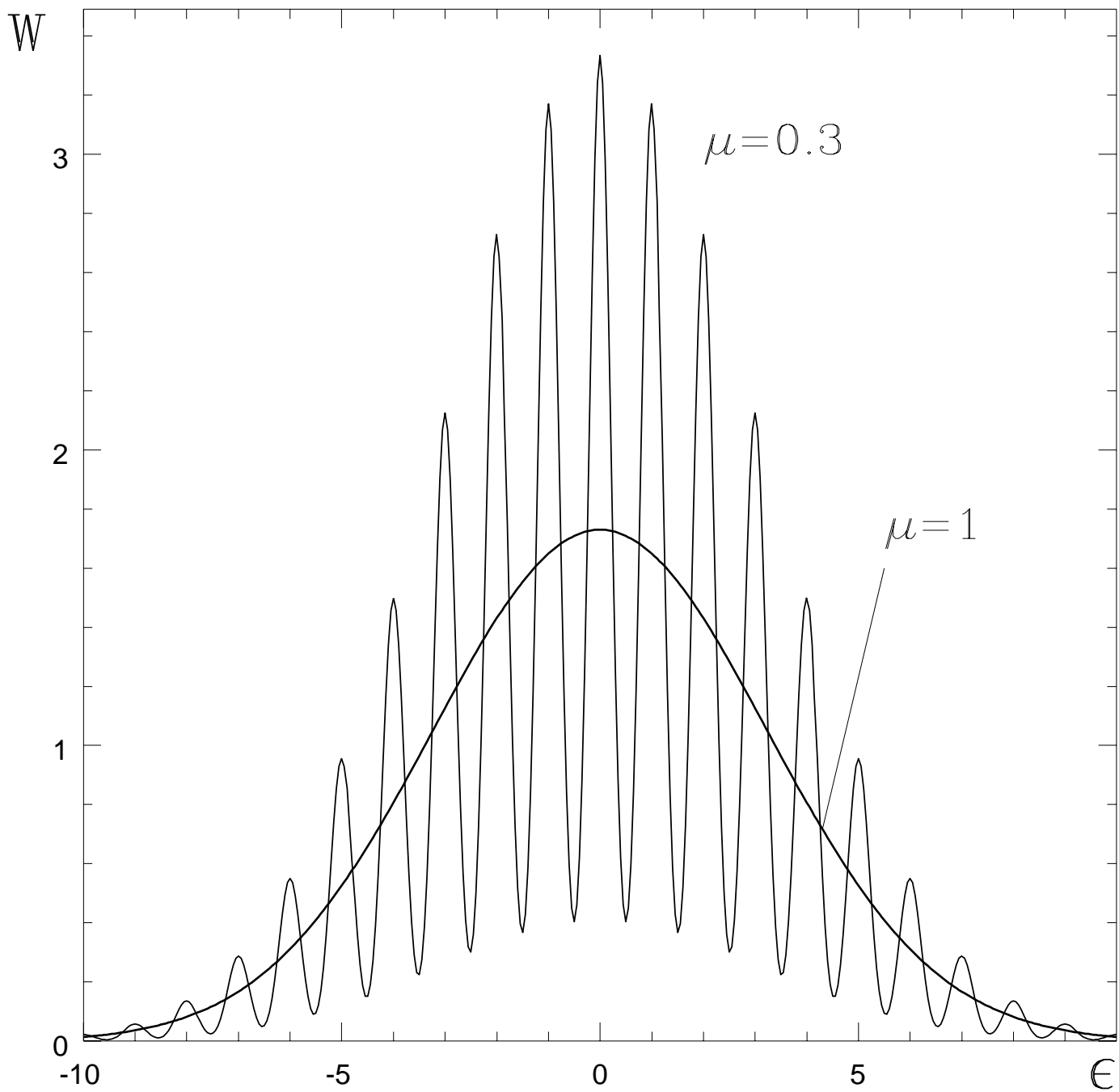
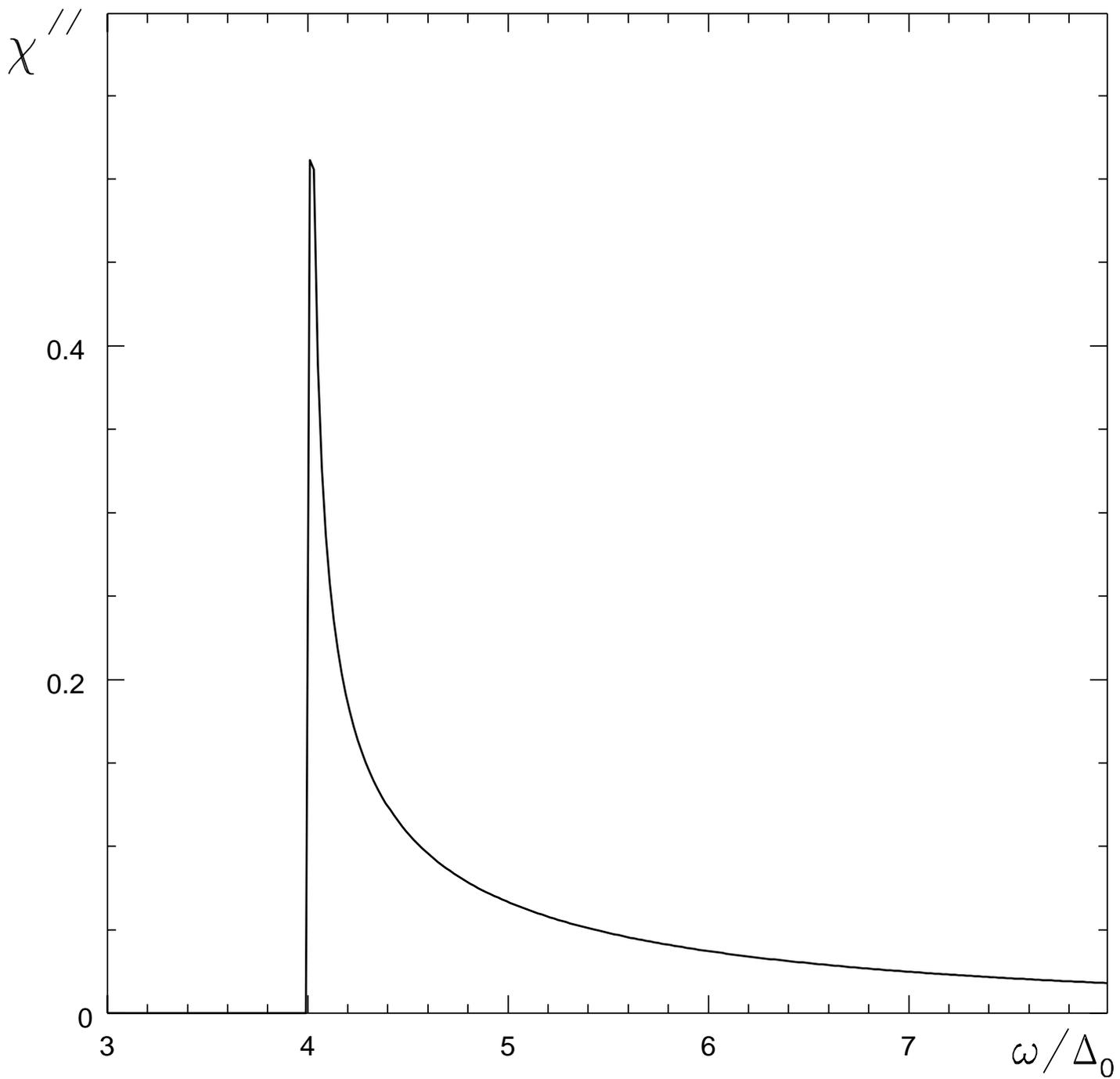
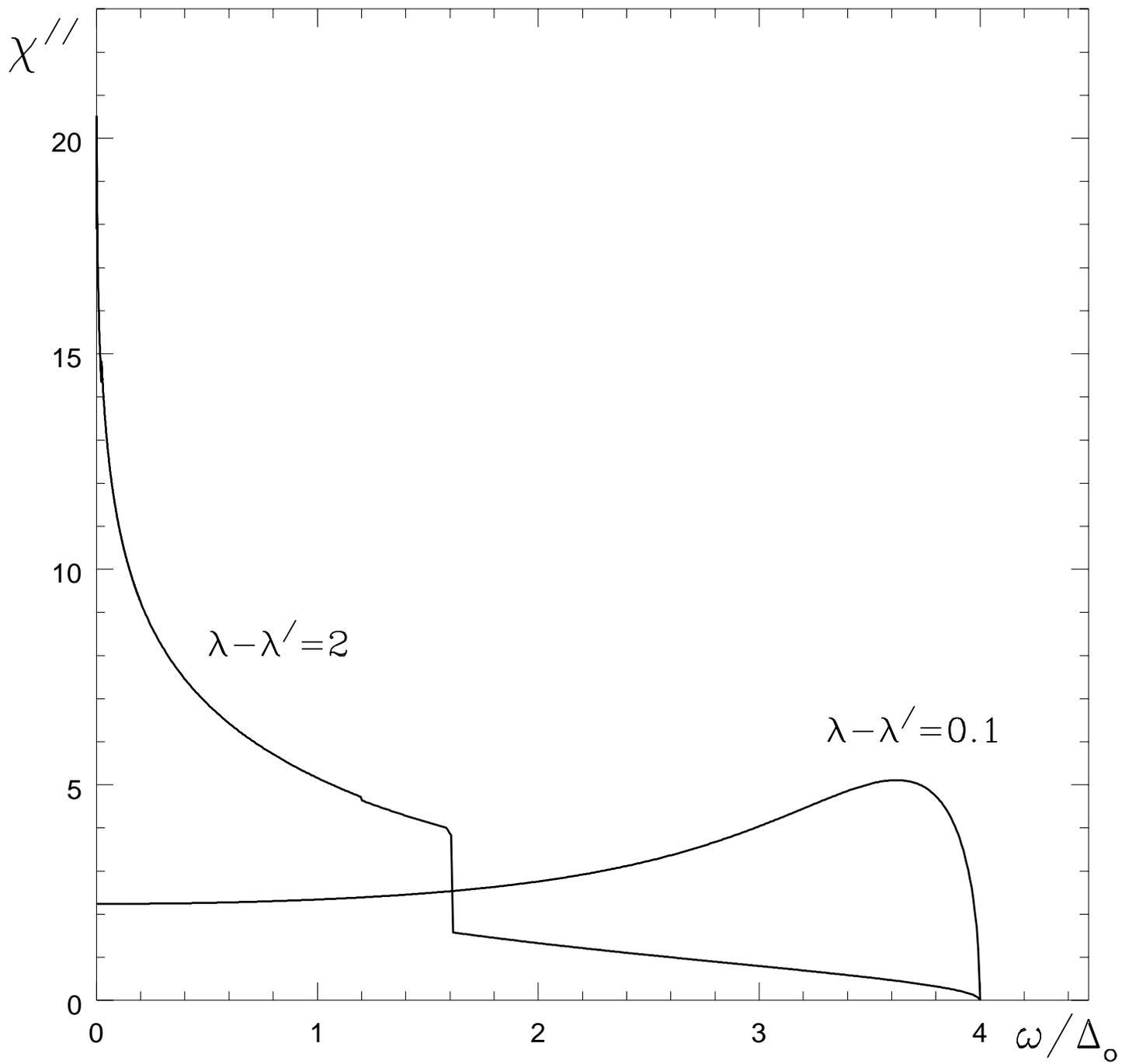
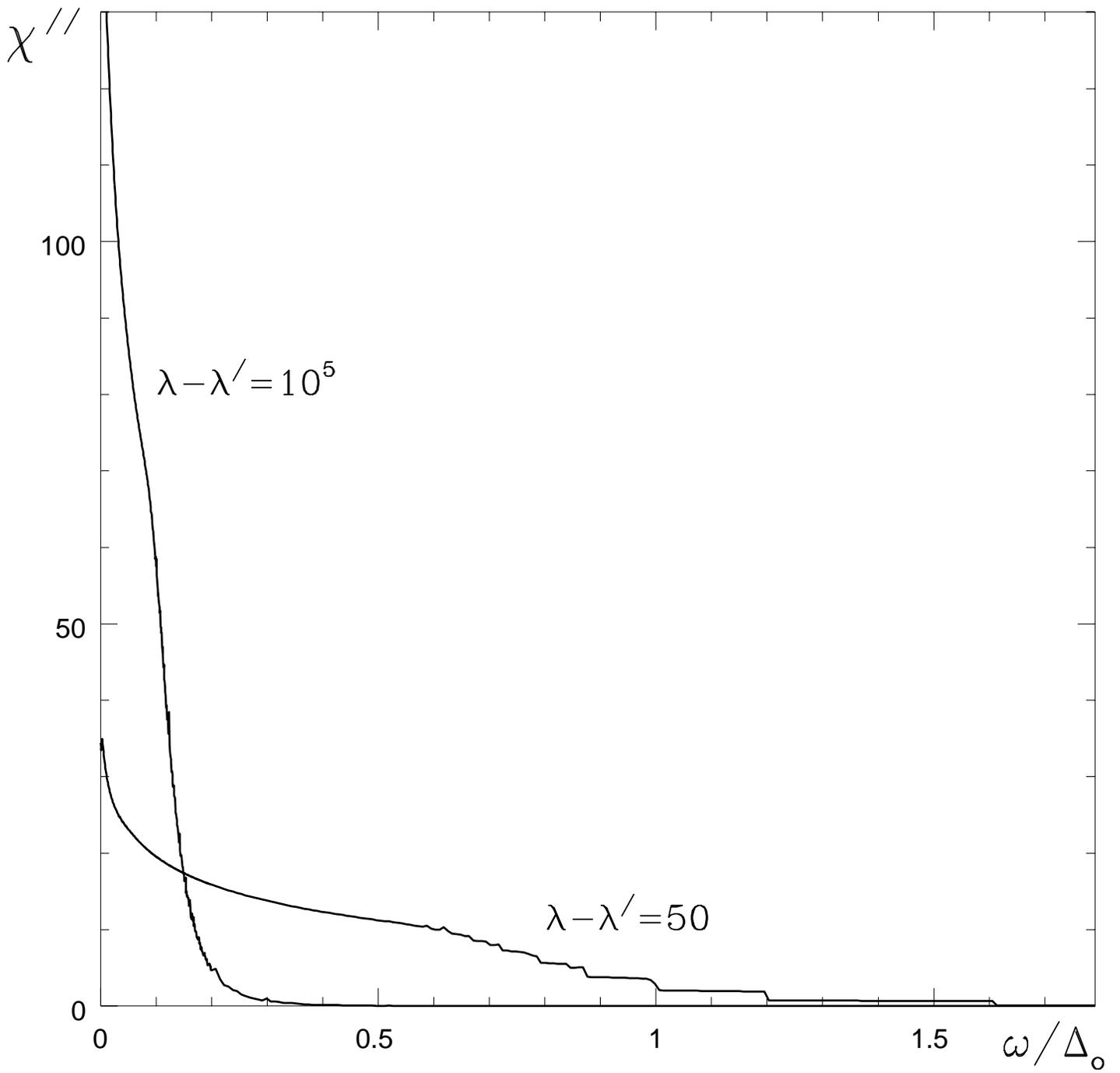
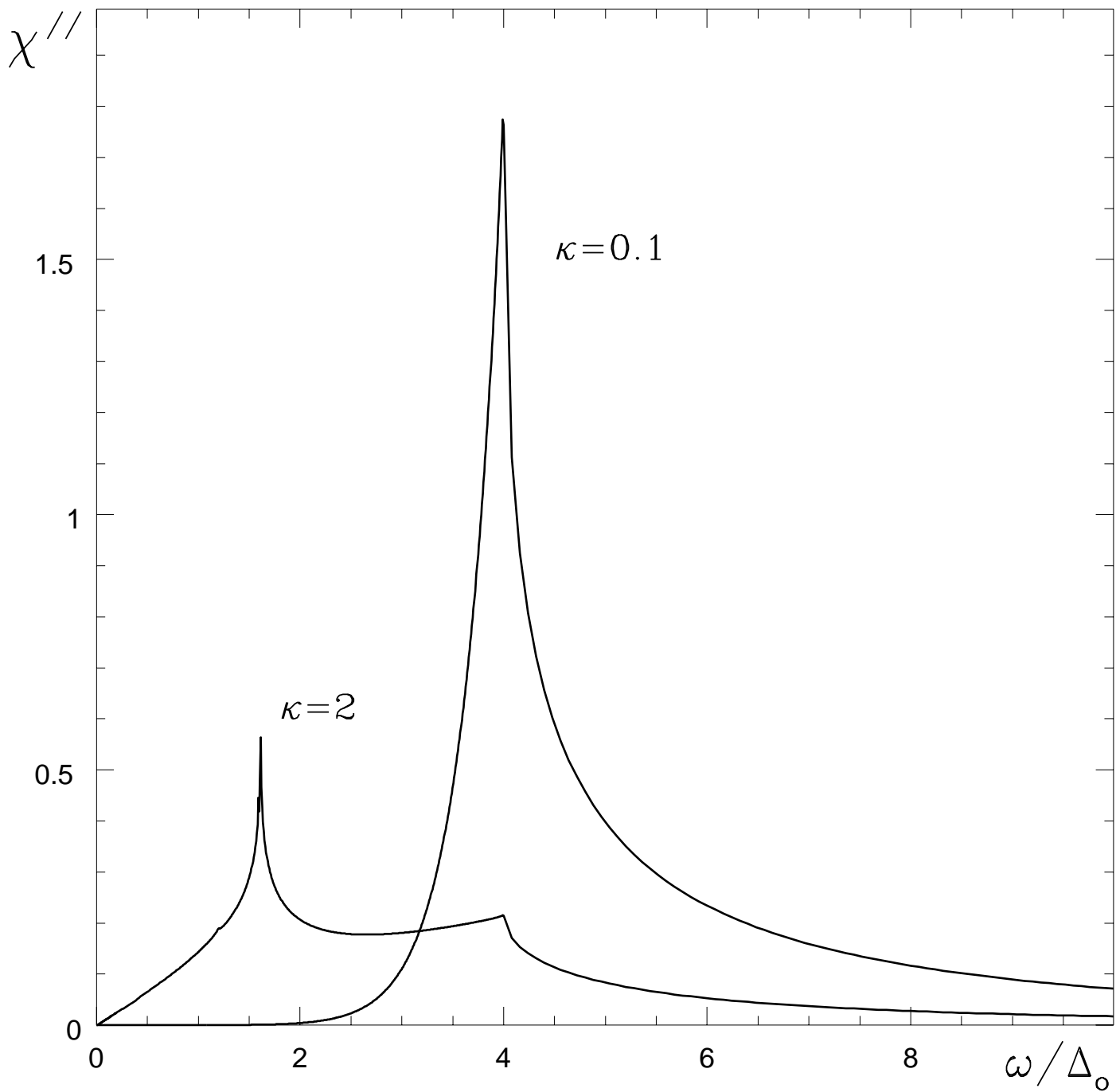
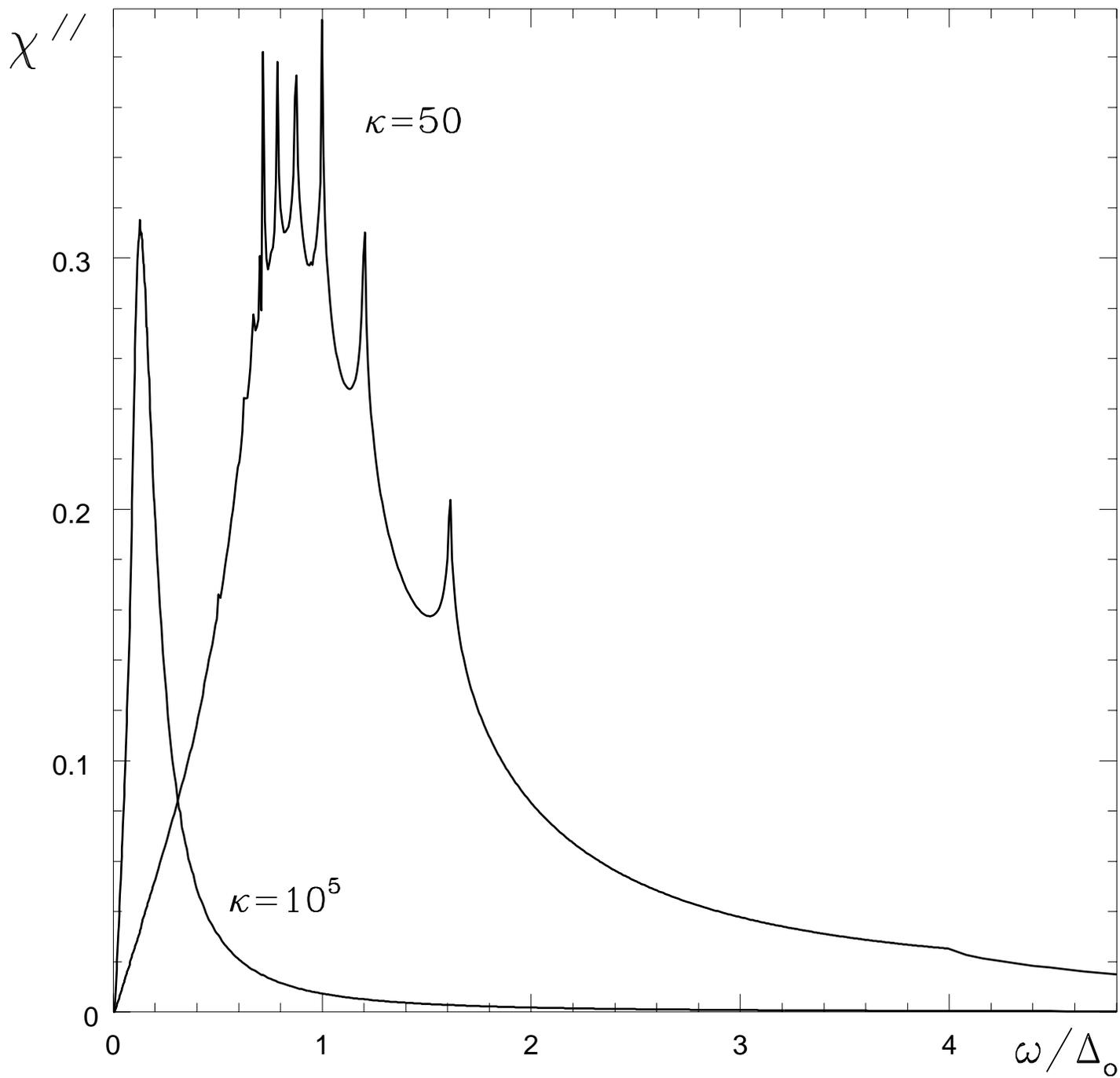
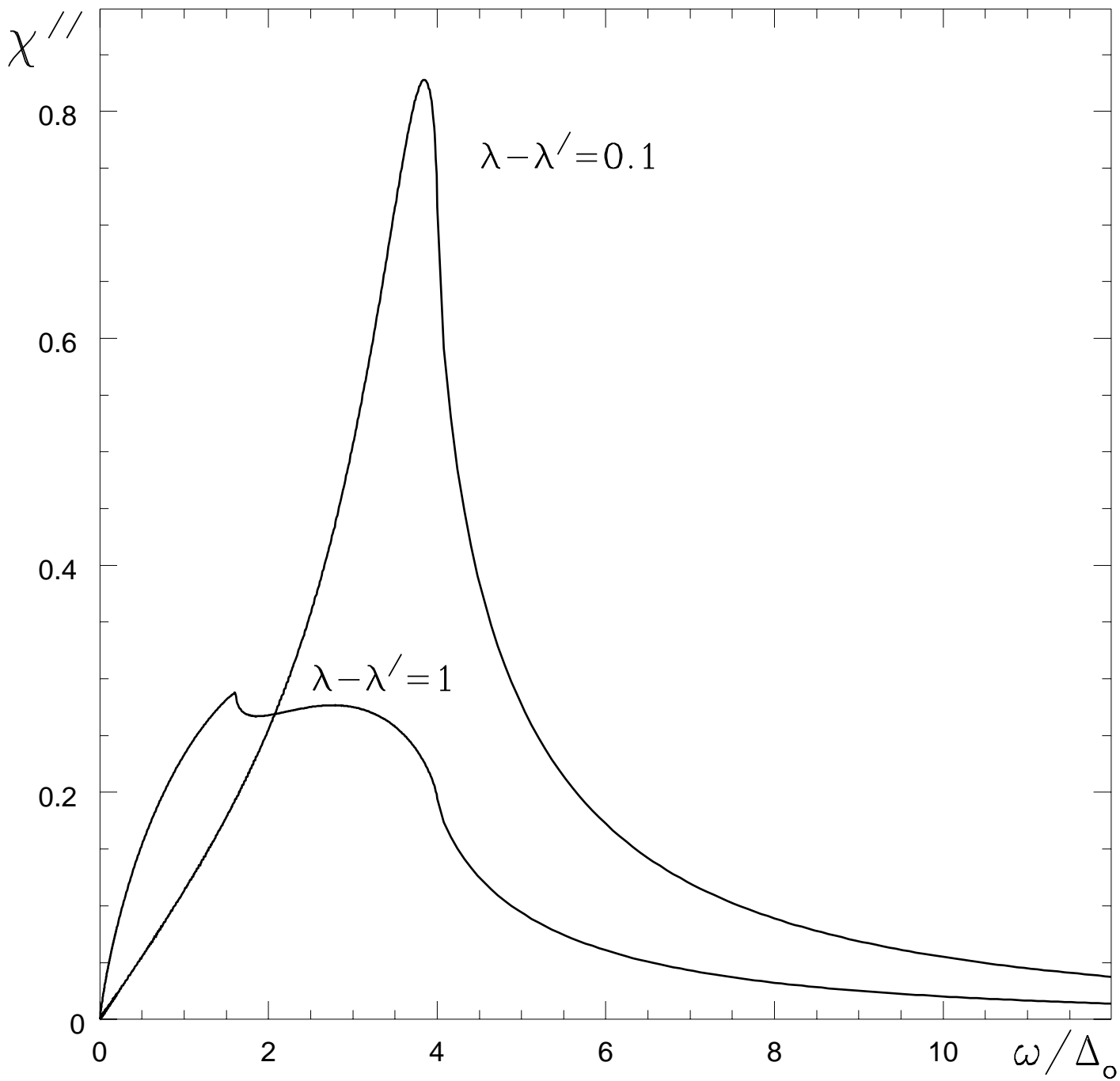
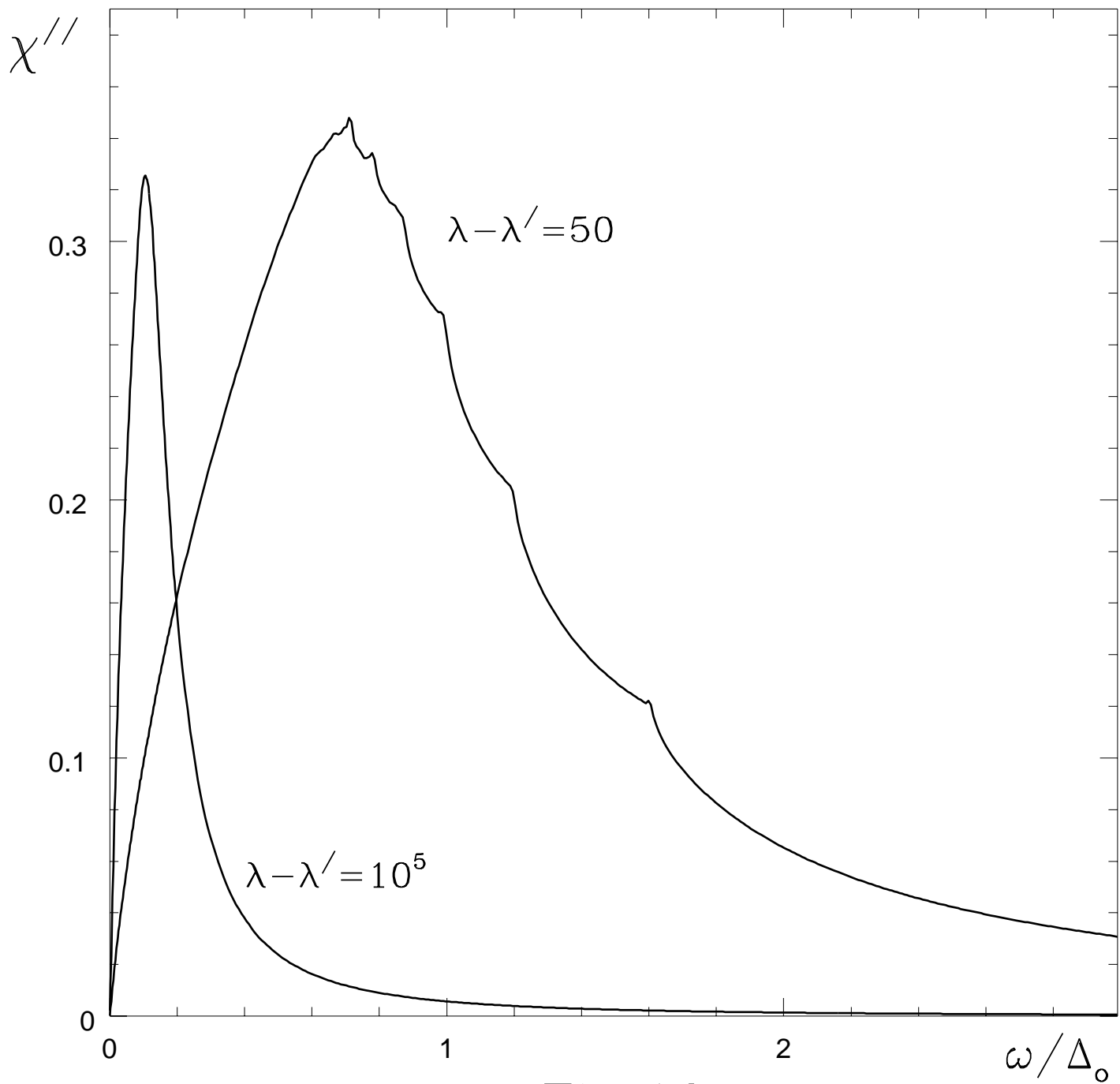
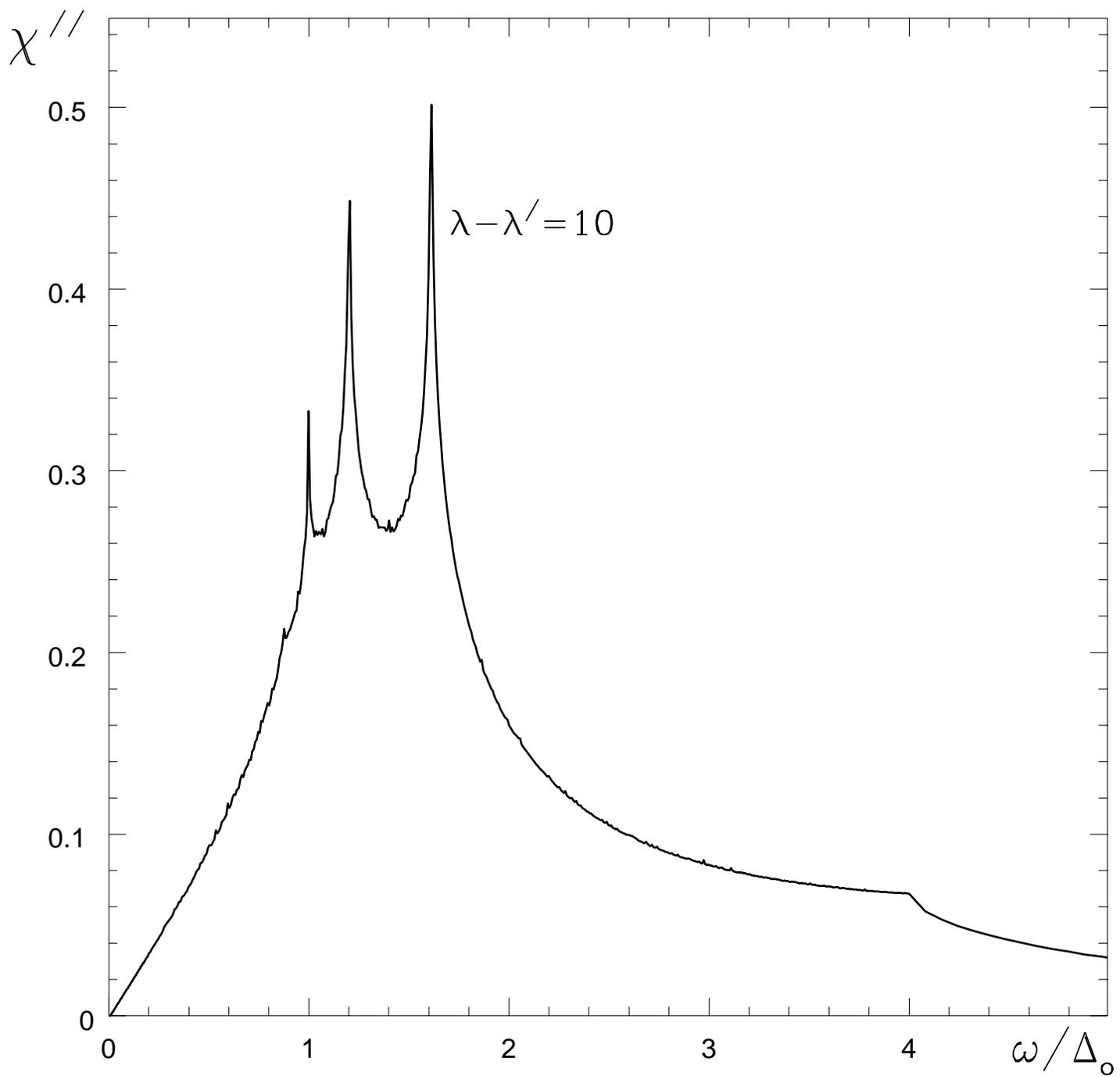
\end{document}